\documentclass[twocolumn]{aastex63}

\usepackage{natbib}
\usepackage{hyperref}
\received{XXX}
\revised{XXX}
\accepted{XXX}
\submitjournal{ApJ}

\shorttitle{low-mass companions of CEMP stars}
\shortauthors{Shejeelammal and Goswami}
\graphicspath{{./}{figures}}

\begin{document}

\title{Probing the nucleosynthetic contribution of low-metallicity, low-mass star companions of CEMP stars \footnote{Based on data collected using HCT/HESP and SUBARU/HDS}}

\correspondingauthor{Aruna Goswami}
\email{aruna@iiap.res.in}

\author[0000-0002-6234-4226]{J. Shejeelammal}
\affiliation{Indian Institute of Astrophysics, Koramangala, Bangalore 560034, India}

\author[0000-0002-8841-8641]{Aruna Goswami}
\affiliation{Indian Institute of Astrophysics, Koramangala, Bangalore 560034, India}

\begin{abstract}
The observed abundance diversities among the CEMP stars can shed light 
on the formation and evolution of elements in the early Galaxy. In this work, 
we present  results obtained from a detailed abundance  analysis of a sample of seven 
extrinsic carbon stars. The analysis is  based on high-resolution spectra obtained with
 HCT/HESP (R$\sim$60,000) and SUBARU/HDS (R$\sim$50,000).
We present, for the first time, the elemental abundance results  for the objects BD$-$19 132,
BD$-$19 290, HE~1304$-$2111, HE~1354$-$2257, and BD+19 3109. 
Abundances of a few elements are available in literature for 
HE~1157$-$0518 and HD~202851, we present an update on the 
abundances of these elements along with  new abundance estimates for 
several other elements. Our analysis confirms the  object HD~202851 to be 
a CH star. While BD$-$19 132, HE~1354$-$2257, and BD+19 3109 are found to be 
CEMP-s stars, the objects  BD$-$19 290, HE~1157$-$0518, and HE~1304$-$2111 are found to 
belong to  CEMP-r/s group. The observed abundance patterns  of the three  CEMP-r/s stars 
are  well reproduced with the i-process model predictions. While the objects BD+19 3109 and 
HD~202851 are confirmed binaries, the binary status of the remaining objects are not known.
Analysis based on different elemental abundance ratios confirms low-mass 
former AGB companions for all the objects. Kinematic analysis shows that BD$-$19 290, 
HE~1157$-$0518, HE~1354$-$2257, and BD+19 3109 belong to the Galactic halo, whereas  
BD$-$19 132, HE~1304$-$2111, and HD~202851 are members of Galactic thin disk.
\end{abstract}
 
 \keywords{stars: Abundances  \,-\, stars: chemically peculiar \,-\, 
stars: carbon  \,-\, stars: individual}
 
\section{Introduction}
Being enriched with carbon (and neutron-capture elements), the 
atmospheres of the low-mass, metal-poor stars in the Galaxy 
bear the fingerprints of the chemical evolution history of the Galaxy.
They are the window to the early Galaxy, as they preserve in their atmosphere 
the chemical imprints of the gas clouds from which they were formed. 
For this reason, metal-poor stars such as CH stars and Carbon-Enhanced 
Metal-Poor (CEMP) stars, have been  extensively studied in the last 
few decades for the information they  provide on the origin and evolution 
of elements in the early Galaxy.
A number of large sky survey programs to identify the most metal-poor 
stars were conducted in the past such as HK survey \citep{Beers_1985, Beers_1992, Beers_2007, Beers_1999}, 
Hamburg/ESO Survey (HES;  \citealt{Christlieb_2001a, Christlieb_2001b, Christlieb_2003, Christlieb_2008}), 
Sloan Digital Sky Survey (SDSS; \citealt{York_2000}), and  LAMOST 
(Large Sky Area Multi-Object Fiber Spectroscopic Telescope) survey 
\citep{Cui_2012, Deng_2012, Zhao_2012}. These surveys have 
revealed that a significant fraction ($\sim$20\%) of Very Metal-Poor 
([Fe/H]$<$$-$2) stars in the Galaxy are CEMP stars 
\citep{Rossi_1999, Christlieb_2003, Lucatello_2006, Carollo_2012}. 
The fraction of carbon-enhanced stars increases with decreasing metallicity; 
$\sim$40\% for [Fe/H]$<$$-$3 and $\sim$75\% for [Fe/H]$<$$-$4 
\citep{Aoki_2013, Lee_2013, Yong_2013b, Placco_2014, Frebel_2015}. 

\par CH stars ($-$2$\leq$[Fe/H]$\leq$$-$0.2) are characterized by the strong CH and C$_{2}$ molecular bands, 
C/O$>$1 and strong features due to the neutron-capture elements \citep{Keenan_1942}. 
CEMP stars are more metal-poor counterparts of CH stars ([Fe/H]$<$$-$1; \citealt{Lucatello_2005, Abate_2016})
which has been traditionally defined as the stars with [C/Fe]$>$1 \citep{Beers_2005}.
However, the definition of CEMP stars is being revised and [C/Fe]$\geq$0.70 \citep{Aoki_2007, Carollo_2012, Lee_2013, Norris_2013a, Skuladottir_2015} and [C/Fe]$>$0.90 \citep{Jonsell_2006, Masseron_2010} are also used.    

\par Based on the level of relative enrichment of neutron-capture elements, CEMP stars 
are classified into different sub-classes; CEMP-s (show enhanced abundances of 
s-process elements), CEMP-r (show strong enhancement of r-process elements),
CEMP-r/s (show simultaneous enhancement of both s- and r-process elements) 
and CEMP-no (does not show any enhanced abundance of neutron-capture elements).
This diverse abundance pattern observed in the CEMP stars points to different formation scenarios. The CEMP-s stars are considered as the metal-poor 
analog of Ba and CH stars, and the binary mass transfer from an evolved low-mass AGB companion is the most accepted  scenario for their abundance peculiarity \citep{Herwig_2005}. The long-term radial velocity monitoring studies have shown that most of the CEMP-s stars are binaries 
\citep{Lucatello_2005, Starkenburg_2014, Jorissen_2016, Hansen_2016c}, supporting the pollution from companion AGB. 
Comparison of  observed abundances in the CEMP-s stars with theoretical 
model predictions  also confirms the binary mass-transfer from the 
AGB companion \citep{Bisterzo_2011, Placco_2013, Hollek_2015}. 

\par There exist several proposed scenarios for the abundance pattern in CEMP-r/s stars that show enhancement in both s- and r-process 
elements that are ascribed to  different astrophysical sites.
The proposed scenarios for CEMP-r/s stars include; binary system formed out of r-enriched ISM, pollution of binary system from a massive tertiary, pollution from the binary companion exploded as Type 1.5 SNe (Accretion Induced Collapse, AIC) etc. (\citealt{Jonsell_2006} and references therein).  
However, none of  these proposed scenarios is able to reproduce the observed abundance trend and frequency of CEMP-r/s stars \citep{Abate_2016}.
Since most of the CEMP-r/s stars are also found to be binaries just like CEMP-s stars, binary mass transfer from the 
AGB companion is thought to be the reason for their origin as well, however the presence of 
r-process component posed a challenge \citep{Jonsell_2006, Herwig_2011, Abate_2016}.
The i-process, intermediate neutron-capture process, with neutron density between 
s- and r- process can produce both s- and r- process elements at a single stellar site \citep{Cowan_1977}. Many studies in literature have 
successfully used  model yields of i-process in low-mass low-metallicity AGB stars to account for the observed abundance patterns in CEMP-r/s stars \citep{Hampel_2016, Hampel_2019, Goswami_2021, Shejeelammal_2021}. Although there are  several suggestions for the i-process sites
(super-massive AGB stars \citep{Doherty_2015, Jones_2016}, low-mass low-meatllicity stars
\citep{Campbell_2008, Campbell_2010, Cruz_2013, Cristallo_2016}, 
Rapidly Accreting White Dwarfs \citep{Herwig_2014, Denissenkov_2017} etc.)
the exact astrophysical site for the i-process is not yet confirmed \citep{Frebel_2018, Koch_2019}.  

\par The origins  of CEMP-no and CEMP-r stars are also not clearly understood. 
CEMP-r stars are rare among all the CEMP sub-classes. Analysis of \cite{Hansen_2011} and \cite{Hansen_2015} 
have shown that the abundance peculiarity of CEMP-r stars is not resulted from a binary mass 
transfer, instead due to the enrichment of their birth cloud from other external sources. Various suggested progenitors of 
the CEMP-r stars that polluted the ISM with r-elements are core-collapse SNe \citep{Qian_2000, Argast_2004, Arcones_2013}, 
fallback SNe \citep{Fryer_2006}, neutron star mergers \citep{Tanvir_2013, Rosswog_2014, Drout_2017, Lippuner_2017} or nuetron star - black hole mergers \citep{Surman_2008}. The observations of more such stars are required to constrain their 
exact origin. Similar to CEMP-r stars, the observed carbon enhancement in CEMP-no stars is also believed to be from the pre-enriched ISM \citep{Norris_2013b, Cooke_2014, Frebel_2015}. The suggested progenitors of CEMP-no stars include faint SNe \citep{Umeda_2005, Nomoto_2013, Tominaga_2014}, spinstar \citep{Meynet_2010, Chiappini_2013}, metal-free massive stars \citep{Heger_2010}. A recent study by 
\cite{Arentsen_2019b} have shown that the binary mass-transfer from an extremely metal-poor AGB companion could also be a possible 
progenitor of the CEMP-no stars.

\par In this study, we have carried out a detailed spectroscopic analysis of seven carbon stars selected from 
various sources in literature, in order to understand their formation history
from a detailed abundance analysis. The structure of this paper is as follows. 
Section 2. describes the stellar sample, source of the spectra and the data reduction.
The derivation of stellar atmospheric parameter, radial velocity estimation and stellar mass determination 
are discussed in Section 3. The details of abundance determination is described in Section 4 and 
a discussion on abundance uncertainties is presented in Section 5. Section 6. provides the details of kinematic analysis. 
In Section 7, the classification of program stars based on the observed abundances is presented.
Section 8 presents a discussion on various elemental abundance ratios and their interpretations. 
A discussion on the binary status of the program stars and on the individual stars along with the 
parametric model based analysis are also given in this section. Finally, conclusions are drawn in Section 9.

\section{STELLAR SAMPLE: SELECTION, OBSERVATION/DATA ACQUISITION AND DATA REDUCTION}
\begin{figure}
\centering
\includegraphics[width=\columnwidth]{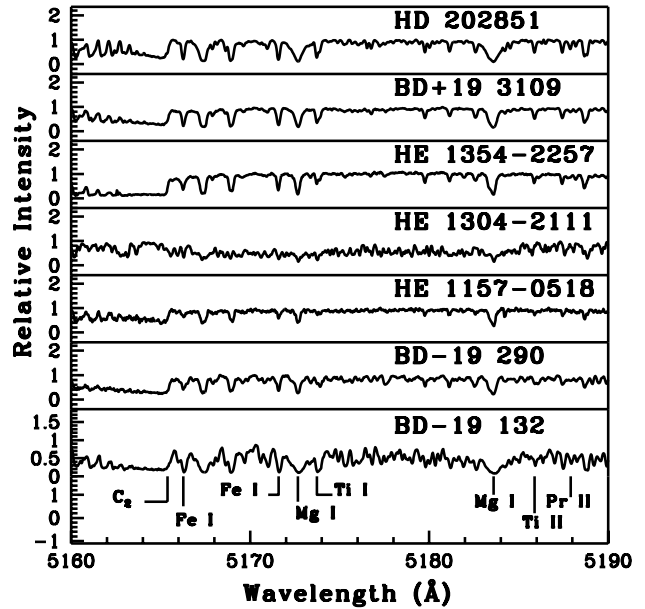}
\caption{ Sample spectra of the program stars in the wavelength region 
5160 to 5190 {\bf  {\rm \AA}}.}\label{sample spectra}
\end{figure}

Out of the total seven objects analyzed in this study, four objects,
BD$-$19 132, BD$-$19 290, BD+19 3109, and HD~202851 are selected from 
the catalog of CH star by \cite{Bartkevicius_1996}. The other three objects, 
HE~1157$-$0518, HE~1304$-$2111, and HE~1354$-$2257 are selected from the
candidate metal-poor stars identified in Hamburg/ESO survey (HES)
\citep{Christlieb_2003} and in the HK survey of \cite{Beers_1992}.
The stars HE~1157$-$0518 and HE~1354$-$2257 are also listed in
the catalog of carbon stars identified from Hamburg/ESO survey (HES)
by \cite{Christlieb_2001a}. The spectra of the objects 
BD$-$19 132, BD$-$19 290, BD+19 3109, and HD~202851 at a resolution 
($\lambda/\delta\lambda$)$\sim$ 60,000 are obtained 
using high-resolution Hanle Echelle SPectrograph (HESP)  
attached to the 2m Himalayan Chandra Telescope 
(HCT) operated at the Indian Astronomical  Observatory, Hanle.
For all these four objects, we have taken three frames each with 
an exposure of 2700 sec and then co-added to improve the quality 
(S/N ratio) of the resulting spectra. This S/N ratio enhanced spectra 
are then used for further analysis. The wavelength span of HCT/HESP spectra 
is 3530 - 9970 {\rm \AA}. The spectra of the objects 
HE~1157$-$0518, HE~1304$-$2111, and HE~1354$-$2257 at a resolution 
($\lambda/\delta\lambda$)$\sim$ 50,000 were taken from 
the SUBARU/HDS archive (\url{http://jvo.nao.ac.jp/portal/v2/}). 
The High Dispersion Spectrograph (HDS) is an echelle spectrograph
of 8.2m SUBARU telescope \citep{Noguchi_2002} at Hawaii, 
operated by National Astronomical Observatory of Japan (NAOJ). 
The SUBARU spectra covers 4100 - 6850 {\rm \AA} 
in wavelength with a wavelength gap between 5440 and 5520 {\rm \AA}
due to the separation between the two CCDs used. 
The data  is further reduced using  
IRAF\footnote{IRAF (Image Reduction and Analysis Facility) 
is distributed by the National Optical Astronomical 
Observatories, which is operated by the Association for Universities 
for Research in Astronomy, Inc., under contract to the National 
Science Foundation} software. The basic information of the 
program stars are given in Table \ref{basic data of program stars} and
some sample spectra of the program stars are shown in Figure \ref{sample spectra}.

 {\footnotesize
\begin{table*}
\caption{\textbf{Primary information of the program stars.}\label{basic data of program stars}}
\resizebox{\textwidth}{!}{
\begin{tabular}{lccccccccccccc}
\hline
Star      &RA$(2000)$       &Dec.$(2000)$    &B       &V       &J        &H        &K     &Exposure       &Date of obs.  & Source       &  & S/N & \\
          &                 &                &        &        &         &         &      &(seconds)      &              & of spectrum  & 4200 \AA &  5500 \AA & 7700 \AA \\
\hline
BD$-$19 132  & 00 50 24.16     & $-$19 04 40.19   & 13.25    & 10.73   & 8.083    & 7.354    & 7.070 & 2700(3)            & 07/11/2017    & HESP  & 11.37 & 30.71 & 40.93  \\      
BD$-$19 290  & 01 40 34.11     & $-$18 56 51.52   & 12.41    & 11.07   & 8.710    & 8.102    & 7.914 & 2700(3)            & 08/11/2017    & HESP  & 6.41 & 23.59 & 47.07  \\   
HE~1157$-$0518  & 12 00 08.06     & $-$05 34 43.12   & 16.273    & 15.120    & 13.418    & 12.917    & 12.846 & 1800            & 25/05/2003    & SUBARU  & 9.32 & 25.15 & -  \\
HE~1304$-$2111  & 13 07 27.26     & $-$21 27 34.24   & 14.547   & 13.061    & 9.651    & 9.009    & 8.764 & 1200            & 25/05/2003    & SUBARU  & 5.21 & 21.53 & -  \\
HE~1354$-$2257  & 13 57 43.30     & $-$23 12 34.55   & 15.22   & 14.00    & 11.809    & 11.290    & 11.110 & 1800            & 26/05/2003    & SUBARU  & 6.39 & 35.56 & -  \\
BD+19 3109  & 16 29 26.99     & +19 30 34.46   & 11.88   & 10.29   & 8.051    & 7.40    & 7.253 & 2700(3)            & 23/05/2018    & HESP  & 7.80 & 48.27 & 95.90  \\  
HD~202851  & 21 18 43.49     & $-$01 32 03.34   & 10.89    & 9.67    & 7.711    & 7.174    & 7.029 & 2700(3)            & 07/11/2017    & HESP  & 17.01 & 31.82 & 54.57  \\ 
\hline

\end{tabular}}

The number of frames taken are given with exposure time, in parenthesis.
\end{table*}
}

\section{STELLAR ATMOSPHERIC PARAMETERS AND RADIAL VELOCITY}
The radial velocities of the program stars are calculated from the 
Doppler equation using the measured wavelength shift of a set of clean spectral 
lines of several elements. Our radial velocity estimates along with that 
from Gaia DR2 \citep{Gaia_2018} are presented in 
Table \ref{atmospheric parameters}. Radial velocity data are not available in literature for the stars  HE~1157$-$0518 and HE~1354$-$2257. 
While the objects BD$-$19 290, HE~1157$-$0518, HE~1354$-$2257 and
BD+19 3109 are found to be high velocity objects with V$\rm_{r}$ in the range $-$181.8 to +289.4 km s$^{-1}$, 
BD$-$19 132, HE~1304$-$2111, and HD~202851 are low velocity objects with  
V$\rm_{r}$ in the range 2.1 to +22.2 km s$^{-1}$.
The objects BD+19 3109 and HD~202851 are confirmed binaries 
with periods of 2129$\pm$13 and 1295$\pm$6 days respectively \citep{Sperauskas_2016}.
The estimated radial velocities of these two objects differ by 
$\sim$ 5 km s$^{-1}$ from  the literature values available in Gaia archive.
For the objects BD$-$19 132, BD$-$19 290, and HE~1304$-$2111 
our estimates show a difference of $\sim$ 13, 6, and 17 km s$^{-1}$
respectively from those listed in Gaia archive. This may be an indication that 
these stars could possibly be binaries.

\begin{figure}
\centering
\includegraphics[width=\columnwidth]{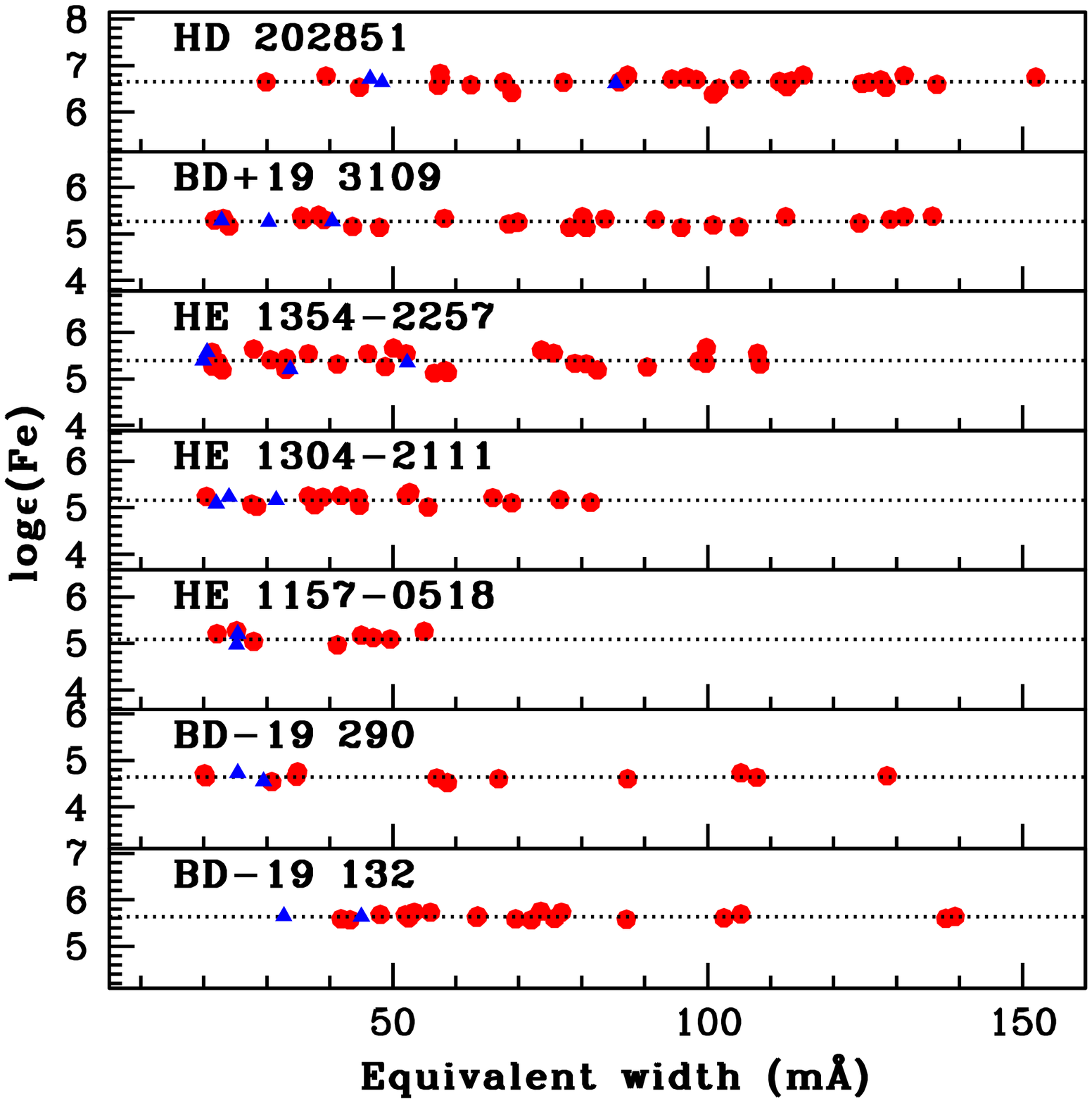}
\includegraphics[width=\columnwidth]{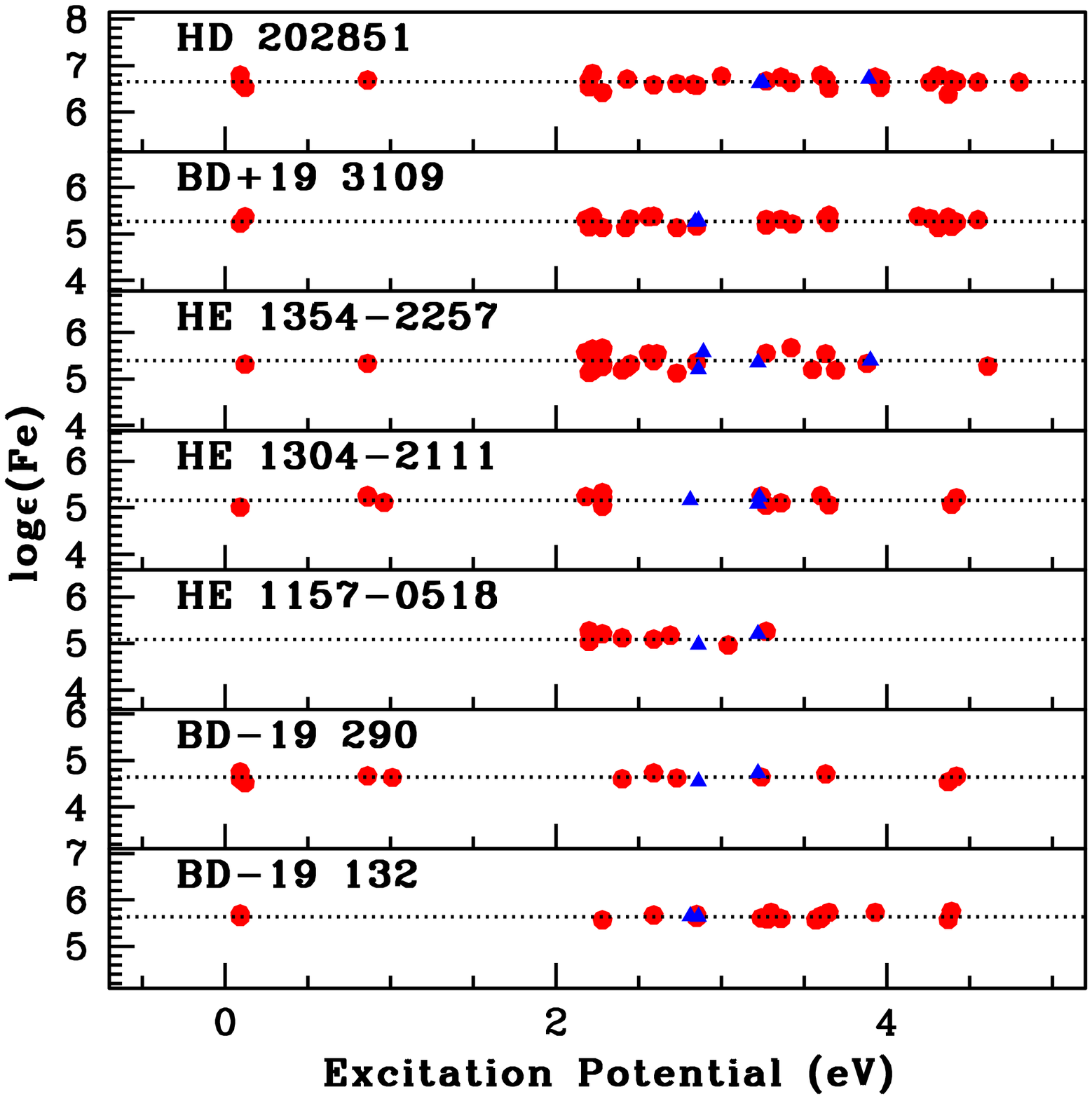}
\caption{Iron abundances of the program stars derived from individual Fe I and Fe II 
lines as function of excitation potential (lower panel),
and equivalent width (upper panel). The dotted lines correspond to the derived and adopted Fe abundance for each star.
Solid circles correspond to Fe I and solid 
triangles correspond to Fe II lines. } \label{ep_ew}
\end{figure}

The  procedure adopted for deriving  the atmospheric parameters of the program stars is 
explained in detail in our previous papers \cite{Shejeelammal_2020, Shejeelammal_2021}. 
Here we briefly mention a few relevant points. 
The equivalent width measured from a set of clean Fe I and Fe II lines are 
used for the derivation of the stellar atmospheric parameters. 
The lines are selected such that they have equivalent width and excitation potential
in the range 8 - 180 m{\rm \AA} and 0 - 6 eV respectively. We have used 
the most recent version of the radiative transfer code MOOG by Sneden \citep{Sneden_1973}
employing Local Thermodynamic Equilibrium (LTE) for our analysis.
The photometric temperatures of the program stars are calculated using the 
color-temperature calibration equations of \cite{Alonso_1999, Alonso_2001} and 
used as initial guess of effective temperatures for model calculation. With this temperature estimate
and a guess of log g value typical of giants, the initial model atmospheres are selected from the 
Kurucz grid of model atmosphere with no convective overshooting (\url{http://kurucz.harvard.edu/grids.html}). 
The excitation balance of Fe I lines is employed to fix the final effective temperature. 
The zero slope between the measured equivalent widths and the abundances of Fe I lines gives the 
microturbulent velocity. Finally, the ionization balance between the Fe I and Fe II 
gives the surface gravity. The final model atmosphere is adopted from the initial 
one through an iterative process until these three conditions are satisfied simultaneously.
Once this is achieved, [Fe/H] is estimated from the derived  iron abundance. 
The abundances derived from Fe I and Fe II lines as functions of excitation potentials 
and equivalent widths are shown in Figure \ref{ep_ew}. The derived atmospheric parameters
along with the available literature values of the program stars are 
given in Table \ref{atmospheric parameters}.

{\footnotesize
\begin{table*}
\caption{Derived atmospheric parameters of the program stars.} \label{atmospheric parameters}
\begin{tabular}{lcccccccccc}
\hline
Star                &T$\rm_{eff}$  & log g      &$\zeta$         & [Fe I/H]          &[Fe II/H]          & V$_{r}$             &  V$_{r}$    & Reference \\
                    &    (K)       & cgs        &(km s$^{-1}$)   &                   &                   & (km s$^{-1}$)       & (km s$^{-1}$)  &  \\
                    & $\pm$100     & $\pm$0.2   & $\pm$0.2       &                   &                   & (This work)         & (Gaia)           &    \\  
\hline
BD$-$19 132          & 4005      & 1.13  & 2.02           & $-$1.86$\pm$0.06  & $-$1.86$\pm$0.01  & +3.94$\pm$0.81      & +16.97$\pm$0.79    & 1 \\
BD$-$19 290          & 4315      & 0.61  & 3.10           & $-$2.86$\pm$0.07  & $-$2.86$\pm$0.13  & +106.28$\pm$0.90    & +111.99$\pm$0.73    & 1\\
HE~1157$-$0518      & 5050      & 2.52  & 2.15           & $-$2.42$\pm$0.15  & $-$2.42$\pm$0.16  & +116.02$\pm$0.70    & -             & 1  \\
                    & 4900      & 2.00  & --             & $-$2.40           & -                 & -                   & -             & 2  \\
HE~1304$-$2111      & 4325      & 1.06  & 0.63           & $-$2.34$\pm$0.10  & $-$2.34$\pm$0.07  & +2.14$\pm$0.34      & +19.33$\pm$0.59  & 1\\
HE~1354$-$2257      & 4700      & 1.13  & 1.86           & $-$2.11$\pm$0.17  & $-$2.11$\pm$0.15  & +289.46$\pm$1.80    & -                & 1 \\
BD+19 3109          & 4035      & 0.75  & 2.05           & $-$2.23$\pm$0.09  & $-$2.23$\pm$0.02  & $-$181.89$\pm$1.38  & $-$187.09$\pm$1.55 & 1\\
HD~202851           & 4900      & 2.20  & 1.54           & $-$0.85$\pm$0.11  & $-$0.85$\pm$0.05  & +22.22$\pm$0.90     & +17.38$\pm$1.22   & 1 \\
                    & 4800      & 2.10  & 1.50           & $-$0.70           & -                 & -                   & -                 & 3 \\
                    & 4733      & 1.60  & -              & $-$0.88           & -                 & -                   & -               & 4 \\
\hline
\end{tabular}

References: 1. This work, 2. \cite{Aoki_2007}, 3. \cite{Sperauskas_2016}, 
4. \cite{Arentsen_2019} \\
\end{table*}
}

\begin{figure}
\centering
\includegraphics[width=\columnwidth]{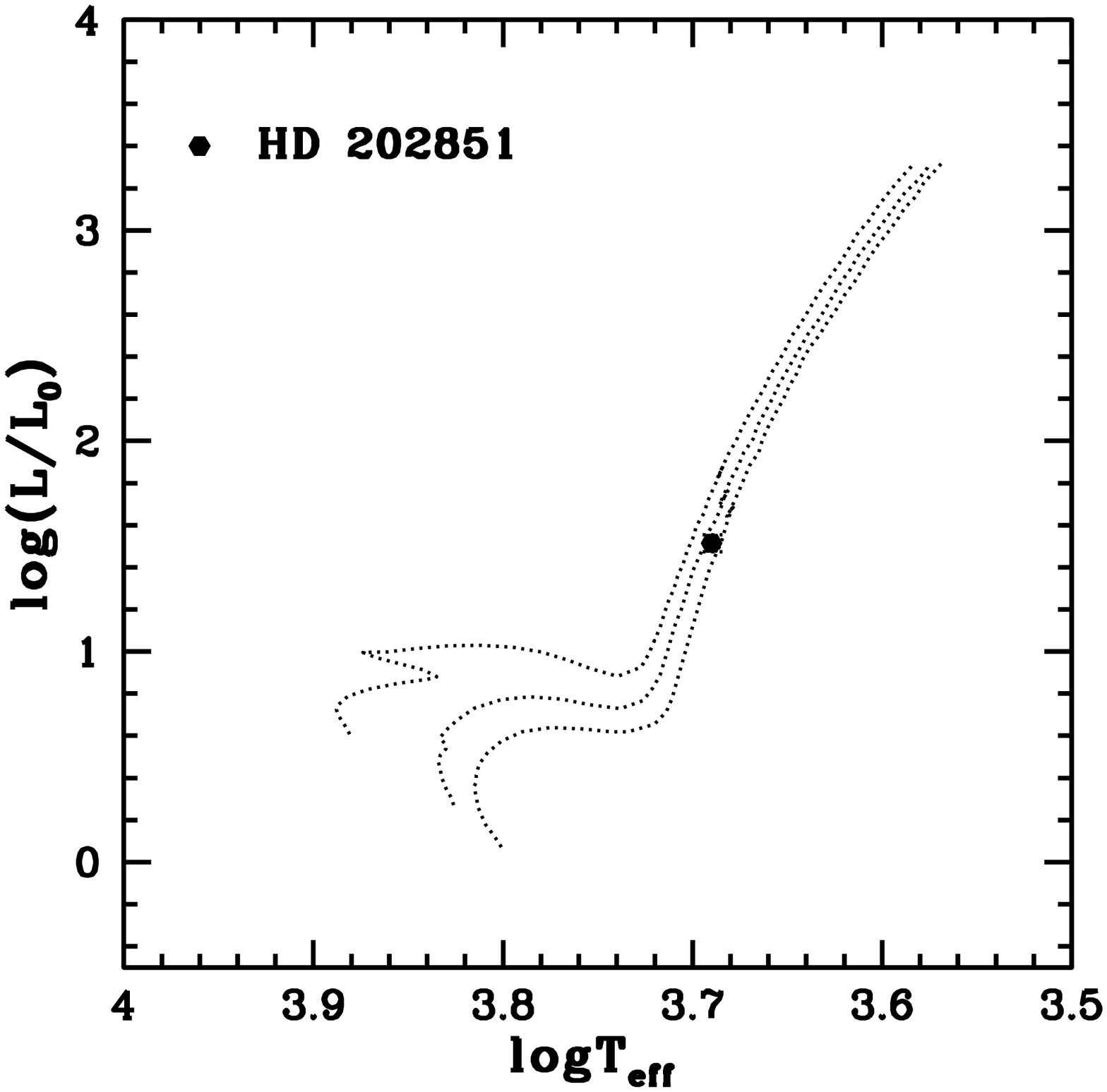}
\includegraphics[width=\columnwidth]{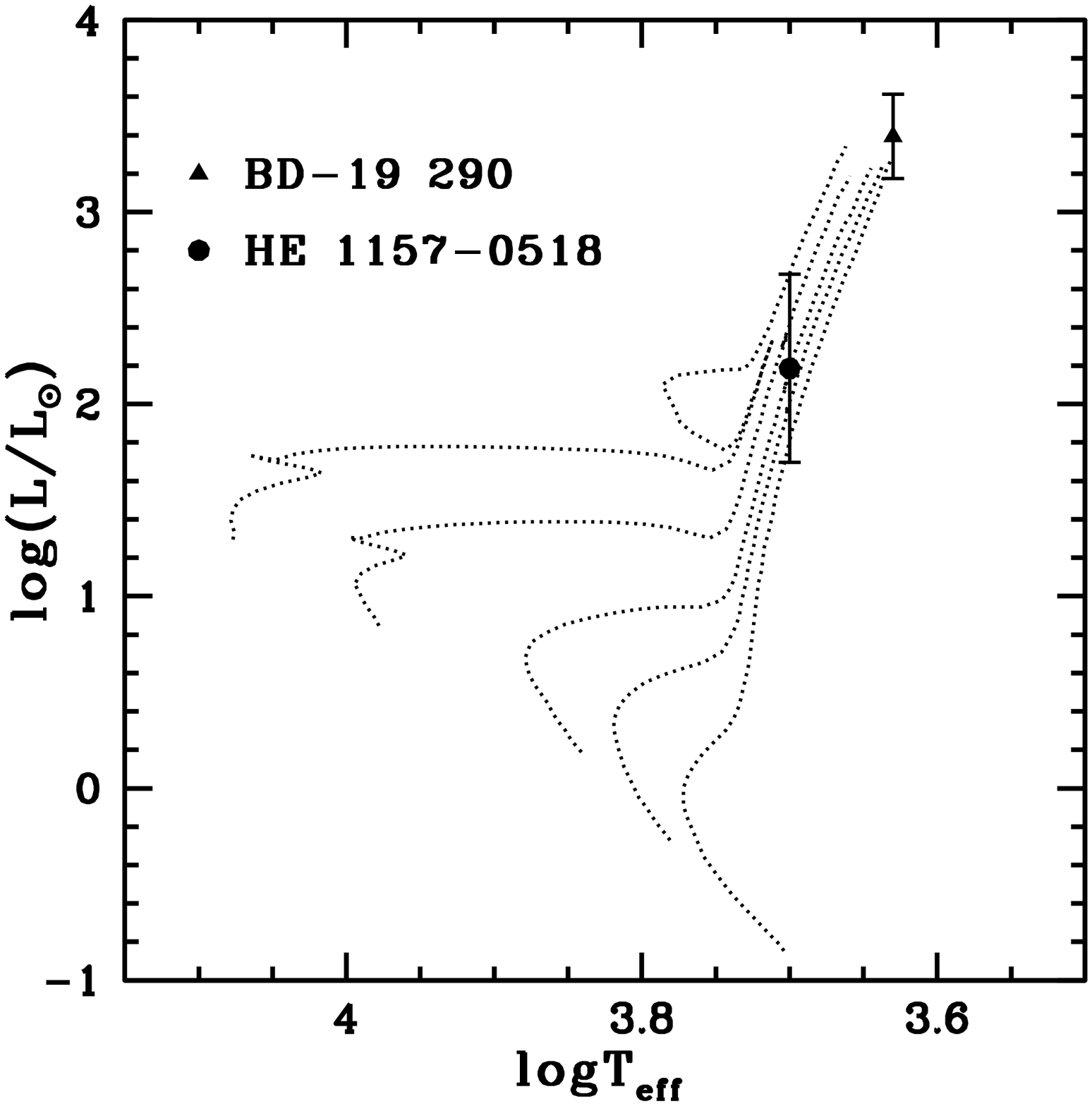}
\caption{The evolutionary tracks for 1.0, 1.1, and 1.2 M$_{\odot}$ for z = 0.004 (upper panel)  
and for 0.6, 0.8, 1.0, 1.4, and 1.8 M$_{\odot}$ for z = 0.0004 (lower panel)
are shown from bottom to top.} \label{tracks}
\end{figure}

We have also estimated the log g value using the standard equation \\
log (g/g$_{\odot}$)= log (M/M$_{\odot}$) + 4log (T$\rm_{eff}$/T$\rm_{eff\odot}$) - log (L/L$_{\odot}$)\\
We have adopted the solar values log g$_{\odot}$ = 4.44, T$\rm_{eff\odot}$ = 5770K and M$\rm_{bol\odot}$ = 4.74 mag.
The detailed procedure is given in \cite{Shejeelammal_2020}. 
The luminosity, log (L/L$\rm_{\odot}$), is calculated using the 
V magnitudes taken from SIMBAD astronomical database and the parallaxes ($\pi$) taken from 
Gaia DR2 (\citealt{Gaia_2018}, \url{https://gea.esac.esa.int/archive/}).  
The bolometric correction and the interstellar extinction required for the calculation 
of luminosity are estimated using the calibration equation of \cite{Alonso_1999} 
and \cite{Chen_1998} respectively. With this calculated luminosity and spectroscopic 
temperature estimate, the mass is found from the position of the star 
on Hertzsprung - Russell (HR) diagram (log T$\rm_{eff}$ v/s log (L/L$_{\odot}$)).
The HR diagram is generated using the stellar evolutionary tracks of \cite{Girardi_2000}.
Thus log g is  calculated with this mass using the above equation. 
We have used z = 0.0004 tracks for the objects BD$-$19 290 and HE~1157$-$0518 
and z=0.004 tracks for HD~202851. These HR diagrams are shown 
in Figure \ref{tracks}. For the objects BD$-$19 132, HE~1304$-$2111,
HE~1354$-$2257, and BD+19 3109, we could not determine their masses as the tracks corresponding to their 
luminosity and temperatures are not available. The mass estimates and log g values 
estimated from the parallax method are given in Table \ref{mass}. 
For our analysis, we have used the spectroscopic log\, g  values.

{\footnotesize
\begin{table*}
\caption{Mass and log\,{g} estimated from parallax method} \label{mass}
\begin{tabular}{lcccccc}
\hline                       
 Star name         & Parallax             & $M_{bol}$           & log(L/L$_{\odot}$)  & Mass(M$_{\odot}$) & log g          & log g (spectroscopic)   \\
                   & (mas)                &                     &                     &                   & (cgs)          & (cgs)                    \\
\hline
BD$-$19 132        & 0.3556$\pm$0.0560    & $-$2.395$\pm$0.345  & 2.854$\pm$0.138     & --                & --             & 1.13                      \\
BD$-$19 290        & 0.1449$\pm$0.0417    & $-$3.745$\pm$0.334  & 3.390$\pm$0.260     & 1.00$\pm$0.40     & 0.54$\pm$0.07  & 0.61                   \\
HE~1157$-$0518     & 0.0778$\pm$0.0397    & $-$0.725$\pm$1.222  & 2.186$\pm$0.490     & 1.00$\pm$0.60     & 2.02$\pm$0.25  & 2.52                  \\
HE~1304$-$2111     & 22.7558$\pm$0.040    & 9.229$\pm$0.004     & $-$1.797$\pm$0.002  & --                & --             & 1.06                  \\
HE~1354$-$2257     & 0.0150$\pm$0.0514    & 5.638$\pm$3.231     & 4.151$\pm$1.292     & --                & --             & 1.13                  \\
BD+19 3109         & 0.3764$\pm$0.0284    & $-$2.765$\pm$0.165  & 3.000$\pm$0.060     & --                & --             & 0.75                  \\
HD~202851          & 2.2160$\pm$0.0982    & 0.955$\pm$0.999     & 1.514$\pm$0.040     & 1.05$\pm$0.05     & 2.66$\pm$0.02  & 2.20                  \\
\hline
\end{tabular}
\end{table*}
}

\section{Abundance determination}
The elemental abundances are derived by two methods; from the 
measured equivalent widths of spectral lines due to several elements and
the spectral synthesis calculation. The spectral lines of different elements
are identified by comparing the spectrum of Arcturus with the spectra 
of the program stars. The line parameters, excitation potential and oscillator
strength (log~gf) values, are taken from Kurucz database of atomic line lists (\url{https://lweb.cfa.harvard.edu/amp/ampdata/kurucz23/sekur.html}).
The abundances from the molecular bands are derived using the spectral synthesis
calculation. For the elements showing the hyper-fine splitting (HFS),
spectral synthesis calculation is performed in addition to the abundance derived
from the equivalent width measurements. In the case of elements with NLTE effects, 
appropriate NLTE corrections have been  applied to the estimated abundances. The details of the abundance
derivation of each element along with the details of HFS and the NLTE corrections
used are discussed in the following. 
The derived elemental abundances of the program stars are presented 
in Table \ref{abundance_table1} and \ref{abundance_table2} and the lines used 
are given in Table \ref{linelist1} and \ref{linelist2}.
The abundance ratios are calculated with respect to solar value \citep{Asplund_2009}.

{\footnotesize
\begin{table*}
\caption{Elemental abundances in BD$-$19 132, BD$-$19 290, HE~1157$-$0518, and HE~1304$-$2111 } \label{abundance_table1}
\resizebox{\textwidth}{!}
{\begin{tabular}{lcccccccccccccccccc}
\hline
       &    &                              &                    & BD$-$19 132 &            &                    & BD$-$19 290 &        &                    & HE~1157$-$0518 &      &                    & HE~1304$-$2111 &       \\ 
\hline
       & Z  & solar log$\epsilon^{\ast}$   & log$\epsilon$      & [X/H]    & [X/Fe]     & log$\epsilon$      & [X/H]    & [X/Fe] & log$\epsilon$      & [X/H]    & [X/Fe]\\ 
\hline 
C (C$_{2}$ band 5165 {\rm \AA})      & 6  & 8.43                         & -          & -  & -       & 7.35(syn)          & $-$1.08   & 1.78 & 8.18(syn)          & $-$0.25   & 2.17   & --                 & --       & --   \\
C (C$_{2}$ band 5635 {\rm \AA})      & 6  & 8.43                         & 7.27(syn)          & $-$1.16  & 0.70       & 7.58(syn)          & $-$0.85   & 2.01     & 8.45(syn)          & 0.02    & 2.44     & 8.33(syn)          & $-$0.10  & 2.24 \\
$^{12}$C/$^{13}$C      & --  & --                         & --       & --  & 18       & --       & --  & 4    & --                 & --       & --  & --    & --                 & --             \\
N      & 7  & 7.83                         & 7.43(syn)(3)       & $-$0.40  & 1.46       & 6.86(syn)(3)       & $-$0.97  & 1.89    & 6.93(syn)       & $-$0.90  & 1.52    & --                 & --       & --   \\
O      & 8  & 8.69                         & 7.06(syn)(1)       & $-$1.63  & 0.23       & 7.17(syn)(1)       & $-$1.52  & 1.34    & --                 & --       & --   & --                 & --       & --\\
Na I   & 11 & 6.24                         & 5.47$\pm$0.07(2)   & $-$0.77  & 1.09       & 4.56$\pm$0.13(2)   & $-$1.68  & 1.18    & --                 & --       & --      & 6.73$\pm$0.12(2)   & 0.49     & 2.83  \\
Mg I   & 12 & 7.60                         & 6.32$\pm$0.02(2)   & $-$1.28  & 0.58       & 5.35(1)            & $-$2.25  & 0.61    & 5.45(1)            & $-$2.15  & 0.27    & 5.46(1)   & $-$2.12     & 0.22 \\
Si I   & 14 & 7.51                         & --                 & --       & --         & --                 & --       & --      & --                 & --       & --      & 5.91$\pm$0.01(2)   & $-$1.60  & 0.74 \\
Ca I   & 20 & 6.34                         & 5.21$\pm$0.11(4)   & $-$1.13  & 0.73       & 3.93$\pm$0.05(4)   & $-$2.41  & 0.45     & 4.07$\pm$0.13(4)   & $-$2.27  & 0.15    & 4.49$\pm$0.16(2)   & $-$1.85  & 0.49 \\
Sc II  & 21 & 3.15                         & 2.09(syn)(1)       & $-$1.06  & 0.80       & 1.10$\pm$0.07(syn)(2)          & $-$2.05  & 0.81    & 1.35(syn)(1)       & $-$1.80  & 0.62    & 1.45(syn)(1)          & $-$1.70  & 0.64 \\
Ti I   & 22 & 4.95                         & 3.68$\pm$0.14(7)   & $-$1.27  & 0.59       & 2.28$\pm$0.0(2)    & $-$2.67  & 0.19     & 2.36(1)            & $-$2.59  & $-$0.17 & 3.07$\pm$0.14(3)  & $-$1.88  & 0.46  \\
Ti II  & 22 & 4.95                         & 3.42$\pm$0.13(7)   & $-$1.53  & 0.33       & 2.17$\pm$0.03(3)   & $-$2.78  & 0.08     & 2.59$\pm$0.17(3)   & $-$2.36  & 0.06    & 2.54(1)      & $-$2.41 & $-$0.07  \\
V I    & 23 & 3.93                         & 1.93(syn)(1)       & $-$2.00  & $-$0.14    & 2.13(syn)(1)       & $-$1.80  & 1.06    & 1.75(syn)(1)       & $-$2.18  & 0.24    & 2.99$\pm$0.23(syn)(2)        & $-$0.94  & 1.40  \\
Cr I   & 24 & 5.64                         & --                 & --       & --         & --                 & --       & --      & 3.18$\pm$0.11(2)   & $-$2.46  & $-$0.04  & 4.43$\pm$0.11(4)   & $-$1.21  & 1.13 \\
Cr II  & 24 & 5.64                         & --                 & --       & --         & 3.80$\pm$0.10(2)   & $-$1.84  & 1.02    & --                 & --       & --       & --                 & --       & --  \\
Mn I   & 25 & 5.43                         & 3.50(syn)(1)       & $-$1.93  & $-$0.07    & 3.68(syn)(1)       & $-$1.75  & 1.11    & 3.25$\pm$0.08(syn)(2)       & $-$2.18  & 0.24    & 3.70(syn)(1)       & $-$1.73  & 0.61 \\
Fe I   & 26 & 7.50                         & 5.64$\pm$0.06(19)  & $-$1.86  & -          & 4.64$\pm$0.07(12)  & $-$2.86  & -       & 5.08$\pm$0.15(8)  & $-$2.42  & -       & 5.16$\pm$0.10(16)  & $-$2.34  & -   \\
Fe II  & 26 & 7.50                         & 5.64$\pm$0.01(2)   & $-$1.86  & -          & 4.64$\pm$0.13(2)   & $-$2.86  & -       & 5.08$\pm$0.16(2)   & $-$2.42  & -       & 5.16$\pm$0.07(3)   & $-$2.34  & - \\
Co I   & 27 & 4.99                         & 3.01(syn)(1)       & $-$1.98  & $-$0.12    & --                 & --       & --      & --                 & --       & --      & --                 & --       & --   \\
Ni I   & 28 & 6.22                         & 4.22$\pm$0.18(4)  & $-$2.00  & $-$0.14     & 4.12$\pm$0.04(3)   & $-$2.10  & 0.76    & 3.97$\pm$0.10(3)   & $-$2.25  & 0.17    & 4.15$\pm$0.15(4)   & $-$2.07     & 0.27 \\
Cu I   & 29 & 4.19                         & --                 & --       & --         & --                 & --       & --      & --                 & --       & --      & 1.40(syn)(1)          & $-$2.81  & $-$0.47  \\
Zn I   & 30 & 4.56                         & 3.00(1)            & $-$1.56  & 0.30       & 2.33(1)            & $-$2.23  & 0.63    & --                 & --       & --    & --                 & --       & --     \\
Rb I   & 37 & 2.52                         & 0.75(syn)(1)       & $-$1.77  & 0.09       & 0.72(syn)(1)       & $-$1.80  & 1.06    & --                 & --       & --      & --                 & --       & --    \\
Sr I$_{NLTE}$  & 38 & 2.87                 & 3.07(syn)(1)       & 0.20     & 2.06       & --                 & --       & --      & --                 & --       & --      & --                 & --       & --     \\
Y I    & 39 & 2.21                         & 2.43(syn)(1)       & 0.22     & 2.08       & --                 & --       & --      & --                 & --       & --      & --                 & --       & --   \\
Y II   & 39 & 2.21                         & 2.71$\pm$0.13(3)   & 0.50     & 2.36       & 0.38$\pm$0.10(2)   & $-$1.83  & 1.03    & 0.58$\pm$0.08(4)   & $-$1.63  & 0.79    & 0.75$\pm$0.12(2)  & $-$1.46     & 0.88  \\
Zr I   & 40 & 2.58                         & 2.88(syn)(1)       & 0.30     & 2.16       & 1.03$\pm$0.10(2)   & $-$1.55  & 1.31    & --                 & --       & --      & 1.77$\pm$0.03(2)          & $-$0.81     & 1.53 \\
Zr II  & 40 & 2.58                         & 2.29$\pm$0.07(2)   & $-$0.29  & 1.57       & 0.78(syn)(1)       & $-$1.80  & 1.06    & 1.20(syn)(1)       & $-$1.38  & 1.04      & 1.32(syn)(1)       & $-$1.26  & 1.08  \\
Ba II$_{LTE}$   & 56 & 2.18                & 2.66(syn)(1)       & 0.48     & 2.34       & 1.03(syn)(1)       & $-$1.15  & 1.71       & 1.50(syn)(2)       & $-$0.68  & 1.74    & 1.42$\pm$0.05(syn)(1)                 & $-$0.76       & 1.58     \\
Ba II$_{NLTE}$  & 56 & 2.18                & 2.58(syn)(1)       & 0.40     & 2.26          & --                 & --       & --      & 1.51(syn)(1)       & $-$0.67  & 1.75    & --                 & --       & --    \\
La II  & 57 & 1.10                         & 1.90(syn)(2)       & 0.80     & 2.66       & 0.22(syn)(1)       & $-$0.88  & 1.98    & --                 & --       & --     & 0.30(syn)(2)          & $-$0.80     & 1.54  \\
Ce II  & 58 & 1.58                         & 2.20$\pm$0.13(3)   & 0.62     & 2.48       & 0.63$\pm$0.13(4)   & $-$0.95  & 1.91    & 1.00$\pm$0.14(4)   & $-$0.58  & 1.84    & 0.75$\pm$0.12(2)  & $-$0.83     & 1.51  \\
Pr II  & 59 & 0.72                         & 1.43$\pm$0.11(3)   & 0.71     & 2.57       & $-$0.14$\pm$0.12(2)   & $-$0.86  & 2.00    & 0.18$\pm$0.11(4)   & $-$0.54  & 1.88    & $-$0.06(1)   & $-$0.90    & 1.44 \\
Nd II  & 60 & 1.42                         & 2.20$\pm$0.14(6)   & 0.78     & 2.64       & 0.53$\pm$0.13(6)   & $-$0.89  & 1.97    & 0.99$\pm$0.14(13)  & $-$0.43  & 1.99    & 0.69$\pm$0.03(2)  & $-$0.73     & 1.61 \\
Sm II  & 62 & 2.41                         & 1.17$\pm$0.11(6)   & 0.21     & 2.07       & 0.19$\pm$0.02(3)   & $-$0.77  & 2.09    & 0.59$\pm$0.19(2)   & $-$0.37  & 2.05    & 0.06$\pm$0.02(2)   & $-$0.90     & 1.44  \\
Eu II$_{LTE}$  & 63 & 0.52                 & $-$0.05(syn)(1)    & $-$0.57  & 1.29       & $-$1.02(syn)(1)    & $-$1.54  & 1.32    & $-$0.62(syn)(1)    & $-$1.14  & 1.28    & $-$0.70(syn)(1)          & $-$1.22 & 1.12 \\
\hline
\end{tabular}}

$\ast$  \cite{Asplund_2009}, The number of lines used for abundance determination are given within the parenthesis.
\end{table*}
}

{\footnotesize
\begin{table*}
\caption{Elemental abundances in HE~1354$-$2257, BD+19 3109, and HD~202851} \label{abundance_table2}
\resizebox{\textwidth}{!}
{\begin{tabular}{lccccccccccccccc}
\hline
       &    &                                &                    & HE~1354$-$2257 &        &                    & BD+19 3109 &                    & HD~202851       &       \\ 
\hline
       & Z  & solar log$\epsilon^{\ast}$   & log$\epsilon$      & [X/H]    & [X/Fe]     & log$\epsilon$      & [X/H]    & [X/Fe] & log$\epsilon$      & [X/H]    & [X/Fe]    \\ 
\hline 
C (C$_{2}$ band 5165 {\rm \AA})      & 6  & 8.43                         & 8.40(syn)          & $-$0.03    & 2.08   & 8.69                 & 0.26       & 2.49     & 8.30(syn)          & $-$0.13  & 0.72  \\
C (C$_{2}$ band 5635 {\rm \AA})      & 6  & 8.43                         & 8.30(syn)          & $-$0.13    & 1.98     & 8.65                 & 0.22       & 2.45     & 8.45(syn)       & 0.02     & 0.87 \\
$^{12}$C/$^{13}$C      & --  & --                          & --       & --  & --    & --                 & --       & 9      & --       & --  & 42  \\
N      & 7  & 7.83                         & 6.23(syn)       & $-$1.60  & 0.51    & 7.83(syn)(3)       & 0.00  & 2.23    & 8.40$\pm$0.02(syn)(3)       & $-$0.57  & 1.42  \\
O      & 8  & 8.69                         & 7.09(syn)(1)          & $-$1.60       & 0.51      & --                 & --       & --      & 8.19(syn)(1)          & $-$0.50       & 0.35   \\
Na I   & 11 & 6.24                         & 4.71$\pm$0.06(2)   & $-$1.53  & 0.58    & 4.13$\pm$0.02(2)   & $-$2.11     & 0.12        & 5.92$\pm$0.08(3)   & $-$0.32  & 0.53   \\
Mg I   & 12 & 7.60                         & 5.47(1)            & $-$2.13  & $-$0.02 & 6.13$\pm$0.14(3)   & $-$1.47     & 0.76           & 7.02$\pm$0.10(2)   & $-$0.58  & 0.27 \\
Si I   & 14 & 7.51                         & 5.89$\pm$0.16(2)   & $-$1.62  & 0.49    & --                 & --       & --         & 6.63(1)   & $-$0.88  & $-$0.03   \\
Ca I   & 20 & 6.34                         & 4.45$\pm$0.13(7)   & $-$1.89  & 0.22    & 4.20$\pm$0.11(7)   & $-$2.14  & 0.09      & 5.46$\pm$0.14(8)   & $-$0.88  & $-$0.03 \\
Sc II  & 21 & 3.15                         & 1.05(syn)(1)       & $-$2.10  & 0.01    & 1.45(syn)(1)          & $-$1.70  & 0.53       & 2.86(syn)(1)       & $-$0.29  & 0.56  \\
Ti I   & 22 & 4.95                         & 2.60(1)            & $-$2.35  & $-$0.24 & 3.45$\pm$0.10(6)  & $-$1.50  & 0.73        & 4.37$\pm$0.15(13)  & $-$0.58  & 0.27  \\
Ti II  & 22 & 4.95                         & 2.70$\pm$0.07(4)   & $-$2.25  & $-$0.14 & 3.14$\pm$0.13(5)      & $-$1.81 & 0.42           & 4.40$\pm$0.17(9)   & $-$0.55  & 0.30  \\
V I    & 23 & 3.93                         & 2.95(syn)(2)       & $-$0.98  & 1.13    & 1.83(syn)(1)        & $-$2.10  & 0.13        & 2.75(syn)(1)       & $-$1.18  & $-$0.33   \\
Cr I   & 24 & 5.64                         & --                 & --       & --      & 2.83$\pm$0.09(7)   & $-$2.81  & $-$0.58      & 5.25$\pm$0.13(9)   & $-$0.39  & 0.46    \\
Cr II  & 24 & 5.64                         & 4.30$\pm$0.08(2)   & $-$1.34  & 0.77    & --                 & --       & --     & 5.20$\pm$0.19(3)   & $-$0.44  & 0.41  \\
Mn I   & 25 & 5.43                         & 3.59(syn)(1) & $-$1.84  & 0.27    & 3.13(syn)(2)       & $-$2.30  & $-$0.07     & 4.71(syn)(1) & $-$0.72  & 0.13   \\
Fe I   & 26 & 7.50                         & 5.39$\pm$0.17(20)  & $-$2.11  & -       & 5.27$\pm$0.09(27)  & $-$2.23  & -    & 6.65$\pm$0.01(30)  & $-$0.85  & -   \\
Fe II  & 26 & 7.50                         & 5.39$\pm$0.15(4)   & $-$2.11  & -       & 5.27$\pm$0.02(3)   & $-$2.23  & -    & 6.65$\pm$0.07(3)  & $-$0.85  & -    \\
Co I   & 27 & 4.99                         & 3.59(syn)(1)       & $-$1.40  & 0.71    & 3.20$\pm$0.07(3)       & $-$1.79  & 0.44     & 4.30(syn)(1)       & $-$0.69  & 0.16  \\
Ni I   & 28 & 6.22                         & 3.81$\pm$0.20(3)   & $-$2.41  & $-$0.30 & 4.46$\pm$0.12(3)   & $-$1.76     & 0.47            & 5.80$\pm$0.12(11)  & $-$0.42  & 0.43   \\
Cu I   & 29 & 4.19                         & --                 & --       & --      & --                 & --       & --    & 2.85(syn)(1)       & $-$1.34  & $-$0.49   \\
Zn I   & 30 & 4.56                         & 2.50(1)            & $-$2.06  & 0.05    & 2.48(1)                & $-$2.08       & 0.15      & 3.48(1)            & $-$1.08  & $-$0.23  \\
Rb I   & 37 & 2.52                         & --                 & --       & --      & 0.20(syn)(1)                 & $-$2.32       & $-$0.09      & 1.49(syn)(1)       & $-$1.03  & $-$0.18   \\
Sr I$_{NLTE}$  & 38 & 2.87                 & 2.44(syn)(1)       & $-$0.43  & 1.68    & --                 & --       & --     & 3.12(syn)(1)       & 0.25     & 1.10  \\
Y I    & 39 & 2.21                         & 2.18(syn)(1)       & $-$0.03  & 2.08    & 1.16(syn)(1)          & $-$1.05     & 1.18      & 2.44(syn)(1)       & 0.23     & 1.08  \\
Y II   & 39 & 2.21                         & --                 & --       & --      & 1.42$\pm$0.16(3)  & $-$0.79     & 1.44      & 3.05$\pm$0.08(4)   & 0.84  & 1.69  \\
Zr I   & 40 & 2.58                         & 2.27(syn)(1)       & $-$0.37  & 1.80    & 0.98(syn)(1)          & $-$1.60     & 0.63       & 3.50(syn)(1)       & 0.92     & 1.77  \\
Zr II  & 40 & 2.58                         & --                 & --       & --      & 1.28(1)(1)                 & $-$1.30       & 0.93     & 3.79(syn)(1)       & 1.21     & 2.06     \\
Ba II$_{LTE}$   & 56 & 2.18                & 1.70(syn)(2)       & $-$0.48  & 1.63    & 1.68(syn)(1)                 & $-$0.60       & 1.73     & 3.50(syn)(1)       & 1.32     & 2.17     \\
Ba II$_{NLTE}$  & 56 & 2.18                & --                 & --       & --      & --                 & --       & --      & 3.15(syn)(1)       & 0.97     & 1.82   \\
La II  & 57 & 1.10                         & 0.40(syn)(1)       & $-$0.70  & 1.41    & 0.50(syn)(1)          & $-$0.60     & 1.63   & 1.85(syn)(2)       & 0.75     & 1.60   \\
Ce II  & 58 & 1.58                         & 1.17$\pm$0.07(4)   & $-$0.41  & 1.70    & 0.92$\pm$0.11(7)  & $-$0.66     & 1.57    & 2.62$\pm$0.14(8)   & 1.04     & 1.89 \\
Pr II  & 59 & 0.72                         & 0.14$\pm$0.09(3)   & $-$0.58  & 1.53    & 0.23$\pm$0.12(4)   & $-$0.49    & 1.74         & 2.04$\pm$0.14(4)   & 1.32     & 2.17 \\
Nd II  & 60 & 1.42                         & 0.98$\pm$0.13(9)   & $-$0.44  & 1.67    & 0.82$\pm$0.09(14)  & $-$0.60     & 1.63         & 2.38$\pm$0.11(9)   & 0.96     & 1.81  \\
Sm II  & 62 & 2.41                         & 0.50$\pm$0.08(3)   & $-$0.46  & 1.65    & 0.38$\pm$0.11(8)   & $-$0.58     & 1.65        & 1.83$\pm$0.09(6)   & 0.87     & 1.72  \\
Eu II$_{LTE}$  & 63 & 0.52                 & $-$1.12(syn)(1)    & $-$1.64  & 0.47    & $-$0.99(syn)(1)          & $-$1.51 & 0.72    & 0.56(syn)(1)       & 0.04     & 0.89  \\
Eu II$_{NLTE}$  & 63 & 0.52                & $-$1.16(syn)(1)    & $-$1.68  & 0.43   & --                 & --       & --       & --                 & --       & --   \\
\hline
\end{tabular}}

$\ast$  \cite{Asplund_2009}, The number of lines used for abundance determination are given within the parenthesis.
\end{table*}
}

\subsection{Light elements: C, N, O, $^{12}$C/$^{13}$C, 
Na, $\alpha$-, and $Fe$-peak elements}
The abundance of oxygen is derived from the spectral synthesis calculation 
of [O I] 6300.304 {\rm \AA} line. This line is insensitive to NLTE effect.
The other oxygen forbidden line at 
6363.776  {\rm \AA} is not usable for the abundance determination in 
any of the program stars. The O I triplet lines around 7770 {\rm \AA}
was not good in the stars BD$-$19 132, BD$-$19 290, BD+19 3109, and HD~202851. 
In the other three stars this region is absent as the SUBARU spectra is limited
up to 6850 {\rm \AA}. We could determine the oxygen abundance only in four stars:
BD$-$19 132, BD$-$19 290, HE~1354$-$2257, and HD~202851, with BD$-$19 290 showing the
highest enhancement of [O/Fe]$\sim$1.34. In other three stars [O/Fe] ranges from 
0.23 to 0.51. 

\par
\begin{figure}
\centering
\includegraphics[width=\columnwidth]{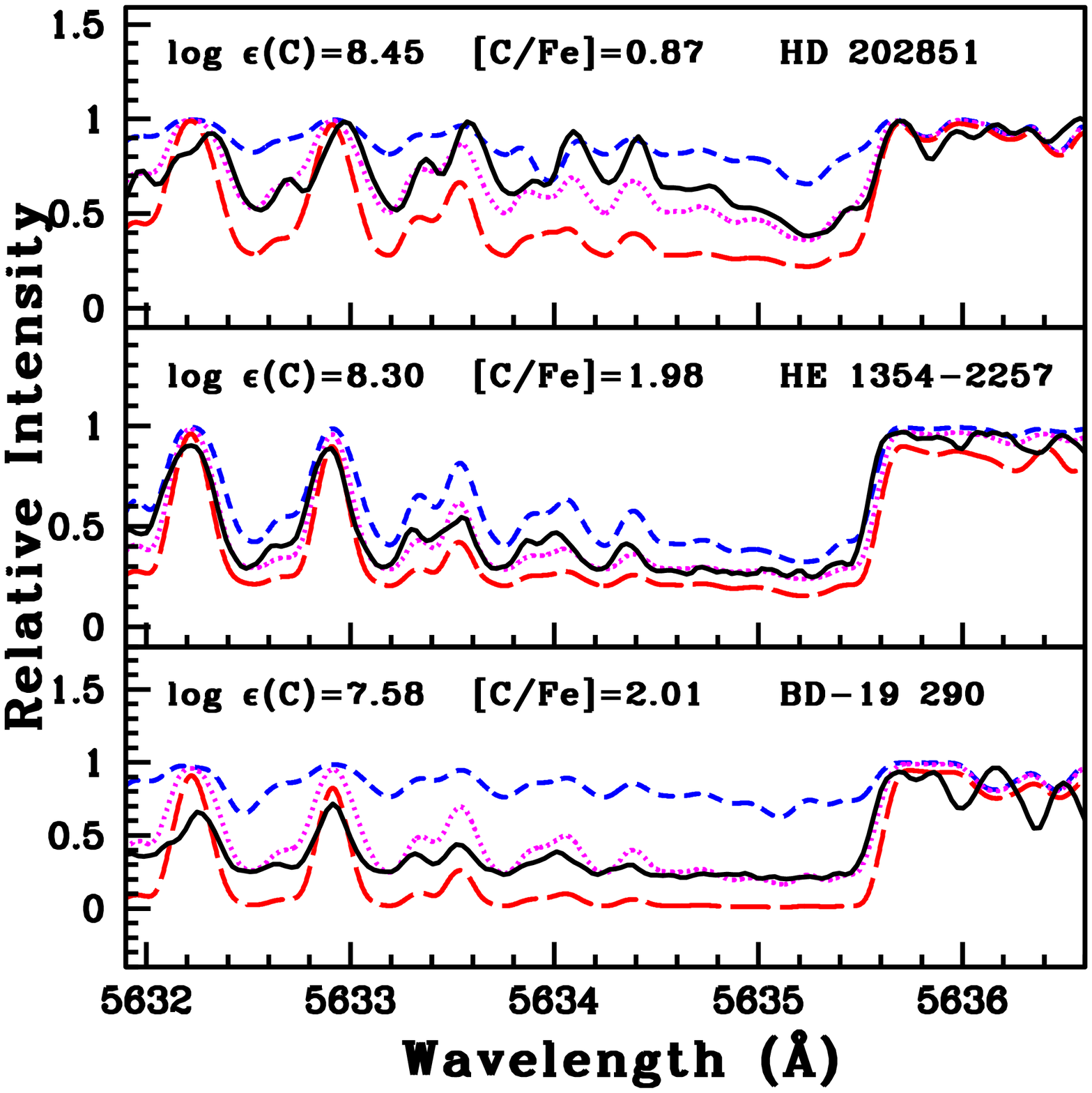}
\includegraphics[width=\columnwidth]{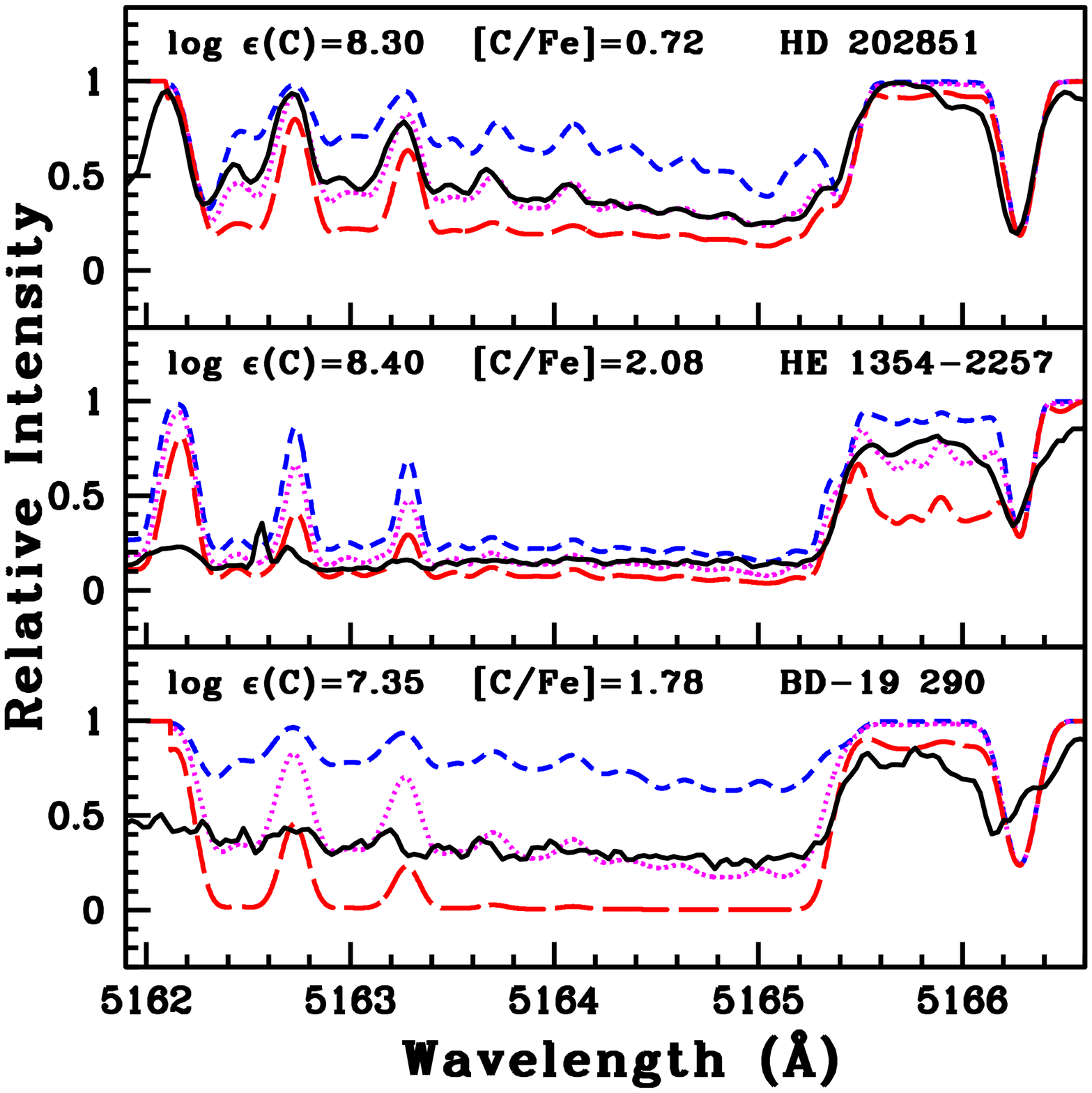}
\caption{ Synthesis of C$_{2}$ band around 5165 {\rm \AA} (lower panel) and 5635 {\rm \AA} (upper panel). 
Dotted and solid lines represent synthesized and observed spectra respectively. 
Short dashed and long dashed lines represent the synthetic spectra 
for $\Delta$ [C/Fe] = $-$0.3 and +0.3 respectively.} \label{carbon}
\end{figure}

 Once the oxygen abundance is determined, the carbon abundance is derived 
from the C$_{2}$ molecular bands at 5165 and 5635 {\rm \AA} by spectral synthesis
calculation. The spectral synthesis fits for these two regions for a few program stars
are shown in Figure \ref{carbon}. The C$_{2}$ bands at 5165 {\rm \AA} is noisy 
in the spectra of BD$-$19 132 and HE~1304$-$2111; so we could not use this region 
in these stars. In all other stars, we derived the carbon abundance from both the
regions. The abundances derived from these two bands are  consistent within 0.3 dex.
The average abundance from these two C$_{2}$ regions is taken as the final carbon abundance. 
All the stars except BD$-$19 132 and HD~202851 show [C/Fe]$>$1, with BD+19 3109 
being the most carbon enhanced with [C/Fe]$\sim$2.47. 

\par The nitrogen abundance is derived from the $^{12}$CN lines at 8000 {\rm \AA} region
in the program stars that are observed with HESP and from $^{12}$CN band at 4215 {\rm \AA} 
region in the spectra obtained with SUBARU. We could determine nitrogen abundance 
in all the stars except HE~1304$-$2111. Except HE~1354$-$2257, all other stars show 
[N/Fe]$>$1. 
Once the carbon, nitrogen, and oxygen abundances are derived, we re-derived the oxygen abundance.
Then with this oxygen abundance carbon and then nitrogen abundances are re-estimated. 
This iterative process is done until a convergence is obtained. 

\par The $^{12}$CN lines at 8003.292, 8003.553, 8003.910  {\rm \AA},  and $^{13}$CN lines at 8004.554, 
8004.728, 8004.781 {\rm \AA} are used to derive the carbon isotopic ratio, $^{12}$C/$^{13}$C in
the HSEP spectra of the stars that cover these regions. The SUBARU spectra used for three objects, however extend only till 6800 \AA\, in the wavelength region. The values obtained for this ratio in BD$-$19 132, BD$-$19 290, 
BD+19 3109, and HD~202851 are 18, 4, 9, and 42 respectively. These estimates are consistent with 
what is expected from the evolutionary stages of the program stars and also consistent with the values in the 
range 2.5 - 40 \citep{Bisterzo_2011} for CEMP-s and CEMP-r/s stars.

\par
\begin{figure}
\centering
\includegraphics[width=\columnwidth]{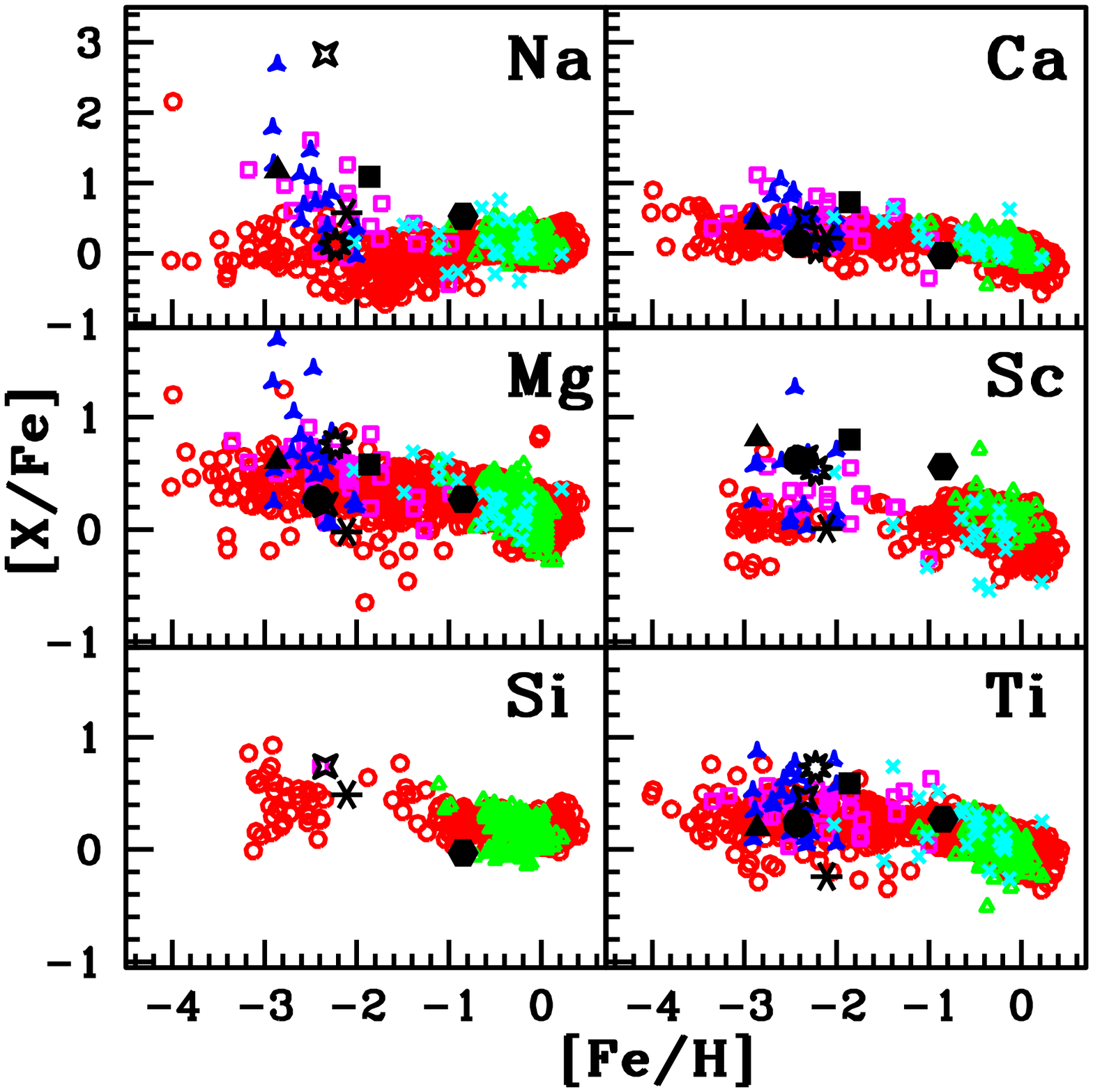}
\includegraphics[width=\columnwidth]{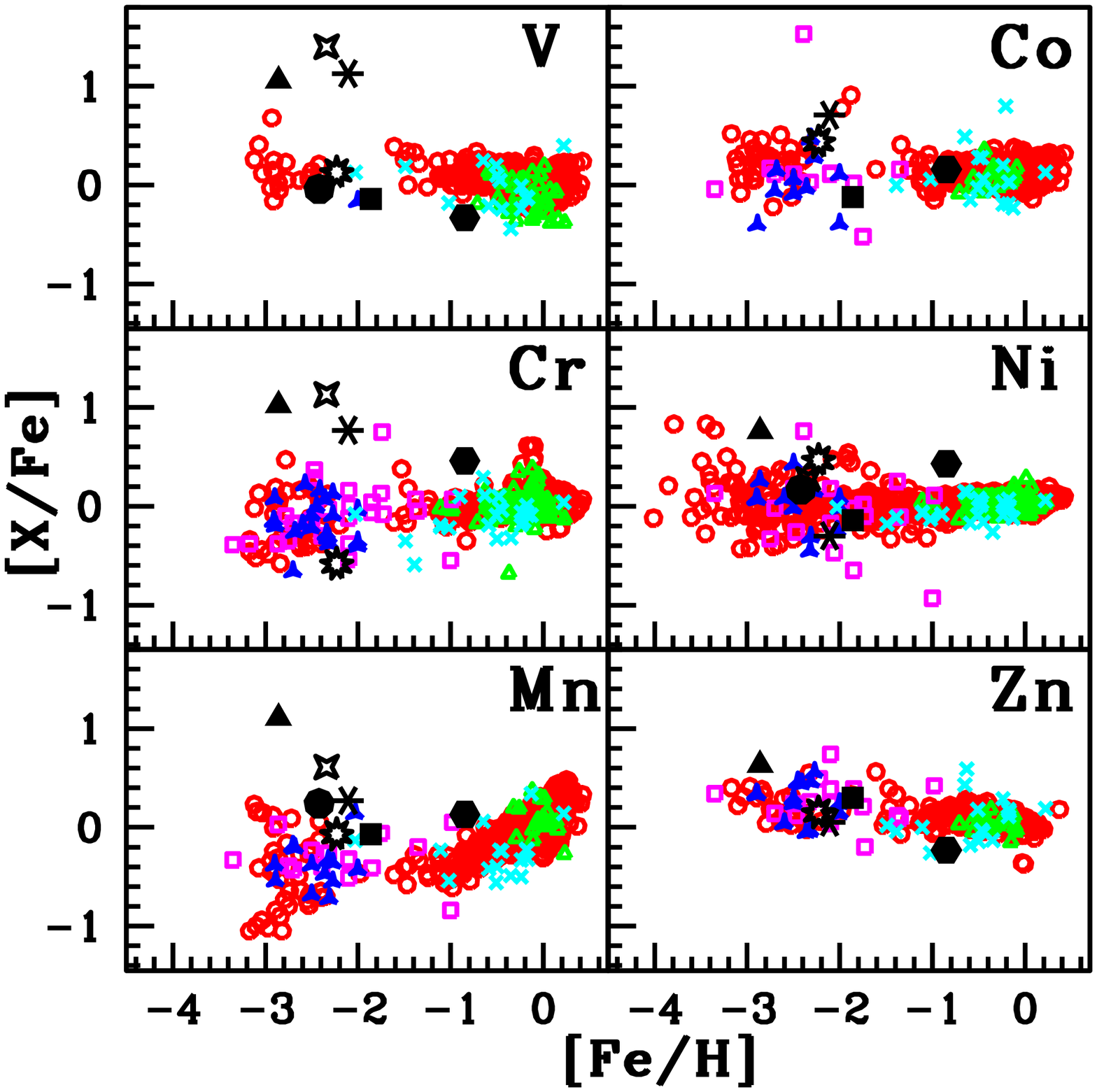}
\caption{Observed [X/Fe] ratios of the light elements in the  
program stars with respect to metallicity [Fe/H].   
Red open circles correspond to normal giants from literature \citep{Honda_2004, Venn_2004, 
Aoki_2005, Aoki_2007, Reddy_2006, Luck_2007, Hansen_2016a, Yoon_2016}.
Magenta open squares and blue starred triangles represent CEMP-s and
CEMP-r/s stars respectively from literature \citep{Masseron_2010, Purandardas_2019, 
Karinkuzhi_2021, Shejeelammal_2021}. Cyan crosses represent CH stars from literature 
\citep{Vanture_1992, Karinkuzhi_2014, Karinkuzhi_2015, Goswami_2016, Shejeelammal_2021}.
Green open triangles are Ba stars from literature \citep{Allen_2006a, deCastro_2016,
Yang_2016, Karinkuzhi_2018, Shejeelammal_2020}.   
BD$-$19 132 (filled square), BD$-$19 290 (filled triangle), HE~1157$-$0518 (filled circle), 
HE~1304$-$2111 (four-sided star), HE~1354$-$2257 (six-sided cross), BD+19 3109 (nine-sided star), 
and HD~202851 (filled hexagon).} \label{light_elements}
\end{figure}

 The equivalent width measurement of several lines listed in Table \ref{linelist2}
are used to derive the abundance of Na, Mg, Si, Ca, Ti, Cr, Ni, and Zn. 
We could not estimate sodium abundance in HE~1157$-$0518. 
The star HE~1304$-$2111 shows the largest enhancement of Na with 
[Na/Fe]$\sim$2.83. The lines used to derive the Na abundance in this star 
were very strong, so this abundance may not be reliable. The NLTE corrections 
of these lines were not available. The stars, 
BD$-$19 132 and BD$-$19 290 have [Na/Fe]$>$1. The Mg abundance of the 
program stars are in the range $-$0.02$\leq$[Mg/Fe]$\leq$0.61, 
similar to that observed in the normal stars in the Galaxy (Figure \ref{light_elements}).  
  
\par The elements Sc, V, Mn, Co, and Cu show  HFS. The abundances of these 
elements are derived using the spectral synthesis calculation, taking the hyper-fine 
components into account. The hyper-fine components of these elements are taken from 
\cite{Prochaska_MW_2000} and \cite{Prochaska_2000}.
Sc abundance is derived mainly from the Sc II lines at 6245.637, 6309.920, and 6604.601 {\rm \AA}.
It shows solar value in the star HE~1354$-$2257, while it is enhanced in all other stars
with values ranging from 0.53 to 0.81.
Spectral synthesis calculation of the V I lines at 4864.731, 5727.048, and 6251.827 {\rm \AA}
are used to derive the vanadium abundance. Three objects in our sample, BD$-$19 290, HE~1304$-$2111,
and HE~1354$-$2257 show [V/Fe]$>$1, whereas it is slightly under-abundant in HD~202851 and near-solar 
in the other stars. The Mn abundance is estimated from the Mn I lines at 4761.530, 6013.513, and 6021.89 {\rm \AA}. 
We have derived the Co abundance from  Co I lines at 5342.695, and 5483.344 {\rm \AA} and Cu abundance from
Cu I line at 5105.537 {\rm \AA}. In the case of BD+19 3109, cobalt abundance is derived from the lines 
Co I 4792.846, 4867.872, and 4792.846 {\rm \AA}. 
A comparison of the light element abundances observed in our program stars with their counterparts 
in normal stars and other classes of chemically peculiar stars are shown in Figure \ref{light_elements}.
While the estimated abundances of the elements Mg, Si, Ca, Sc, Ti, Co, Ni, and Zn in all the program 
stars follow the Galactic trend (similar to that observed in normal stars), Na, V, Cr, and Mn are enhanced
in a few program stars. 
Such enhancements of Na and Fe-peak elements are observed in a few  CEMP stars such as HE~1327$-$2326 \citep{Aoki_2006b, Frebel_2008} and SMSS~J031300.36$-$670839.3 \citep{Bessel_2015}. 
The models of spinstar \citep{Maeder_2015a, Maeder_2015b}, faint SNe \citep{Ivamoto_2005, Tominaga_2007, Bessel_2015} 
have been used to reproduce their observed abundance pattern. According to \cite{Aoki_2006b}, either faint SNe or AGB-binary mass transfer 
along with the accretion of metals from ISM might explain the observed abundance pattern of HE~1327$-$2326.
\cite{Choplin_2017} have suggested that spinstar could also have played a role in the formation of some CEMP-s stars and 
the abundance patterns of CEMP-s stars may be resulting from several sources. From these facts, the observed enhanced abundances of 
Na, V, Cr, and Mn in the program stars BD$-$19 290, HE~1304$-$2111 and HE~1354$-$2257 may be attributed to the pre-enriched ISM 
from which they were formed. Also, some of the CEMP-(s\&r/s) stars show enhanced abundances of Na at low metallicities (\citealt{Bisterzo_2011} 
and references therein, \citealt{Allen_2012}, \citealt{Karinkuzhi_2021}). 
A discussion on the Na abundance is provided in section 8.3.

\subsection{Heavy elements}
\subsubsection{\textbf{The light s-process elements: Rb, Sr, Y, Zr}}
The spectral synthesis calculation of Rb I 7800.259 {\rm \AA} resonance line
is used to derive the rubidium abundance in the program stars, by taking the 
hyper-fine components from \cite{Lambert_1976}. 
Rb I 7947.597 {\rm \AA} was blended and could not be used for the abundance 
determination. 
We could determine Rb abundance only in the stars whose spectra are from 
HCT/HESP. It is enhanced in BD$-$19 290 with [Rb/Fe]$>$1, shows near-solar value in other 
three stars: BD$-$19 132, BD+19 3109, and HD~202851. 

\par
\begin{figure}
\centering
\includegraphics[width=\columnwidth]{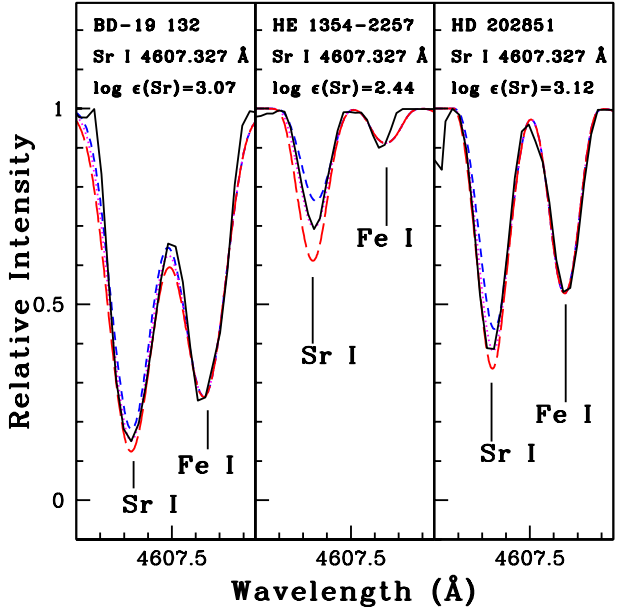}
\caption{ Synthesis of Sr I line at 4607.327 {\rm \AA}. 
Dotted and solid lines represent synthesized and observed spectra respectively.
Short dashed and long dashed lines represent the synthetic spectra 
for $\Delta$[Sr/Fe] = $-$0.3 and +0.3 respectively.} \label{Sr4607}
\end{figure}

 Strontium abundance is derived from the spectral synthesis calculation of 
Sr I 4607.327 {\rm \AA} line with log $gf$ value taken from
\cite{Bergemann_2012}. We could estimate strontium abundance only in
three objects; BD$-$19 132, HE~1354$-$2257, and HD~202851 and they show [Sr/Fe]$>$1. 
It has been proven that the Sr I lines are affected by NLTE effects 
\citep{Barklem_2000, Short_2006}. The NLTE corrections
of Sr I 4607.327 {\rm \AA} line is adopted from \cite{Bergemann_2012}.
They have calculated the NLTE corrections for different combinations 
of stellar atmospheric parameters for four different stars. For the
star HD~122563, the NLTE correction is +0.47 dex, which has got 
similar atmospheric parameters (T$\rm_{eff}$ = 4600 K, 
log g = 1.60, $\zeta$ = 1.80 km s$^{-1}$, [Fe/H] = $-$2.50) as BD$-$19 132 and HE~1354$-$2257. 
For these two stars, we have applied this correction term to our estimated abundance of Sr.
For the star HD~202851, the NLTE correction ($\sim$+0.32 dex) has been taken 
from the Table 3 of \cite{Bergemann_2012} which corresponds to a similar 
combination of atmospheric parameters as HD~202851. 
The spectrum synthesis fit for Sr I 4607.327 {\rm \AA} line for the
program stars are shown in Figure \ref{Sr4607}

\par The yittrium abundance is derived from Y I 6435.004 {\rm \AA} line 
and the equivalent width measurement of the Y II lines listed in Table \ref{linelist2}.  
We could not estimate Y I abundances in BD$-$19 290, HE~1157$-$0518, and HE~1304$-$2111 
and Y II abundance in HE~1354$-$2257 as no good lines are detected  in their spectra due to these species.
The abundance derived from Y II lines are higher than that estimated from Y I lines 
wherever we could use both the species. It is enhanced in all the program stars 
with [X/Fe]$\geq$0.79. 
The zirconium abundance is estimated using the spectral synthesis calculation 
of Zr I 6134.585 {\rm \AA} and Zr II 5112.297 {\rm \AA} and equivalent width measurement
of several Zr I and Zr II lines whenever available. We could not estimate Zr I abundance in 
HE~1157$-$0518  and Zr II abundance in HE~1354$-$2257. 
The abundance estimated from Zr I lines are in the range 0.63 - 2.16 and that from
Zr II lines are  in the range 0.93 - 2.06.

\subsubsection{\textbf{The heavy s-process elements: Ba, La, Ce, Pr, Nd}}
\begin{figure}
\centering
\includegraphics[width=\columnwidth]{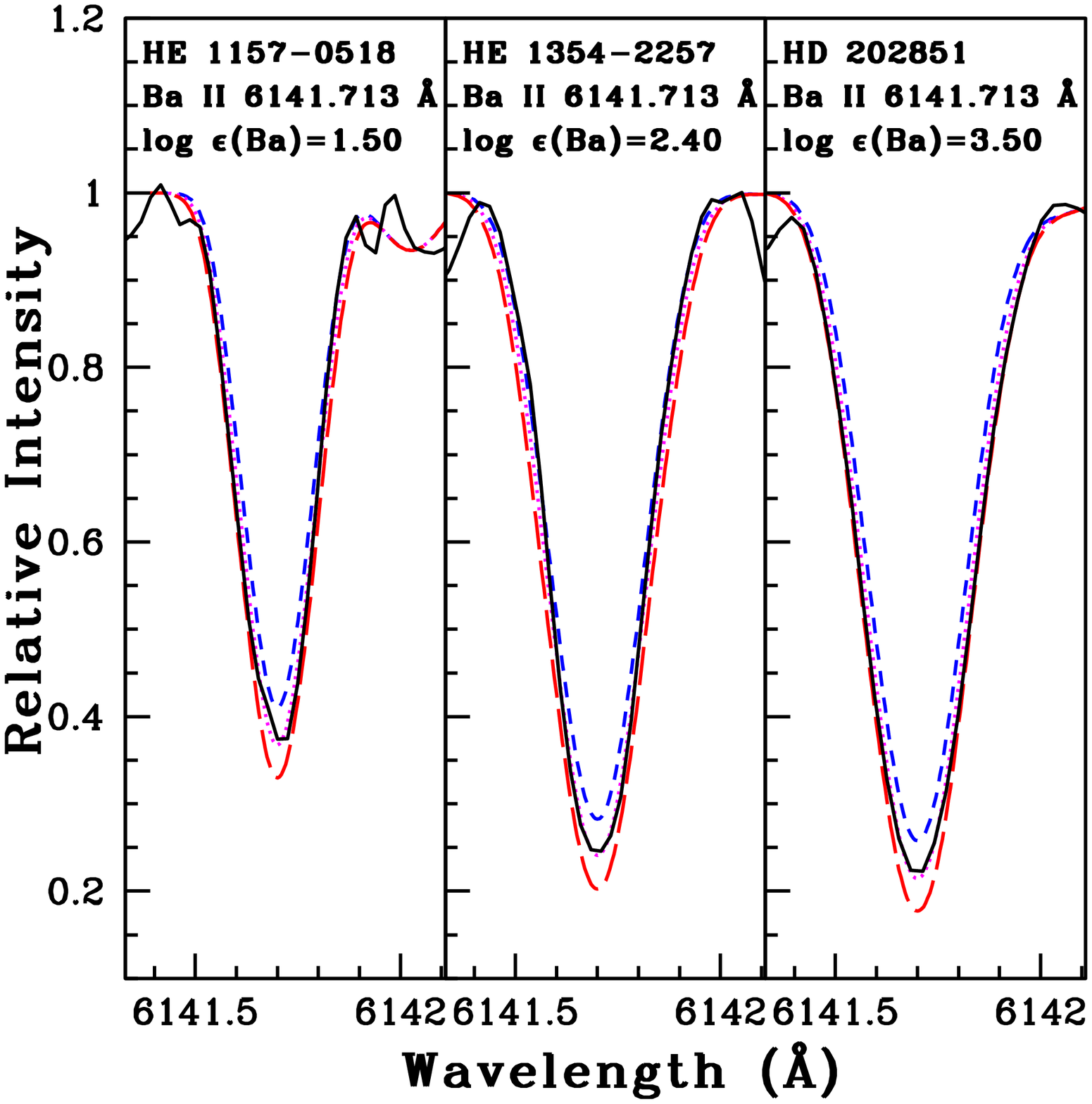}
\includegraphics[width=\columnwidth]{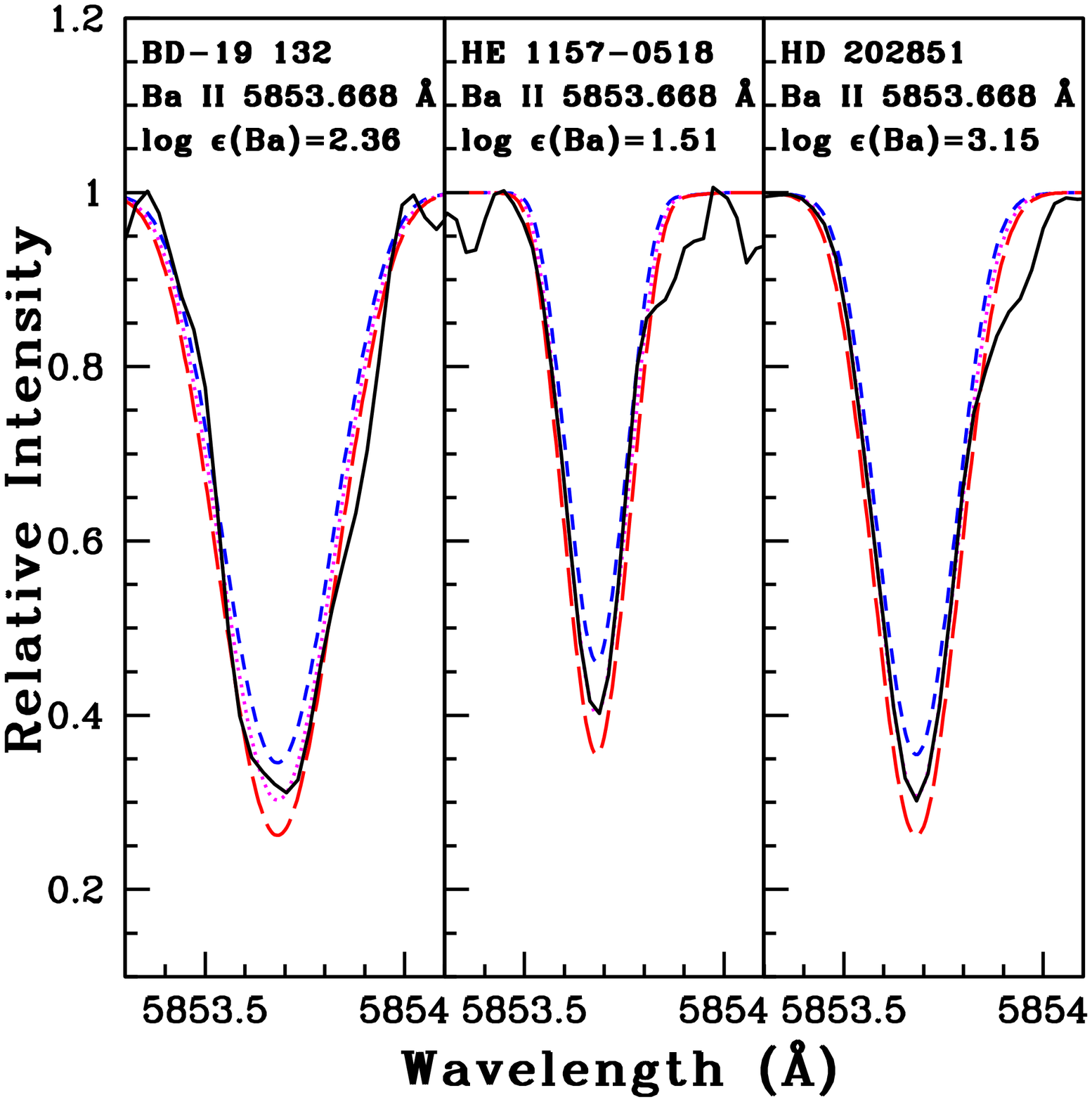}
\caption{ Synthesis of Ba II 5853.668 {\rm \AA} (lower panel) and 6141.713 {\rm \AA} (upper panel)
lines. 
Dotted and solid lines represent synthesized and observed spectra respectively. 
Short dashed and long dashed lines represent the synthetic spectra 
for $\Delta$ [Ba/Fe] = $-$0.3 and +0.3 respectively.} \label{Ba_fits}
\end{figure}

The barium abundance is derived from the spectral synthesis calculation of 
Ba II 5853.668 {\rm \AA} line in BD$-$19 132, HE~1157$-$0518, and HD~202851.
This line is blended in all other stars and could not be used for abundance 
determination. This line is known to be affected by NLTE effect.
The NLTE correction to the abundance derived from this line is adopted from 
\cite{Andrievsky_2009}, which are +0.22, +0.41, and +0.35 dex respectively for 
the above objects.
The Ba II 6141.713, 4934.076, and 6496.897 {\rm \AA} lines are also used for 
the estimation of Ba abundance. The line list for Ba II 6496.897 {\rm \AA} line is
taken from linemake \footnote{linemake  
contains laboratory atomic data (transition probabilities, 
hyperfine and isotopic substructures) published by the Wis-
consin Atomic Physics and the Old Dominion Molecular Physics
groups. These lists and accompanying line list assembly software
have been developed by C. Sneden and are curated by V. Placco at
\url{https://github.com/vmplacco/linemake}.} atomic and molecular line database \citep{Placco_2021}.
The hyper-fine components of other lines are taken from 
\cite{Mcwilliam_1998}. Barium is enhanced in all the program stars with [Ba/Fe]$>$1.
The spectral synthesis fits for Ba II 5853.668 and 6141.713 {\rm \AA} lines
are shown in Figure \ref{Ba_fits}.

\par Spectral synthesis calculation of La II 4748.726, 5259.379, 4921.776,
5301.969, 5303.528 {\rm \AA} lines, whenever available, is used to derive 
the abundance of lanthanum in the program stars. The hyper-fine components 
of La II  4921.776 {\rm \AA} line are taken from
\cite{Jonsell_2006} and 5301.969, 5303.528 {\rm \AA} lines from \cite{Lawler_2001}, 
whereas it is not available for 4748.726 and  5259.379 {\rm \AA} lines. We could not 
find any useful lines due to La in the star HE~1157$-$0518 for abundance determination. 
In all other stars it is enhanced with [La/Fe]$\geq$1.41 with BD$-$19 132 showing the  highest enhancement  of [La/Fe]$\sim$2.66. 
Equivalent width measurement of several spectral lines were used to derive the
abundances of Ce, Pr, and Nd. All these elements are enhanced in all our program stars  with [X/Fe]$>$1.40. 

\par Finally, the [ls/Fe], [hs/Fe], and [hs/ls] ratios of the program stars are
estimated, where ls is the light s-process elements; Sr, Y, and Zr and hs is the 
heavy s-process elements; Ba, La, Ce, and Nd. We have also estimated [s/Fe] ratio, 
an indicator of total s-process content of the star (s refers to the s-process elements Sr, Y, Zr, Ba, La, Ce, and Nd). 
The neutron-density dependent [Rb/Zr] ratio is also estimated. 
The values for these ratios are given in Table \ref{hs_ls}.
These ratios are discussed in section 8.

{\footnotesize
\centering
\begin{table*}
\caption{Estimates of  [ls/Fe], [hs/Fe], [s/Fe], [hs/ls], [Rb/Zr]} \label{hs_ls}
\begin{tabular}{lcccccc}
\hline                       
Star name          & [Fe/H]   & [ls/Fe] & [hs/Fe]  & [s/Fe] & [hs/ls]    & [Rb/Zr]    \\ 
\hline
BD$-$19 132        & $-$1.86  & 1.38    & 2.52     & 2.34   & 1.14       & $-$2.07      \\  
BD$-$19 290        & $-$2.86  & 1.05    & 1.89     & 1.61   & 0.84       & $-$0.25     \\
HE~1157$-$0518     & $-$2.42  & 1.27    & 1.86     & 1.37   & 0.59       & --           \\
HE~1304$-$2111     & $-$2.34  & 0.98    & 1.56     & 1.37   & 0.58       & --         \\
HE~1354$-$2257     & $-$2.11  & 1.85    & 1.61     & 1.71   & $-$0.24    & --          \\
BD+19 3109         & $-$2.23  & 0.91    & 1.64     & 1.40   & 0.73       & $-$0.72      \\   
HD~202851          & $-$0.85  & 1.31    & 1.83     & 1.61   & 0.52       & $-$1.95     \\         
\hline
\end{tabular}
\end{table*}
}

\subsubsection{\textbf{The r-process elements: Sm, Eu}}
\begin{figure}
\centering
\includegraphics[width=\columnwidth]{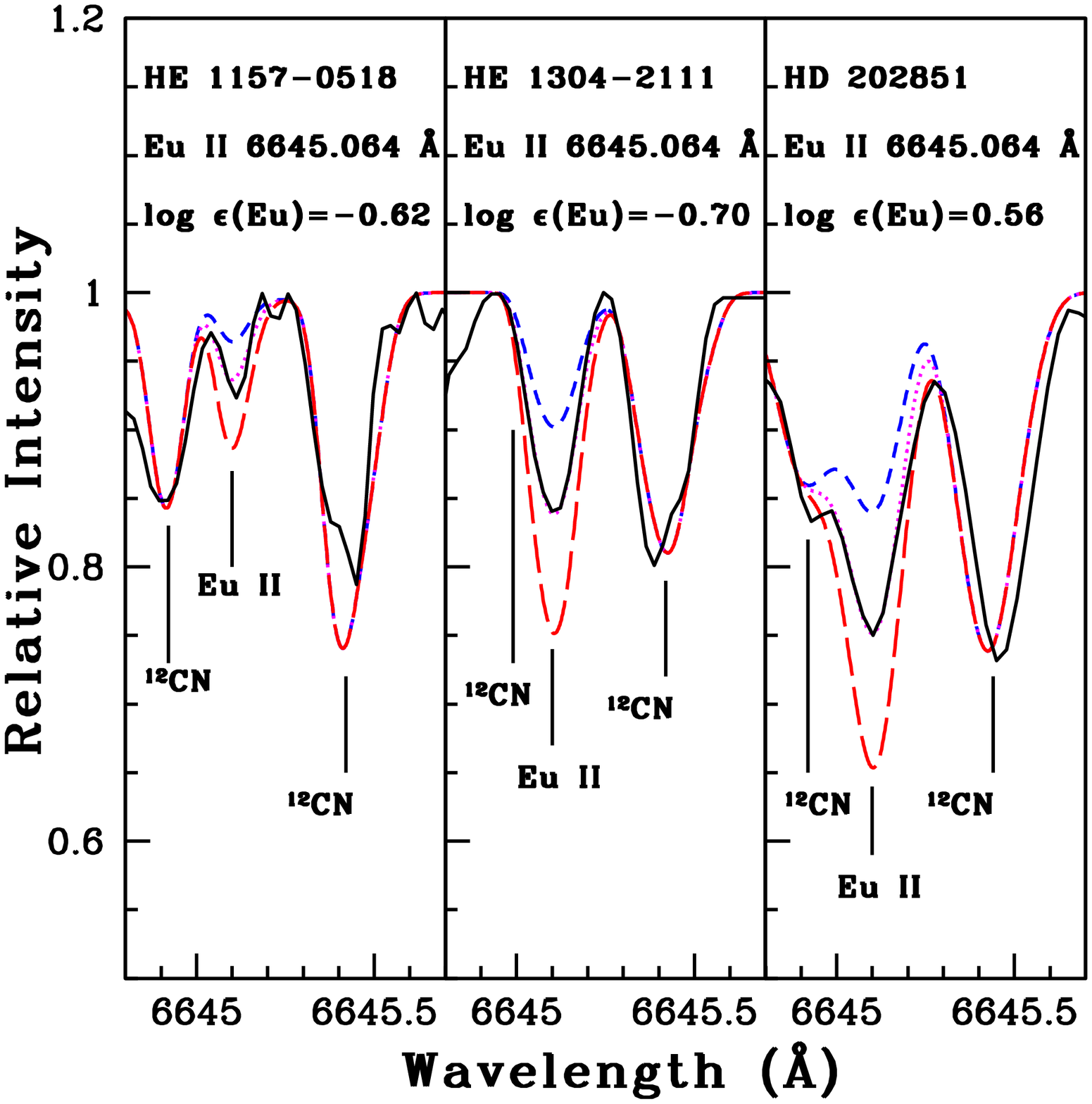}
\caption{ Synthesis of Eu II 6645.064 {\rm \AA} line. 
Dotted and solid lines represent synthesized and observed spectra respectively. 
Short dashed and long dashed lines represent the synthetic spectra 
for $\Delta$ [Eu/Fe] = $-$0.3 and +0.3 respectively.} \label{Eu_fits}
\end{figure}

The abundance of samarium is derived from the equivalent width measurement of 
Sm II lines listed in Table \ref{linelist2}. The [Sm/Fe] values in our program
stars are in the range 1.44 - 2.09. 
The europium abundance is derived from the spectral synthesis calculation of 
Eu II 6645.064 {\rm \AA} line in all our program stars. In HE~1354$-$2257, we could 
use Eu II 4129.725 {\rm \AA} line also. In all other program stars, this line is not usable 
for abundance determination. This line shows NLTE effect and the appropriate correction ($\sim$+0.16)
has been adopted from \cite{Mashonkina_2008}. None of the stars has 
good Eu II 6437.640 {\rm \AA} line usable for abundance determination.
The hyper-fine components of Eu are taken from \cite{Worley_2013}.
All the stars show enhancement of Eu with [Eu/Fe]$>$0.7 except the star HE~1354$-$2257
where it is moderately enhanced with [Eu/Fe]$\sim$0.47. 
The spectral synthesis fits for Eu II 6645.064 {\rm \AA} line
are shown in Figure \ref{Eu_fits}.

\par
\begin{figure}
\centering
\includegraphics[width=\columnwidth]{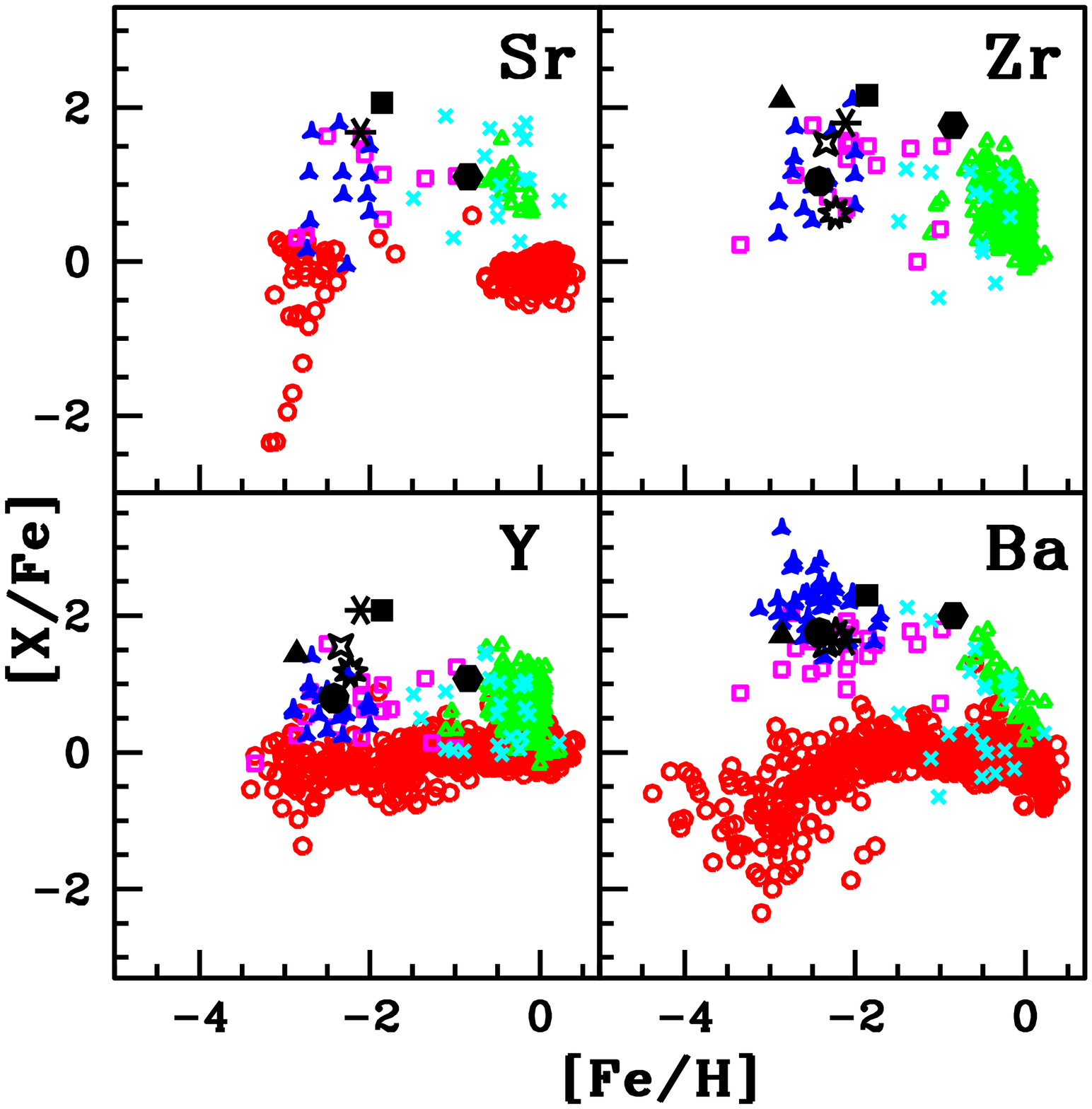}
\includegraphics[width=\columnwidth]{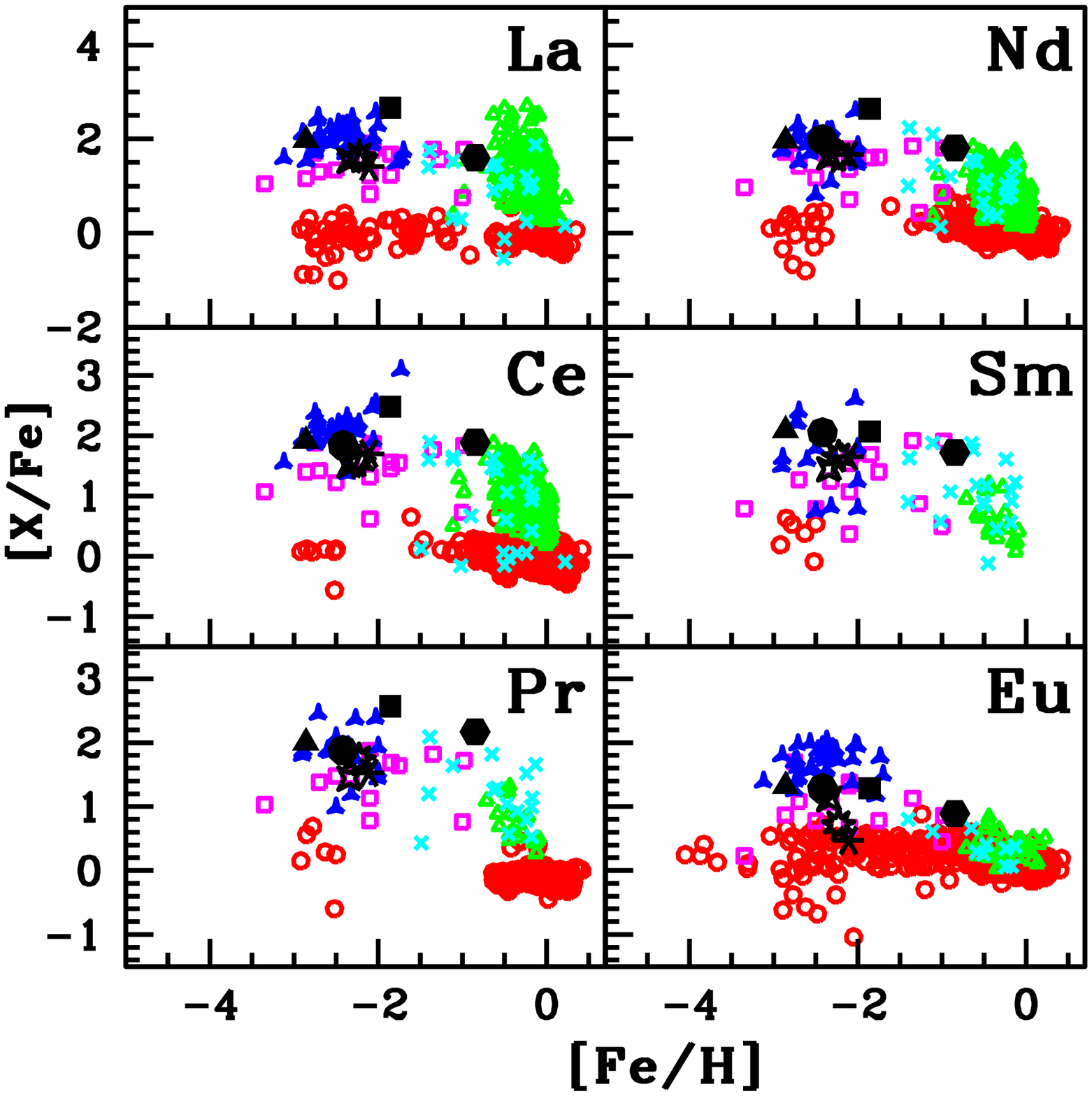}
\caption{Observed [X/Fe] ratios of the heavy elements in the  
program stars with respect to metallicity [Fe/H].   
Symbols have same meaning as in Figure \ref{light_elements}} \label{heavy_elements}
\end{figure}

 In Tables \ref{abundance_table1} ad \ref{abundance_table2}, the abundances derived from the 
lines showing NLTE effect are given separately after the NLTE correction. 
Figure \ref{heavy_elements} shows a comparison of the observed abundances of heavy elements
in our program stars with their counterparts in normal stars and other chemically peculiar 
stars. From the figure, it is clear that our program stars show clear over-abundance of
heavy elements with respect to the normal stars. 
A comparison of the elemental abundances in the program stars with the
literature value is given in Table \ref{abundance_comparison_literature}.

{\footnotesize
\begin{table*}
\caption{Comparison of the elemental abundances of our program stars with the literature values.}  \label{abundance_comparison_literature}
\resizebox{0.6\textheight}{!}{\begin{tabular}{lccccccccc}
\hline                       
Star name           & [Fe /H]   & [C/Fe]    & [N/Fe]  & [Mg/Fe]    & [Ca/Fe]   & [Ti/Fe]   & [Cr/Fe]    & Ref. \\
\hline
HE~1157$-$0518      & $-$2.42   & 2.31      & 1.52     & 0.27      & 0.15      & 0.06      & $-$0.04    & 1    \\
                    & $-$2.40   & 2.15      & 1.56     & 0.50      & 0.16      & 0.09      & $-$0.28    & 2  \\
HD~202851           & $-$0.85   & 0.72      & 1.42     & 0.27      & $-$0.03   & 0.30      & 0.46       & 1 \\ 
                    & $-$0.70   & -         & -        & -         & -         & 0.00      & $-$0.20    & 3  \\ 
\hline                       
\end{tabular}}

\resizebox{0.6\textheight}{!}{\begin{tabular}{lcccccccc}
\hline                       
Star name           & [Fe /H]   & [Y/Fe]    & [Ba/Fe]  & [La/Fe]   & [Ce/Fe]   & [Nd/Fe]    & Ref. \\
\hline
HE~1157$-$0518      & $-$2.42   & 0.79      & 1.75     & -         & 1.84      & 1.99       & 1    \\
                    & $-$2.40   & -         & 2.14     & -         & -         & -          & 2  \\
HD~202851           & $-$0.85   & 1.08      & 2.00     & 1.60      & 1.89      & 1.81       & 1 \\ 
                    & $-$0.70   & 1.30      & 1.60     & 1.40      & 1.30      & 1.60       & 3  \\ 
\hline                       
\end{tabular}}
 
References: 1. This work, 2. \cite{Aoki_2007}, 3. \cite{Sperauskas_2016} \\  
\end{table*}
}

\section{ABUNDANCE UNCERTAINTIES}
The uncertainties on the abundances of all the elements, log$\epsilon$, are 
calculated following the detailed procedure in \cite{Shejeelammal_2020}. \\

\noindent The uncertainty in the abundance of an element is given by; \\
\begin{center}
$\sigma_{log\epsilon}^{2}$ = $\sigma_{ran}^{2}$ + $(\frac{\partial log \epsilon}{\partial T})^{2}$ $\sigma_{T\rm_{eff}}^{2}$ + $(\frac{\partial log \epsilon}{\partial log g})^{2}$ $\sigma_{log g}^{2}$ + 
  $(\frac{\partial log \epsilon}{\partial \zeta})^{2}$ $\sigma_{\zeta}^{2}$ + $(\frac{\partial log \epsilon}{\partial [Fe/H]})^{2}$ $\sigma_{[Fe/H]}^{2}$ \\    
\end{center}

\noindent where $\Delta$T$_{eff}$$\sim$ $\pm$100 K, $\Delta$log g$\sim$ $\pm$0.2 dex, 
$\Delta$$\zeta$$\sim$ $\pm$0.2 km s$^{-1}$, and $\Delta$[Fe/H]$\sim$ $\pm$0.1 dex are the typical uncertainties
in the atmospheric parameters. The $\sigma_{ran}$ = $\frac{\sigma_{s}}{\sqrt{N}}$,
is the random error in the elemental abundance which arises from the line-to-line 
scatter (standard deviation, $\sigma_{s}$) of the abundances of a particular 
element species derived from N number of lines.
 Finally, total uncertainty in [X/Fe] is calculated by;\\
$\sigma_{[X/Fe]}^{2}$ = $\sigma_{X}^{2}$ + $\sigma_{Fe}^{2}$. 

In order to simplify the calculation, we made the assumption that the 
uncertainties due to different parameters are independent, 
following \cite{deCastro_2016}, \cite{Karinkuzhi_2018}, \cite{Cseh_2018}. 
However, since most of the stellar atmospheric parameters are correlated, 
our estimates are to be taken as upper limits. 
The differential abundances, $\Delta$log$\epsilon$, for the variations in each stellar
atmospheric parameters calculated for the star HD~202851 is given in Table \ref{differential_abundance}.

{\footnotesize
\begin{table*}
\caption{Change in the abundances ($\Delta$log$\epsilon$) of different elemental species with the variations
in stellar atmospheric parameters for the representative star HD~202851 (columns 2 - 5). The
rms uncertainty, computed from the
random error ($\sigma_{ran}$) and the differential abundances given in second to fifth columns, is given in sixth column. Total uncertainty 
in [X/Fe] of each elemental species is given in seventh column.}  
\label{differential_abundance}
\resizebox{0.75\textwidth}{!}{\begin{tabular}{lcccccc}
\hline                       
Element & $\Delta$T$_{eff}$  & $\Delta$log g  & $\Delta$$\zeta$       & $\Delta$[Fe/H] & ($\Sigma \sigma_{i}^{2}$)$^{1/2}$ & $\sigma_{[X/Fe]}$  \\
        & ($\pm$100 K)       & ($\pm$0.2 dex) & ($\pm$0.2 kms$^{-1}$) & ($\pm$0.1 dex) &                                   &       \\
\hline
C	        & $\pm$0.07	       & $\pm$0.10	    & $\mp$0.10	            & $\pm$0.10	     & 0.19	  & 0.24   \\
N	        & $\pm$0.13	       & $\pm$0.03	    & 0.00	                & $\pm$0.02	     & 0.14	  & 0.20   \\
O	        & 0.00	           & $\pm$0.04	    & $\pm$0.02	            & 0.00	         & 0.05	  & 0.15   \\
Na I	    & $\pm$0.06	       & $\mp$0.02	    & $\mp$0.06	            & $\mp$0.01	     & 0.10	  & 0.17   \\
Mg I	    & $\pm$0.07	       & $\mp$0.06	    & $\mp$0.07	            & $\pm$0.01	     & 0.10	  & 0.17   \\
Si I	    & $\pm$0.01	       & $\mp$0.01	    & $\mp$0.07	            & $\pm$0.01	     & 0.07	  & 0.16  \\
Ca I	    & $\pm$0.09	       & $\mp$0.02	    & $\mp$0.09	            & $\mp$0.01	     & 0.14	  & 0.20   \\
Sc II	    & $\mp$0.03	       & $\pm$0.08	    & $\mp$0.10	            & $\pm$0.03	     & 0.14	  & 0.21   \\
Ti I	    & $\pm$0.13	       & $\mp$0.01	    & $\mp$0.07	            & $\mp$0.02	     & 0.16	  & 0.21   \\
Ti II	    & $\mp$0.01	       & $\pm$0.07	    & $\mp$0.11	            & $\pm$0.03	     & 0.15	  & 0.21    \\
V I	      & $\pm$0.15	       & $\mp$0.01      & $\mp$0.04	            & $\mp$0.02	     & 0.16	  & 0.21   \\
Cr I	    & $\pm$0.12	       & $\mp$0.01      & $\mp$0.09	            & $\mp$0.01	     & 0.16	  & 0.21   \\
Cr II	    & $\mp$0.07	       & $\pm$0.08	    & $\mp$0.03	            & $\pm$0.02	     & 0.16	  & 0.22   \\
Mn I	    & $\pm$0.10	       & $\mp$0.03	    & $\mp$0.13	            & $\mp$0.01	     & 0.17	  & 0.22   \\
Fe I	    & $\pm$0.09	       & $\pm$0.01	    & $\mp$0.11	            & 0.00  	       & 0.14  	& -- \\
Fe II	    & $\mp$0.09	       & $\pm$0.09	    & $\mp$0.07	            & $\pm$0.05  	   & 0.16  	& -- \\
Co I	    & $\pm$0.09	       & $\pm$0.01	    & $\mp$0.03	            & 0.00	         & 0.10	  & 0.17   \\
Ni I	    & $\pm$0.07	       & $\pm$0.01	    & $\mp$0.08	            & 0.00	         & 0.11	  & 0.18   \\
Cu I	    & $\pm$0.14	       & $\mp$0.01	    & $\mp$0.18	            & $\mp$0.01	     & 0.23	  & 0.27   \\
Zn I	    & $\mp$0.04	       & $\pm$0.05	    & $\mp$0.09	            & $\pm$0.02	     & 0.11	  & 0.18   \\
Rb I	    & $\pm$0.10	       & 0.00	          & $\mp$0.03	            & 0.00	         & 0.11	  & 0.18   \\
Sr I	    & $\pm$0.14	       & $\mp$0.02	    & $\mp$0.16	            & $\pm$0.02	     & 0.22	  & 0.26   \\
Y I	      & $\pm$0.16	       & 0.00	          & $\mp$0.01	            & $\mp$0.01	     & 0.16	  & 0.22   \\
Y II	    & 0.00	           & $\pm$0.05	    & $\mp$0.14	            & $\pm$0.03	     & 0.16	  & 0.22   \\
Zr I	    & $\pm$0.11	       & $\mp$0.01	    & $\mp$0.10	            & $\mp$0.02	     & 0.15	  & 0.21   \\
Zr II	    & $\mp$0.01	       & $\pm$0.06	    & $\mp$0.19	            & $\pm$0.02	     & 0.20	  & 0.26   \\
Ba II	    & $\pm$0.03	       & $\pm$0.02	    & $\mp$0.07	            & $\pm$0.05	     & 0.10	  & 0.19   \\
La II     & $\pm$0.02        & $\pm$0.08      & $\mp$0.09             & $\pm$0.03      & 0.13   & 0.20    \\
Ce II	    & $\pm$0.01	       & $\pm$0.07	    & $\mp$0.13	            & $\pm$0.03	     & 0.16	  & 0.22   \\
Pr II	    & $\pm$0.02	       & $\pm$0.08	    & $\mp$0.11	            & $\pm$0.03	     & 0.16	  & 0.22   \\
Nd II	    & $\pm$0.02	       & $\pm$0.07	    & $\mp$0.14	            & $\pm$0.03	     & 0.16	  & 0.23   \\
Sm II	    & $\pm$0.01	       & $\pm$0.08	    & $\mp$0.08	            & $\pm$0.03	     & 0.12	  & 0.20   \\
Eu II	    & $\mp$0.02	       & $\pm$0.08	    & $\mp$0.02	            & $\pm$0.03	     & 0.09	  & 0.18   \\
\hline
\end{tabular}}

\end{table*}
}

\section{Kinematic  Analysis}
It is equally as important as the chemical composition of a star to understand 
its kinematics and the Galactic population to which it belongs. 
We have calculated the components of the spatial velocities of the program stars 
with respect to the Local Standard of Rest (LSR), U$\rm_{LSR}$, V$\rm_{LSR}$, and W$\rm_{LSR}$,
following the procedures discussed in \cite{Johnson_1987} and \cite{Bensby_2003}.  

The spatial velocity components with respect to the LSR is given by;\\
\begin{center}
  $(U, V, W)_{LSR} =(U,V,W)+(U, V, W)_{\odot}$ km/s.
  \end{center}
(U, V, W)$_{\odot}$ = (11.1, 12.2, 7.3) km/s is the solar motion with respect to LSR \citep{Schonrich_2010}.
Total space velocity is calculated as; \\
 $V\rm_{spa}^{2}=U\rm_{LSR}^{2}+V\rm_{LSR}^{2}+W\rm_{LSR}^{2}$. 
 
 \par The membership of the stars to each Galactic population is determined following the 
 probability calculations given in \cite{Mishenina_2004}, \cite{Bensby_2003, Bensby_2004}, and \cite{Reddy_2006}.
 The proper motions in RA and DEC ($\mu_{\alpha}$ and $\mu_{\delta}$), parallax ($\pi$) 
 are taken from SIMBAD astronomical database and Gaia DR2 (\citealt{Gaia_2018}, \url{https://gea.esac.esa.int/archive/}).
 The detailed procedure can be found in \cite{Shejeelammal_2021}. 
 The components of the spatial velocity, total space velocity and probability 
 that a star belongs to a particular Galactic population are presented in Table \ref{kinematic analysis}.
 While the stars BD$-$19 290, HE~1157$-$0518, HE~1354$-$2257, and BD+19 3109 belong to the Galactic halo,
 BD$-$19 132, HE~1304$-$2111, and HD~202851 belong to the Galactic thin disk.

{\footnotesize
\begin{table*}
\caption{Estimates of spatial velocity and probabilities for 
the membership to the Galactic population of the program stars} \label{kinematic analysis} 
\begin{tabular}{lccccccc} 
\hline                       
Star name             & U$_{LSR}$             & V$_{LSR}$           & W$_{LSR}$         & V$_{spa}$    & P$_{thin}$ & P$_{thick}$ & P$_{halo}$ \\
                      & (kms$^{-1}$)          & (kms$^{-1}$)        & (kms$^{-1}$)      & (kms$^{-1}$) &            &             &       \\
\hline
BD$-$19 132           & 13.66$\pm$1.06        & $-$69.13$\pm$12.91     & $-$6.75$\pm$1.79       & 70.79$\pm$12.56     & 0.90  & 0.10   & 0.00  \\
BD$-$19 290           & $-$252.60$\pm$68.56   & $-$573.19$\pm$168.57   & $-$38.00$\pm$16.68     & 627.56$\pm$182.56   & 0.00  & 0.00   & 1.00 \\
HE~1157$-$0518        & $-$113.96$\pm$69.93   & $-$449.47$\pm$202.19   & $-$154.28$\pm$130.96   & 488.68$\pm$241.24   & 0.00  & 0.00   & 1.00 \\
HE~1304$-$2111        & 2.68$\pm$0.18         & $-$41.47$\pm$0.25      & $-$32.03$\pm$0.27      & 52.47$\pm$0.35      & 0.92  & 0.08   & 0.00 \\
HE~1354$-$2257        & $-$695.99$\pm$3045.67 & $-$1855.03$\pm$5913.33 & $-$295.05$\pm$1636.19  & 2003.14$\pm$4853.64 & 0.00  & 0.00   & 1.00 \\
BD+19 3109            & $-$122.65$\pm$1.87    & $-$253.64$\pm$13.83    & 41.66$\pm$11.43        & 284.80$\pm$11.44    & 0.00  & 0.01   & 0.99 \\
HD~202851             & 64.35$\pm$1.90        & $-$38.17$\pm$2.92      & $-$40.82$\pm$1.68      & 85.23$\pm$0.68      & 0.72  & 0.28   & 0.00 \\
\hline
\end{tabular} 
\end{table*}
}

\section{Classification of the program stars}
All the stars analyzed here are metal-poor objects with [Fe/H]$<$$-$1, except
HD~202851 which has a metallicity of [Fe/H]$\sim$$-$0.85. Five objects 
in  the sample are very metal-poor with [Fe/H]$<$$-$2.  
All the stars in our sample show enhanced abundance of carbon with 
[C/Fe]$\geq$0.70. According to \cite{Beers_2005}, the stars 
with [Fe/H]$\leq$$-$2 and [C/Fe]$>$1 are Carbon Enhanced Metal-Poor (CEMP) stars. 
However, a number of stars with [Fe/H]$<$$-$1 are identified as CEMP stars in literature 
\citep{Jonsell_2006, Goswami_2006, Aoki_2007, Goswami_2010, 
Masseron_2010, Abate_2016, Shejeelammal_2021, Karinkuzhi_2021} etc.. 
Several authors have adopted different lower limits for [C/Fe] ratio to define the CEMP 
stars, for instance, [C/Fe]$\geq$0.7 \citep{Carollo_2012, Lee_2013, Norris_2013b, Skuladottir_2015},
 [C/Fe]$>$0.9 \citep{Jonsell_2006, Masseron_2010}, [C/Fe]$>$1
\citep{Abate_2016, Hansen_2019}. However, in the case of
evolved metal-poor giants, the carbon abundance will be lower compared to their
earlier evolutionary stages due to the internal mixing of the CNO processed 
material from the stellar interiors \citep{Gratton_2000, 
Spite_2005, Spite_2006, Aoki_2007, Placco_2014}. 
This will be discussed in detail in the next section.
So while setting the limit for the [C/Fe] ratio of CEMP stars, the evolutionary stage of the
star should also be taken care of. \cite{Aoki_2007} have taken into consideration the 
luminosity of the star, which represents the evolutionary stage, 
to account for the depletion of surface carbon to 
set a threshold [C/Fe] that defines CEMP star population.
The following empirical definition is given by \cite{Aoki_2007} for the CEMP stars;\\

\noindent (i) [C/Fe]$\geq$+0.7; if log(L/L$_{\odot})$$\leq$2.3 \\

\noindent(ii) [C/Fe]$\geq$+3.0$-$log(L/L$_{\odot})$; if log(L/L$_{\odot})$$>$2.3  \\

\begin{figure}
\centering
\includegraphics[width=\columnwidth]{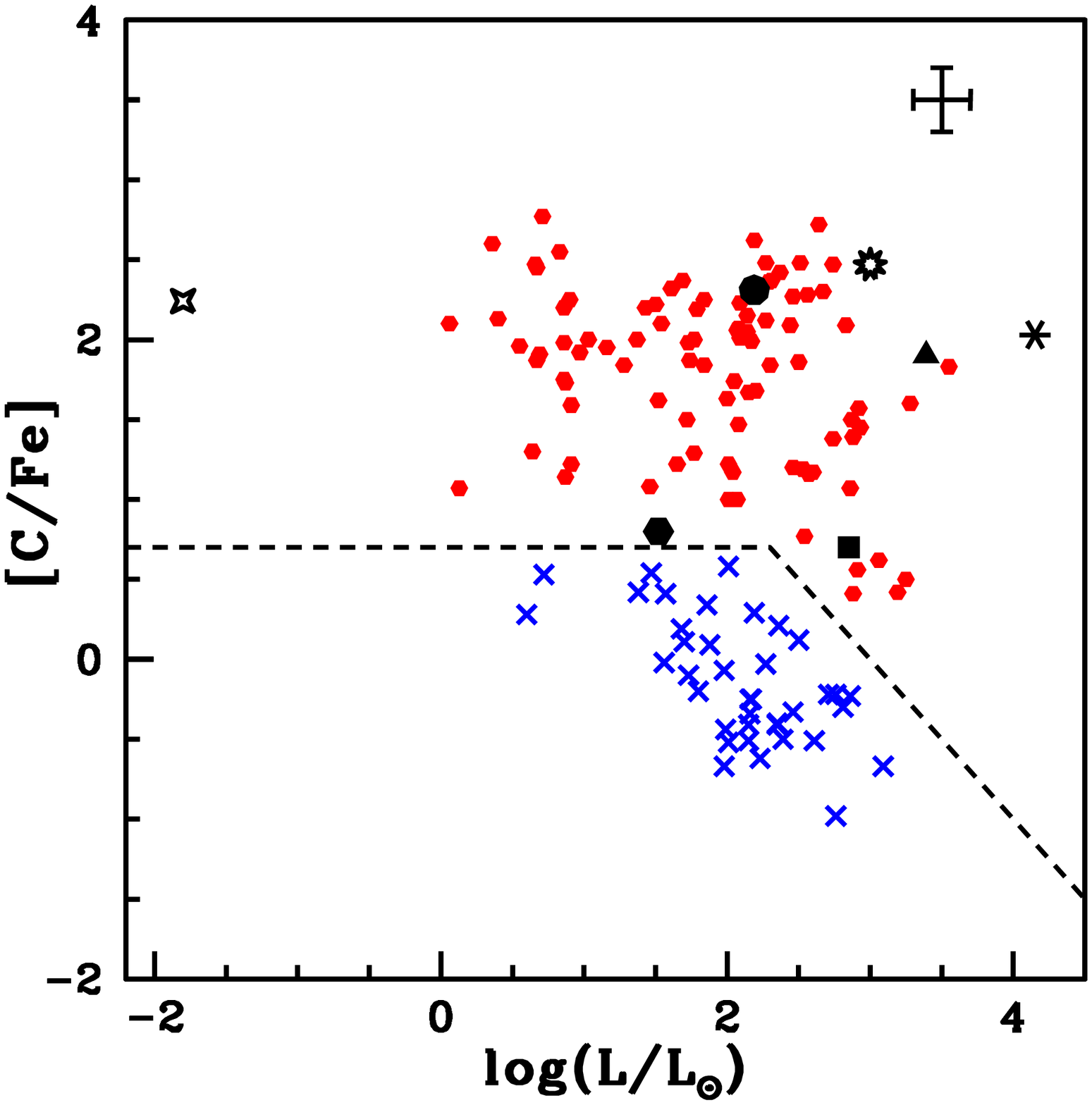}
\caption{Observed [C/Fe] ratios as a function of log(L/L$_{\odot})$.   
Red filled hexagons represent CEMP stars from literature 
(\citealt{Aoki_2007} and references therein, \citealt{Goswami_2016, 
Purandardas_2019, Shejeelammal_2021, Karinkuzhi_2021, Goswami_2021}).
Blue crosses represent carbon-normal metal-poor stars from literature 
\citep{Aoki_2005, Aoki_2007, Cayrel_2004, Honda_2004}. 
BD$-$19 132 (filled square), BD$-$19 290 (filled triangle), HE~1157$-$0518 (filled circle), 
HE~1304$-$2111 (four-sided star), HE~1354$-$2257 (six-sided cross), BD+19 3109 (nine-sided star), 
and HD~202851 (filled hexagon). 
The threshold [C/Fe] is shown by the dashed line. A representative error bar is shown at the top right corner. 
} \label{CEMP_Aoki}
\end{figure}

\noindent We have adopted this criteria to define the CEMP stars in our sample, 
which is demonstrated in Figure \ref{CEMP_Aoki}. It is clear from the figure that 
CEMP stars and non-carbon-enhanced metal-poor stars belong to two different regions
of the diagram. Even though the star HD~202851 shows a [C/Fe] value that resembles 
CEMP stars according to this definition, because of its higher metallicity, [Fe/H]$\sim$$-$0.85,
we classify it as a CH star, whereas all other six stars in our sample are 
CEMP stars. 

There exist a number classification scheme in literature 
for the sub-classification of CEMP stars \citep{Abate_2016, Frebel_2018,
Hansen_2019}, the \cite{Beers_2005} being 
the pioneering one. \cite{Beers_2005} and \cite{Abate_2016} considered
[Ba/Fe], [Eu/Fe], and [Ba/Eu] ratios to define CEMP sub-classes,
\cite{Hansen_2019} considered [Sr/Ba] ratio, and \cite{Frebel_2018} considered [Sr/Ba], [Sr/Eu],
[Ba/Pb], and [La/Eu] ratios. \cite{Goswami_2021} revisited
all these classification schemes and came up with a slightly modified criteria in terms of 
[Ba/Eu] and [La/Eu]. We adopted this scheme to sub-classify our CEMP stars which is as follows; \\

\noindent-CEMP-s; [Ba/Fe]$\geq$1  \\
\indent (i) [Eu/Fe]$<$1, [Ba/Eu]$>$0 and/or [La/Eu]$>$0.5 \\
\indent (ii) [Eu/Fe]$\geq$1, [Ba/Eu]$>$1 and/or [La/Eu]$>$0.7 \\

\noindent-CEMP-r/s; [Ba/Fe]$\geq$1, [Eu/Fe]$\geq$1 \\
\indent (i) 0$\leq$[Ba/Eu]$\leq$1 and/or 0$\leq$[La/Eu]$\leq$0.7 \\

\begin{figure}
\centering
\includegraphics[width=\columnwidth]{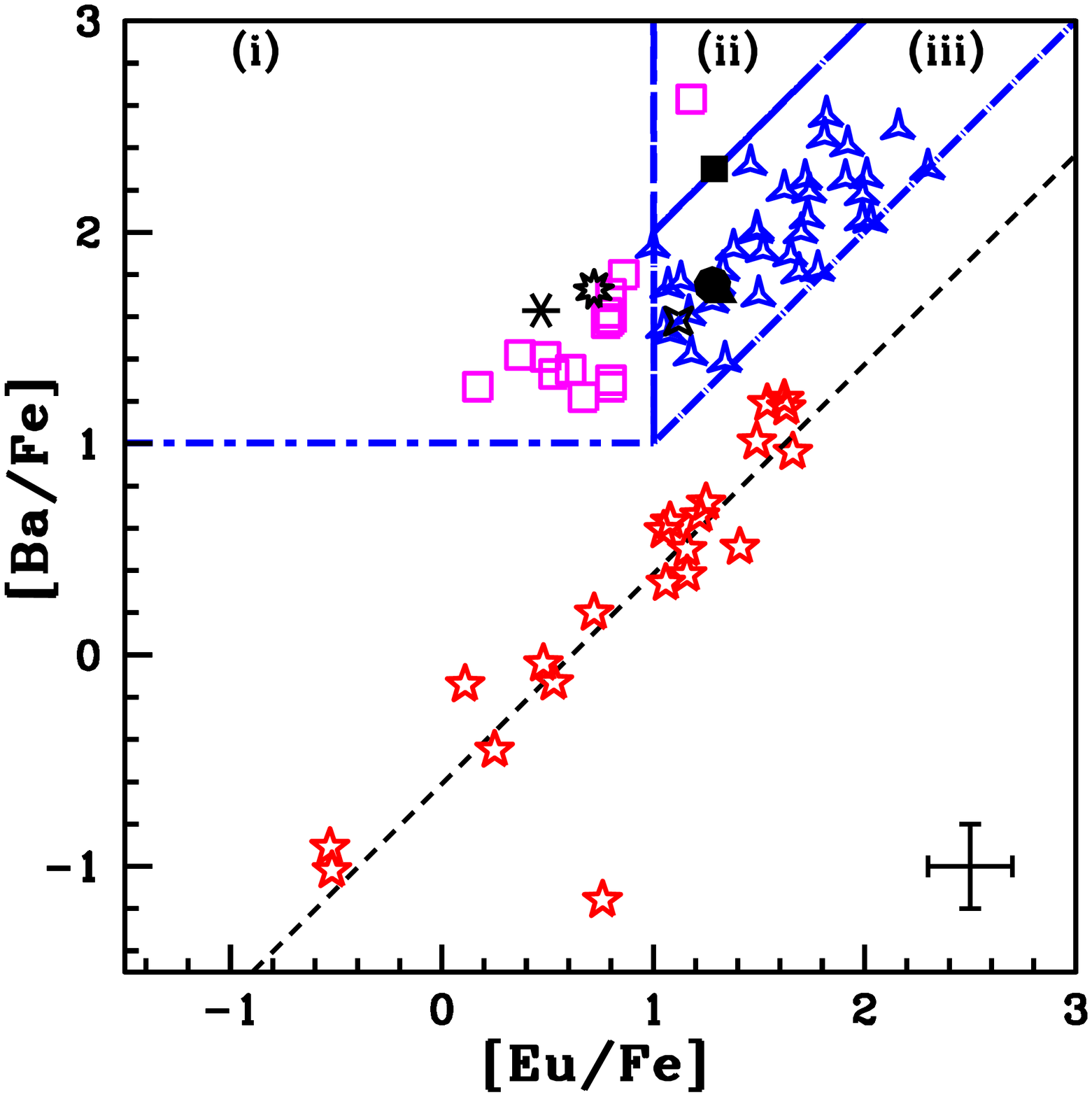}
\includegraphics[width=\columnwidth]{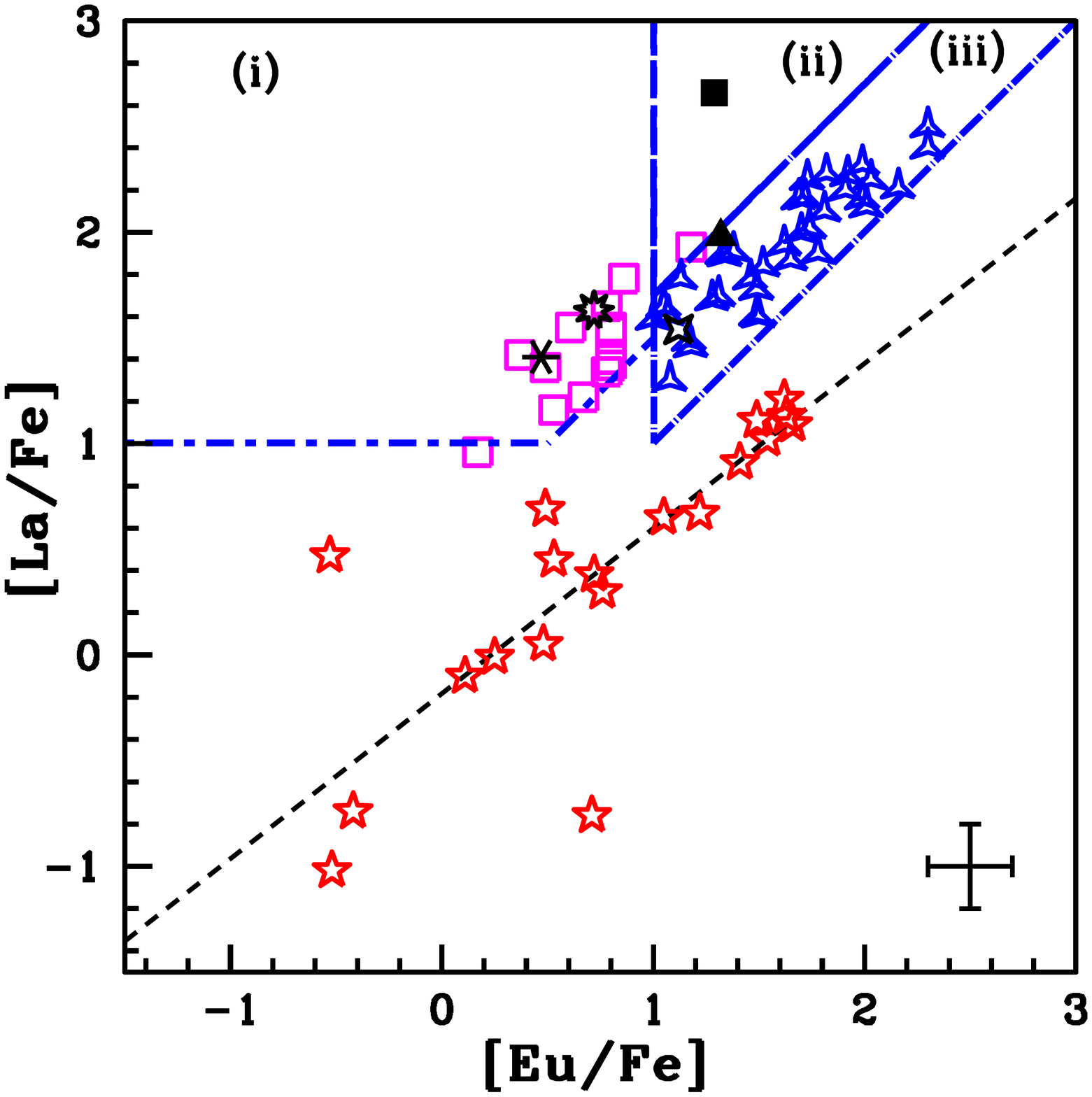}
\caption{Observed [Ba/Fe] (upper panel) and [La/Fe] (lower panel)
as functions of observed [Eu/Fe] for CEMP and r stars.   
Magenta squares, blue starred triangles and red five-sided stars represent CEMP-s
CEMP-r/s and r (including both CEMP-r and rI/rII stars) stars respectively 
from literature \citep{Masseron_2010, Shejeelammal_2021, Karinkuzhi_2021, Goswami_2021}. 
BD$-$19 132 (filled square), BD$-$19 290 (filled triangle), HE~1157$-$0518 (filled circle), 
HE~1304$-$2111 (four-sided star), HE~1354$-$2257 (six-sided cross),
BD+19 3109 (nine-sided star), and HD~202851 (filled hexagon). Region (i) and (ii) are the CEMP-s and (iii) is CEMP-r/s
star region as given by \cite{Goswami_2021}. The dashed line is the least-square 
fit to the observed abundances in r-stars. A representative error bar is shown at the bottom right corner of each panel.} \label{Ba_La_Eu}
\end{figure}

\noindent Figure \ref{Ba_La_Eu} depicts this classification scheme.
The regions marked as (i) and (ii) are two CEMP-s star
regions corresponding to the above mentioned conditions (i) and (ii) respectively
and the region marked as (iii) corresponds to  CEMP-r/s stars.
In this  figure, the stars BD$-$19 290, HE~1157$-$0518, and HE~1304$-$2111 occupy
the region (iii), and they are found to be CEMP-r/s stars. We could not
locate the position of the star HE~1157$-$0518 in the [La/Fe] - [Eu/Fe] 
plot as La abundance could not be estimated in this star.
The objects HE~1354$-$2257 and BD+19 3109 clearly lie in the region (i) in both the
panels and they are CEMP-s stars. The star BD$-$19 132 lies marginally in region (ii) in the [Ba/Fe] - [Eu/Fe] plot, but it clearly 
lies in the region (ii) in the [La/Fe] - [Eu/Fe] plot; hence it is identified as a CEMP-s star.
The estimated [Ba/Eu], [La/Eu] ratios of the program stars are given in Table \ref{Ba_La_Eu_ratio}.

{\footnotesize
\begin{table*}
\caption{The [Ba/Eu] and [La/Eu] ratios in the program stars} \label{Ba_La_Eu_ratio}
\begin{tabular}{lcccccc}
\hline                       
Star name          & [Fe/H]   & [C/Fe]  & [Ba/Fe]  & [Eu/Fe] & [Ba/Eu]    & [La/Eu]    \\ 
\hline
BD$-$19 132        & $-$1.86  & 0.70    & 2.30     & 1.29    & 1.01       & 1.37      \\  
BD$-$19 290        & $-$2.86  & 1.90    & 1.71     & 1.32    & 0.39       & 0.60     \\
HE~1157$-$0518     & $-$2.42  & 2.31    & 1.75     & 1.28    & 0.47       & --         \\
HE~1304$-$2111     & $-$2.34  & 2.24    & 1.58     & 1.12    & 0.46       & 0.42         \\
HE~1354$-$2257     & $-$2.11  & 2.03    & 1.63     & 0.47    & 1.16       & 0.94        \\
BD+19 3109         & $-$2.23  & 2.47    & 1.73     & 0.72    & 1.01       & 0.85      \\   
HD~202851          & $-$0.85  & 0.80    & 2.00     & 0.89    & 1.11       & 0.71     \\         
\hline
\end{tabular}
\end{table*}
}

\section{abundance analysis and discussion}
The detailed analysis of the abundance profiles of the program stars 
and their interpretation  are discussed here.
\subsection{CNO abundances}
As mentioned in the previous section, the surface carbon abundance 
(as well as nitrogen and oxygen abundances) of a star is related to its evolutionary stage. 
When a star evolves into a red giant, the surface carbon abundance decreases 
due to the internal mixing that brings the CNO processed material from the 
stellar interiors to the surface, known as 
First Dredge-Up (FDU). Some extra mixing processes in addition to the 
internal mixing is also found to operate in  more evolved giants. 
Here the mixing between the surface and the hydrogen 
burning layer where C is converted to N through CNO cycle in
the upper RGB stars tends to decrease the carbon abundance further
\citep{Charbonnel_1995}. This has been observed in more evolved metal-poor field giants 
\citep{Charbonnel_1998, Gratton_2000, Spite_2005, Spite_2006} and Globular 
and open cluster giants \citep{Gilroy_1989, Grundahl_2002, Shetrone_2003, Jacobson_2005}.
This internal mixing processes result in an increase of nitrogen abundance at the expense of 
carbon and oxygen (CN and ON cycles).    

\par
\begin{figure}
\centering
\includegraphics[width=\columnwidth]{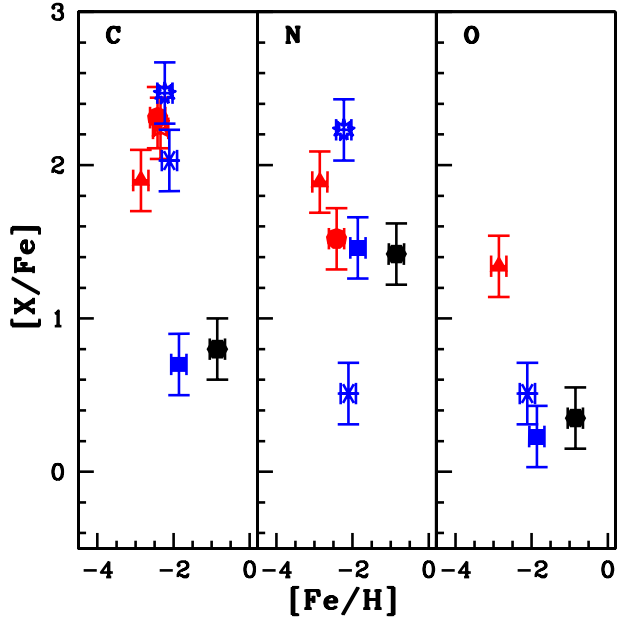}
\caption{Observed [C/Fe], [N/Fe] and [O/Fe] 
as functions of observed [Fe/H] for the program stars. Red symbols are CEMP-r/s stars,
blue symbols CEMP-s stars and black symbol CH star.    
BD$-$19 132 (filled square), BD$-$19 290 (filled triangle), HE~1157$-$0518 (filled circle), 
HE~1304$-$2111 (four-sided star), HE~1354$-$2257 (six-sided cross),
BD+19 3109 (nine-sided star), and HD~202851 (filled hexagon).} \label{CNO_abundance}
\end{figure}

\begin{figure}
\centering
\includegraphics[width=\columnwidth]{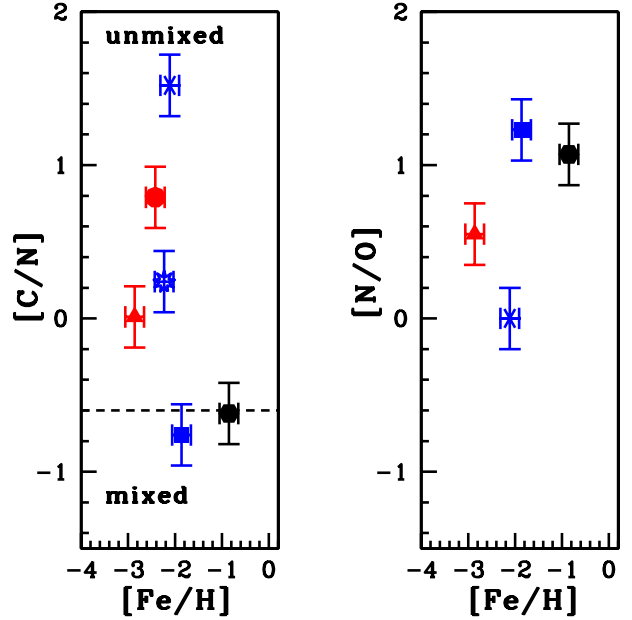}
\caption{Observed [C/N] (left panel) and [N/O] (right panel)
as functions of observed [Fe/H] for the program stars. BD$-$19 132 (filled square), 
BD$-$19 290 (filled triangle), HE~1157$-$0518 (filled circle), 
HE~1304$-$2111 (four-sided star), HE~1354$-$2257 (six-sided cross),
BD+19 3109 (nine-sided star), and HD~202851 (filled hexagon).} \label{CN_NO}
\end{figure}

\begin{figure}
\centering
\includegraphics[width=\columnwidth]{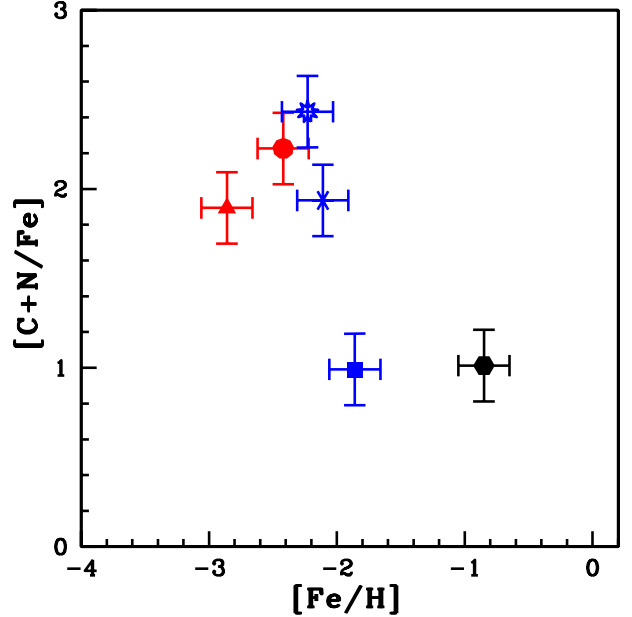}
\caption{Observed [C+N/Fe] ratio as functions of observed [Fe/H] for the 
program stars. BD$-$19 132 (filled square), BD$-$19 290 (filled triangle), HE~1157$-$0518 (filled circle), 
HE~1304$-$2111 (four-sided star), HE~1354$-$2257 (six-sided cross),
BD+19 3109 (nine-sided star), and HD~202851 (filled hexagon).} \label{C+N_Fe}
\end{figure}

 Among our program stars, HE~1157$-$0518 and HD~202851 are found to be on the ascend of 
Red Giant Branch (RGB) (Figure \ref{tracks}, log g = 2.52 and 2.20 respectively), 
and all other stars are evolved giants (0.60$\leq$log g$\leq$1.13).
The observed [C/Fe], [N/Fe], and [O/Fe] ratios of the program stars are shown in 
Figure \ref{CNO_abundance}. Carbon abundance is available for all the program stars, 
nitrogen abundance for six stars and oxygen abundance for four 
stars. Five out of the seven program stars show [C/Fe], [N/Fe]$>$1
and one star show [O/Fe]$>$1. While the five objects in our sample show 
similar [C/Fe] ratio, two objects BD$-$19 132 and HD~202851, 
show lowest (but similar) [C/Fe]. From the Figure \ref{CNO_abundance}, it 
is clear that, these two stars are enhanced in nitrogen and shows lowest [O/Fe],
may be an indication of CNO cycle and internal mixing. If that is the case, these stars are expected to show
low [C/N] and high [N/O] ratio. According to \cite{Spite_2005}, the stars with 
[C/N]$<$$-$0.6 are mixed stars and that with [C/N]$>$$-$0.6 are unmixed stars. 
These two stars lie in the region of mixed stars and show higher [N/O] ratio (Figure \ref{CN_NO}).
The Figure \ref{CNO_abundance} combined with Figure \ref{CN_NO} suggest that 
both CN and ON cycle might have operated in these two stars which altered the surface 
CNO abundances through mixing. If the nitrogen abundance is altered only due to the 
CN processing, then the [C+N/Fe] ratios of the mixed and unmixed stars will be similar,
with lower $^{12}$C/$^{13}$C ratio for the mixed stars and if ON cycle also contributed to the  N enhancement, 
such stars show [C+N] excess compared to the unmixed stars \citep{Spite_2005, Spite_2006}. But, both BD$-$19 132 and HD~202851 show 
lower [C+N/Fe] ratio compared to the other stars (Figure \ref{C+N_Fe}) 
and have relatively higher $^{12}$C/$^{13}$C ratio (18 and 42 respectively) (Table \ref{abundance_table1} and \ref{abundance_table2}).
This suggests that the nitrogen over abundance in these two stars may not have been  resulted from
an in-situ enhancement; it may be either the signature of a massive AGB companion with 
Hot-Bottom Burning (HBB) or the nitrogen enhanced primordial matter from which these stars have formed. 
We explore the possibility of HBB  using different abundance ratios in the next few sections. 

\subsection{The [hs/ls] ratio}
The [hs/ls] ratio, heavy s-process elements to light s-process elements abundance ratio,
is an indicator of s-process efficiency. At higher neutron exposures, second peak (hs) s-process
elements are produced over the first peak (ls), resulting in high [hs/ls] ratio. As the neutron exposure
increases with decreasing metallicity, the [hs/ls] ratio is anti-correlated with metallicity.
Also the neutron source $^{13}$C($\alpha$, n)$^{16}$O is found to be anti-correlated with 
metallicity \citep{Clayton_1988, Wallerstein_1997}.  Hence positive values 
for [hs/ls] ratio could be seen in low-mass, low metallicity AGB stars \citep{Goriely_2000, Busso_2001}. 
Whereas, in the case of massive AGB stars (5 - 8M$_{\odot}$), where the neutron source is $^{22}$Ne($\alpha$, n)$^{25}$Mg,
negative values for this ratio is observed \citep{Goriely_2005, Karakas_2010, Karakas_2012, vanRaai_2012, 
Karakas_2014}. 

\par
\begin{figure}
\centering
\includegraphics[width=\columnwidth]{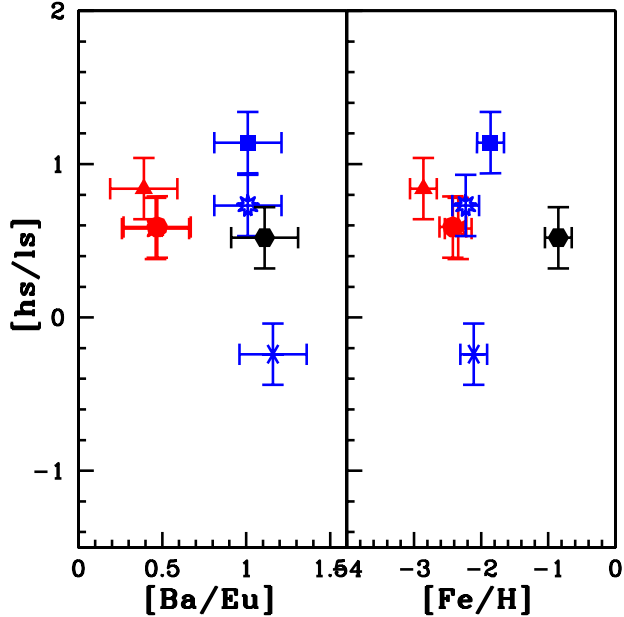}
\caption{The [hs/ls] ratio as a function of [Ba/Eu] (left panel) and [Fe/H] (right panel) 
for the program stars . 
Red symbols are CEMP-r/s stars, blue symbols CEMP-s stars and black symbol CH star.    
BD$-$19 132 (filled square), BD$-$19 290 (filled triangle), HE~1157$-$0518 (filled circle), 
HE~1304$-$2111 (four-sided star), HE~1354$-$2257 (six-sided cross),
BD+19 3109 (nine-sided star), and HD~202851 (filled hexagon).} \label{hs_ls_Ba_Fe}
\end{figure}

 The CEMP-r/s stars generally show higher [hs/ls] value compared to the CEMP-s stars
\citep{Abate_2015a, Hollek_2015}, however there is no clear distinction observed 
between the [hs/ls] ratios of the CEMP-r/s and CEMP-s stars. Analysis of \cite{Goswami_2021}
have shown that CEMP-s stars and CEMP-r/s stars peak at [hs/ls] values $\sim$0.65$\pm$0.35
and 1.06$\pm$0.32 respectively with an overlap between them in the range 0$<$[hs/ls]$<$1.5.
\cite{Karinkuzhi_2021} also noted in their sample of CEMP stars the overlap of [hs/ls] ratio between
CEMP-s and CEMP-r/s stars. 
The [hs/ls] ratio observed in our program stars are below 1.50 (Table \ref{hs_ls}), and we could not see any 
distinction between  these two classes of stars in terms of [hs/ls] ratio (Figure \ref{hs_ls_Ba_Fe}). 
All the three CEMP-r/s stars in our sample show positive values of [hs/ls] and they 
are more metal-poor than the CEMP-s stars (right panel, Figure \ref{hs_ls_Ba_Fe}), 
as also noted by \cite{Karinkuzhi_2021}.  All the stars except HE~1354$-$2257 show positive 
[hs/ls] ratio indicating a low-mass AGB companion. The star HE~1354$-$2257 has [hs/ls]$\sim$$-$0.24, 
indicating a clear over production of ls elements over hs elements, a trend observed in massive AGB stars.
There are a few CEMP-s stars in literature that show negative values for [hs/ls] ratio; for instance,
\cite{Hansen_2019} present a star HE~2158$-$5134 that shows [hs/ls]$\sim$$-$0.15 and \cite{Aoki_2002}
noted a value $\sim$$-$0.36 for the object CS~22942$-$019. Analysis of \cite{Bisterzo_2011} 
have shown that the abundance pattern in CS~22942$-$019 could be reproduced with 1.5 M$_{\odot}$ AGB model.

\subsection{Na, Mg, and heavy elements}
\begin{figure}
\centering
\includegraphics[width=\columnwidth]{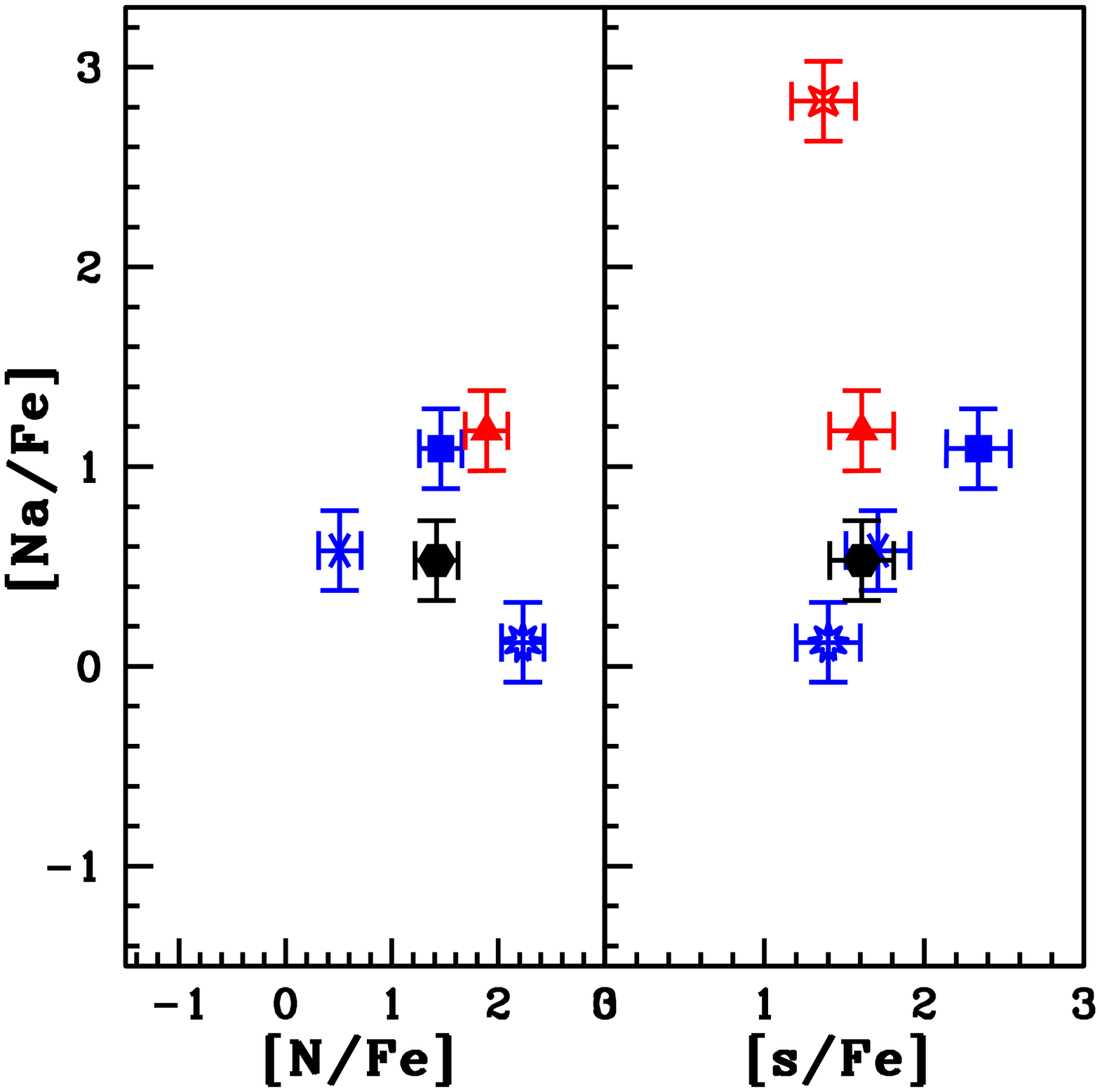}
\includegraphics[width=\columnwidth]{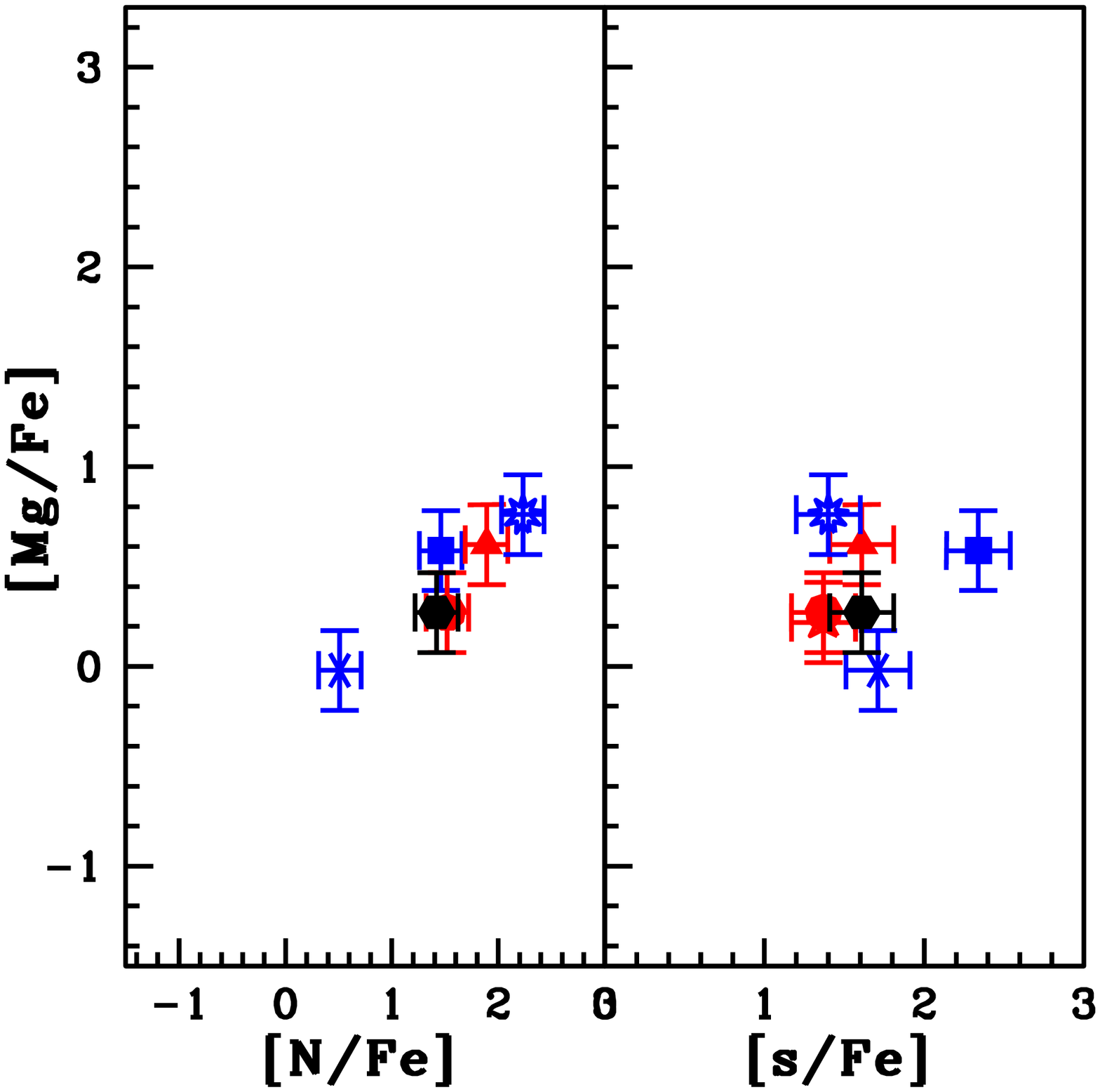}
\caption{The observed [Na/Fe] (upper panel) and [Mg/Fe] (lower panel) ratios
as a function of [N/Fe] and [s/Fe] for the program stars . 
Red symbols are CEMP-r/s stars, blue symbols CEMP-s stars and black symbol CH star.    
BD$-$19 132 (filled square), BD$-$19 290 (filled triangle), HE~1157$-$0518 (filled circle), 
HE~1304$-$2111 (four-sided star), HE~1354$-$2257 (six-sided cross),
BD+19 3109 (nine-sided star), and HD~202851 (filled hexagon).} \label{Na_Mg}
\end{figure}

Three out of the six program stars for which we could estimate the 
Na abundance are enhanced in Na with [Na/Fe]$>$1 and 
HE~1304$-$2111 being the most enhanced with [Na/Fe]$\sim$2.83.
Similar value of sodium enhancement is observed in a CEMP-r/s star 
CS~29528$-$028 which shows [Na/Fe]$\sim$2.68 \citep{Aoki_2007}.
They considered this star to be an extreme case of Ba-enhanced CEMP star.
The Na enhancement through Ne-Na cycle is expected in massive AGB stars 
where HBB operates. HBB also results in an increased surface abundance of N
through the CN cycle \citep{Sugimoto_1971, Lattanzio_1996, Goriely_2005, Karakas_2003, Karakas_2014,
Ventura_2005}. Since the HBB is responsible for the over-production of
N and Na in massive stars, they should be correlated if the abundance peculiarity 
is the result of massive-AGB stars nucleosynthesis. But we could not see any 
such trend in our sample of program stars (Figure \ref{Na_Mg}, top panel, left column).
The diffusive mixing and H-burning in these massive-AGB stars can reduce the 
s-process efficiency and it should be anti-correlated with the products of HBB \citep{Goriely_2005}.
However, we have  not noticed  any such behaviour among our program stars (Figure \ref{Na_Mg}, top panel, right column).
Enhanced Mg is also expected in them, resulting from the $\alpha$-capture reaction of $^{22}$Ne, and hence,
similar behaviour as seen in the case of Na,  is also expected for Mg. 
A slight trend could be noted in [Mg/Fe] v/s [N/Fe] plot (Figure \ref{Na_Mg}, bottom panel, left column), however
the scatter in the [Mg/Fe] v/s [s/Fe] plot (Figure \ref{Na_Mg}, bottom panel, right column) suggests, that the trend may not be real. 
So these facts together with the positive [hs/ls] ratio observed in  three Na enhanced 
stars as discussed in the previous section, rules out the possibility of the HBB  
and massive AGB star pollution.

\par The object HE~1354$-$2257 with negative [hs/ls], as discussed in section 8.2,
does not show any N, Na, and Mg enhancement. So this discards the possibility of a massive AGB star companion.
Similarly,  for the objects BD$-$19 132 and HD~202851 discussed in section 8.1, we rule out the possibility of
massive AGB star companions. 

The AGB models of \cite{Bisterzo_2010} predict higher Na abundances at lower metallicities. 
They have studied the nucleosyntheis in low-mass AGB stars (M$\rm_{AGB}^{ini}\sim$ 1.3 - 2 M$_{\odot}$)
of metalliicities $-3.6\leq$[Fe/H]$\leq-$1 for different $^{13}$C pocket efficiencies.  
The major neutron source here is $^{13}$C($\alpha$, n)$^{16}$O reaction during the 
inter-pulse phase.  The source  $^{22}$Ne($\alpha$, n)$^{25}$Mg  is also considered to be partially activated
during the TP. This models include the production of primary $^{23}$Na produced through a chain of TP/TDU/IP events 
from the $^{12}$C.
The primary $^{12}$C produced by the 3$\alpha$-reaction are brought to the surface 
from the inter-shell region by the TDU. During the following inter-pulse, the $^{12}$C is 
transformed into $^{14}$N through CNO cycle by the H-burning shell and accumulated in the top 
layers of the inter-shell-region. This $^{14}$N is converted to $^{22}$Ne during the early phase 
of next thermal instability via $^{14}$N($\alpha$,$\gamma$)$^{18}$F($\beta^{+}\nu$)$^{18}$O($\alpha$,$\gamma$)$^{22}$Ne \citep{Mowlavi_1999, Gallino_2006}.
This $^{22}$Ne results in the primary production of the $^{23}$Na via the n-capture reaction $^{22}$Ne(n, $\gamma$)$^{23}$Ne($\beta^{-}\nu$)$^{23}$Na  
that are significant at low-metallicities \citep{Gallino_2006, Bisterzo_2011}. 
The amount of the primary Na produced  increases with decreasing metallicity and/or increasing 
M$\rm_{AGB}^{ini}$ (Figure 15 of \citealt{Bisterzo_2010}).  
\cite{Bisterzo_2011} performed a comparison of abundance patterns of a sample of 100 CEMP-s stars 
collected from literature with the AGB nucleosynthesis models of \cite{Bisterzo_2010}. 
Among this sample, nine stars show [Na/Fe]$\geq$1 with two stars showing 
[Na/Fe]$\geq$2, CS~29528$-$028 ([Fe/H]$\sim-$2.86, [Na/Fe]$\sim$2.68  \citealt{Aoki_2007}) 
and SDSS~1707+58 ([Fe/H]$\sim-$2.52, [Na/Fe]$\sim$2.71  \citealt{Aoki_2008}).
Their analysis have shown that, the AGB models with M$\rm_{AGB}^{ini}\sim$1.5 M$_{\odot}$
could reproduce the Na abundances observed in them, but not the entire observed abundances.

\subsection{The [Rb/Zr] ratio}
\begin{figure}
\centering
\includegraphics[width=\columnwidth]{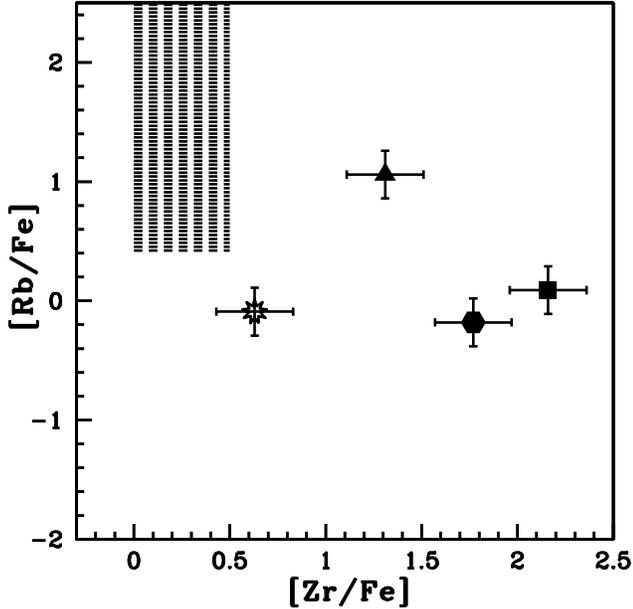}
\caption{The observed [Rb/Fe] and [Zr/Fe] in the program stars.
BD$-$19 132 (filled square), BD$-$19 290 (filled triangle), 
BD+19 3109 (nine-sided star), and HD~202851 (filled hexagon). 
The observed ranges of [Rb/Fe] and [Zr/Fe] in 
intermediate-mass AGB stars of the 
Galaxy and the Magellanic Clouds \citep{vanRaai_2012} are 
shown as shaded region.} \label{Rb_Zr}
\end{figure}

The [Rb/Zr] ratio is an important indicator of neutron density at the s-process site
and the mass of the companion AGB stars. In massive AGB stars (M$\geq$ 4 M$_{\odot}$), 
where the neutron source is $^{22}$Ne($\alpha$, n)$^{25}$Mg reaction (n$\sim$10$^{8}$ neutrons/cm$^{3}$), 
a higher Rb abundance and hence a positive value for [Rb/Zr] ratio are predicted \citep{Abia_2001, vanRaai_2012}, 
whereas, in the low-mass AGB stars (M$\leq$ 3 M$_{\odot}$) where $^{13}$C($\alpha$, n)$^{16}$O 
is the major neutron source (n$\sim$10$^{13}$ neutrons/cm$^{3}$), a negative [Rb/Zr] 
ratio is predicted \citep{Karakas_2012}. This has been supported by the observations of 
low- and intermediate-mass AGB stars of the Galaxy and Magellanic Clouds. 
\citep{Plez_1993, Lambert_1995, Abia_2001,  Garcia_2006, Garcia_2007, Garcia_2009}.
We have presented a detailed discussion on this ratio  in \cite{Shejeelammal_2020}.
The estimated values of the neutron density dependent [Rb/Zr] ratio for our program stars 
are given in Table \ref{hs_ls}. All the four stars where we could estimate this ratio show 
a negative value, indicating a low-mass AGB companion. The observed range of [Rb/Fe] and [Zr/Fe] 
in our program stars along with that in the massive AGB stars is shown in Figure \ref{Rb_Zr} for a comparison.

\subsection{Binary status of the program stars}
The analysis of \cite{Spite_2013} for a sample of dwarf and main-sequence turn-off
CEMP stars ($\sim$50) from literature ([Fe/H]$<$$-$1.80) have shown that the CEMP 
stars show bimodality in the absolute carbon abundance, that is, they show two distinct 
absolute carbon abundance A(C), values. 
The stars with [Fe/H]$\geq$$-$3.4 (CEMP-s and CEMP-r/s) populate 
the high carbon region at A(C)$\sim$8.25 and the stars with [Fe/H]$<$$-$3.4 (CEMP-no) 
populate the low carbon region at A(C)$\sim$6.80. This analysis did not consider any giants
so as to avoid any carbon dilution resulting from any mixing processes. 
Later, \cite{Bonifacio_2015} confirmed the bimodality of CEMP stars from an extended sample 
of CEMP stars that included dwarfs and main-sequence turn-off stars along with a few sub-giants 
and lower RGB giants. In both the studies, they noted a clear separation between the 
two carbon bands. Their interpretation of this bimodality is that, the carbon in  
the stars of these two carbon bands have different astrophysical origin. The carbon in the high carbon 
band stars are of extrinsic origin - resulting from the mass transfer from a low-mass AGB companion and 
that in the low carbon band stars owe to an intrinsic origin - C-enriched ISM from which they 
were formed despite a few of them are binaries as noted by \cite{Starkenburg_2014} \citep{Bonifacio_2015}. 
Subsequently, \cite{Hansen_2015a} also noted the 
bimodal carbon distribution for their sample of ($\sim$64) CEMP stars. 
However they found a smooth transition between the 
two carbon bands, in contrast to the well separated bands reported  by \cite{Spite_2013} 
and \cite{Bonifacio_2015}, with three CEMP-no stars occupying the high carbon band.  
This observation pointed to the importance of the knowledge of the binary status of the 
CEMP stars in order to better constrain the accurate origin of the observed abundances 
in the CEMP sub-classes. 

\par In an attempt to address this question, \cite{Yoon_2016} have
compiled an extensive set of 305 CEMP stars from literature with [Fe/H]$<$$-$1 and [C/Fe]$\geq$0.70. 
The sample comprised of 147 CEMP-s and CEMP-r/s stars (together they denoted as CEMP-s stars) 
and 127 CEMP-no stars, out of which, the binary status of 35 CEMP-s and 22 CEMP-no stars are known.
They also noted the carbon bimodality, but at A(C) values lower than that noted by \cite{Spite_2013}; 
the low- and high- carbon bands peaking at A(C)$\sim$6.28 and 7.96 respectively. 
Based on the behaviour of A(C) with respect to [Fe/H], these stars were grouped into 
three groups; CEMP-s  stars being the members of Group I  and CEMP-no stars being Group II and III objects. 
Their analysis have confirmed the interpretation put forward by \cite{Spite_2013} and
\cite{Bonifacio_2015} for the origin of carbon. Also they presented A(C) - [Fe/H] diagram as a powerful tool to distinguish
between the CEMP-no and CEMP-s/(r/s) sub-classes. According to this study, the absolute carbon abundance A(C)$\sim$7.1 
separates majority of binary stars, which lie above this carbon abundance, from single stars despite of a few outliers. 
Also, stars with A(C)$\leq$7.1 are classified as CEMP-no stars and that with A(C)$>$7.1 are 
classified as CEMP-s/(r/s) stars.

\par
\begin{figure}
\centering
\includegraphics[width=\columnwidth]{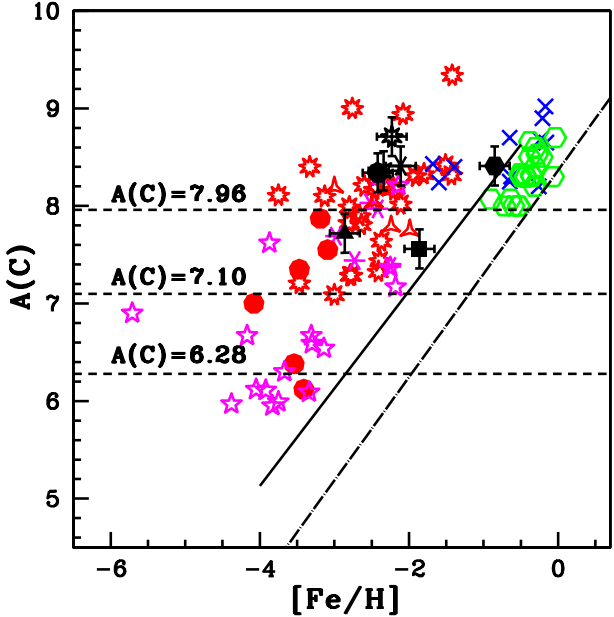}
\caption{Distribution of A(C) as a function of [Fe/H] for known/likely
binary and single stars.     
Red nine-sided stars, magenta six-sided crosses, red starred triangles, red filled circles
and magenta five-sided stars represent binary CEMP-s, single CEMP-s, binary CEMP-r/s, 
binary CEMP-no, and single CEMP-no stars respectively from literature \citep{Yoon_2016}.
All the red symbols corresponds to the binary CEMP stars and magenta symbols to single CEMP stars.
Blue crosses represent the binary CH stars from literature \citep{Purandardas_2019, 
Karinkuzhi_2014, Karinkuzhi_2015, Luck_2017}. Binary Ba stars from literature 
\citep{Shejeelammal_2020, Karinkuzhi_2018} are represented by green open hexagons. 
BD$-$19 132 (filled square), BD$-$19 290 (filled triangle), HE~1157$-$0518 (filled circle), 
HE~1304$-$2111 (four-sided star), HE~1354$-$2257 (six-sided cross),
BD+19 3109 (nine-sided star) and HD~202851 (filled hexagon). The binary and single stars 
are separated by the dashed line at A(C) = 7.10. The solid line corresponds 
to [C/Fe] = 0.70 and long-dash dot line to [C/Fe] = 0.} \label{binarity}
\end{figure}

 We have used A(C) - [Fe/H] plot to understand the binary nature of our program stars. 
Before using this diagram for this purpose, the carbon abundance should be corrected to account for
any internal mixing. \cite{Placco_2014} have calculated the corrections to the carbon abundance
for 0.0$\leq$log g $\leq$5.0, $-$5.5$\leq$[Fe/H]$\leq$0.00, and $-$1.0$\leq$[C/Fe]$\leq$3.0.
We have obtained the corrections to the estimated carbon abundances of our program stars
using the public online tool (\url{http://vplacco.pythonanywhere.com/}) developed by \cite{Placco_2014}. 
The necessary corrections have been applied to our estimates of carbon abundance and
then used in the Figure \ref{binarity}. The star HD~202851 lie among the CH stars and other program 
stars lie among the CEMP-s/(r/s) stars, with all the stars 
lying in the high carbon band. It is also noted that all our program stars lie in the region of binary stars that
bear the signatures of binary mass transfer from the low-mass AGB companions according to 
\cite{Bonifacio_2015}. As discussed in section 3, the stars BD+19 3109 and HD~202851 are confirmed binaries.
The difference in the estimated radial velocities of the stars BD$-$19 132, BD$-$19 290, and HE~1304$-$2111 
from the literature values discussed in section 3 along with their position in the A(C) - [Fe/H]
diagram suggest that these stars are possible binaries.

\subsection{Parametric model based analysis}
CEMP-s stars are known to be the products of binary mass transfer from low-mass 
AGB companion which now evolved into an invisible White Dwarf. However, a number of 
scenarios have been proposed till date to explain the origin of CEMP-r/s stars, enriched in
both s- and r- process elements. Detailed discussions on several scenarios are available 
in many places (i.e., \citealt{Jonsell_2006, Goswami_2021} and references therein). 
Most of these scenarios suggested an independent 
origin for the s- an r- process enrichment in CEMP-r/s stars. 
Analysis of \cite{Abate_2016} have shown that these proposed scenarios have failed to 
reproduce the observed frequency ($\sim$54\%) and higher [hs/ls] ratio of the CEMP-r/s stars.
From the analysis of a sample of CEMP stars ($\sim$18), \cite{Allen_2012} argued that CEMP-r/s stars 
have the same astrophysical origin as CEMP-s stars. 

\par The overlap of [hs/ls] ratio between the CEMP-s and CEMP-r/s stars indicates
that the origin of both the groups of stars owe a common astrophysical site and process 
but under different conditions. The higher [hs/ls] ratio, more 
precisely the higher [hs/Fe] ratio of the CEMP-r/s stars 
compared to the CEMP-s stars indicate a high neutron density process for their 
formation than the classical s-process. The tight correlation between the 
observed [Eu/Fe] and [hs/Fe] ratios in CEMP-r/s stars (Figure 13(c) of \citealt{Goswami_2021})
also points towards a single stellar site where both the s- and r- process elements could be produced simultaneously. 
The low-metallicity and high [hs/Fe] ratio suggest the extreme conditions of low-mass and low-metallicity 
s-process. The intermediate neutron-capture process, i-process, originally proposed by \cite{Cowan_1977},
has been explored recently by several authors to explain the CEMP-r/s phenomena 
\citep{Dardelet_2014, Hampel_2016, Hampel_2019, Hansen_2016c}. 
A higher neutron densities than that required for the classical s-process, of the order of 10$^{15}$ - 10$^{17}$ cm$^{-3}$, 
intermediate between the s- and r-process neutron densities, could be achieved as a result of proton Ingestion Episodes (PIEs; mixing 
of hydrogen rich material to the intershell region) in evolved red giants \citep{Cowan_1977}. 
There are several proposed sites for the PIEs that include low-metallicity low-mass AGB stars (z $\leq$ 10$^{-4}$, M $\leq$ 2M$_{\odot}$) 
\citep{Fujimoto_2000, Campbell_2008, Lau_2009, Cristallo_2009, 
Campbell_2010, Stancliffe_2011, Cruz_2013} etc., massive super-AGB stars \citep{Doherty_2015, Jones_2016},
post-AGB stars \citep{Herwig_2011, Herwig_2014, Bertolli_2013, Woodward_2015}, Rapidly Accreting White Dwarfs 
(RAWD; \citealt{Denissenkov_2019}) etc. A discussion on the several proposed sites of PIEs is given in 
\cite{Goswami_2021}. 

\par \cite{Hampel_2016} successfully used the simulations based on the i-process occurring in the 
low-mass low-metallicity (1 M$_{\odot}$, z$\sim$10$^{-4}$) AGB stars 
to reproduce the observed abundance pattern of 20 CEMP-r/s stars. Later, this method has been extended 
successfully to a sample of seven low-Pb post-AGB stars in the magellanic cloud \citep{Hampel_2019}. 
We have used the i-process model yields [X/Fe], from \cite{Hampel_2016} and compared with the 
observed [X/Fe] of our program stars for the neutron densities ranging from 10$^{9}$ to 10$^{15}$ cm$^{-3}$.
The observed abundances are fitted with the parametric model function;\\

\noindent X = X$_{i}$ . (1-d) + X$_{\odot}$ . d \\  

\noindent where X represents the final abundance, X$_{i}$ the i-process abundance, d is the dilution factor and
X$_{\odot}$ is the solar-scaled abundance. Here d is a free parameter that can be varied to find the best fit
between the model and the observational data for each constant neutron density. 
The best fit  model is found using $\chi^{2}$$\rm_{min}$ method, where $\chi^{2}$ is defined as; \\

\begin{figure}
\centering
\includegraphics[width=\columnwidth]{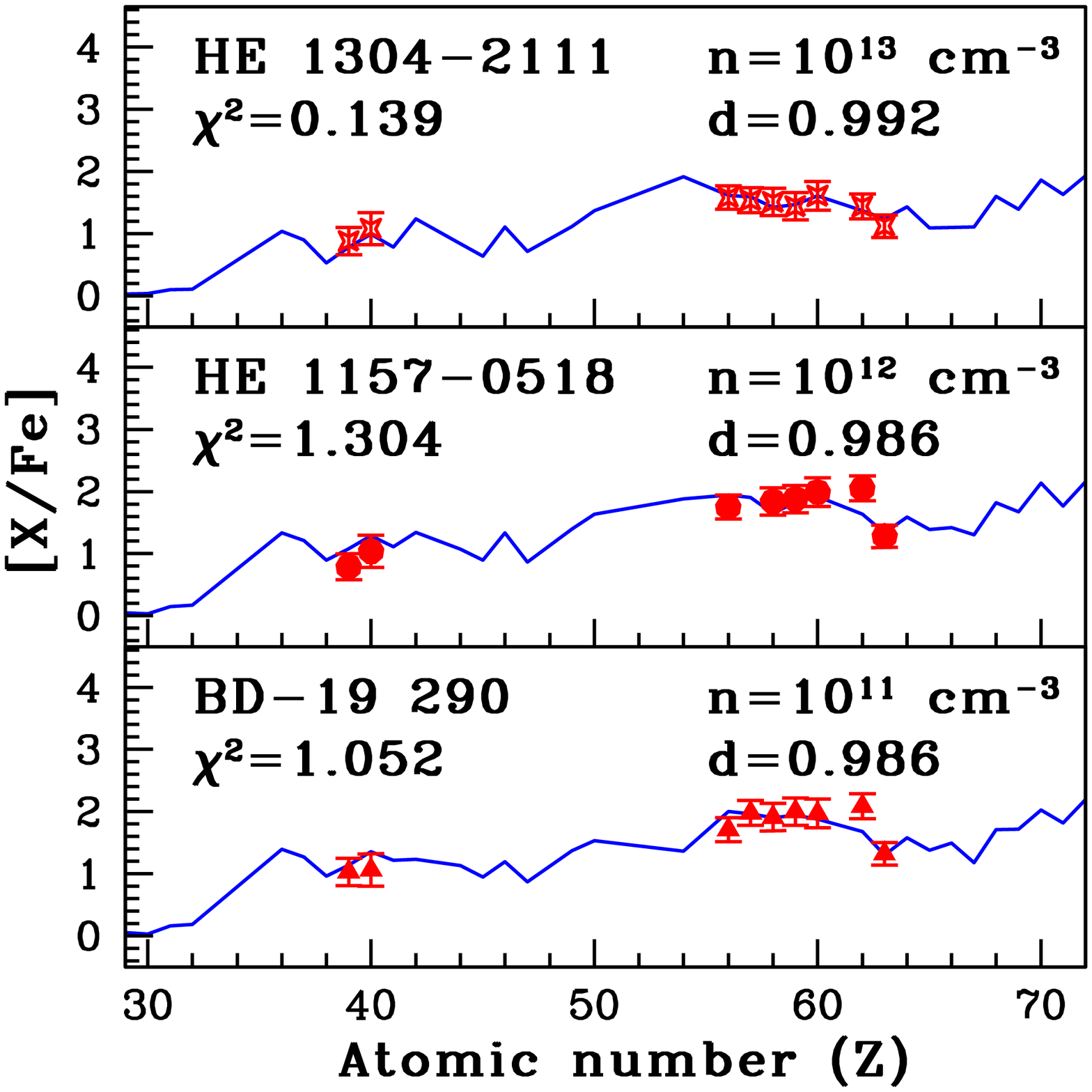}
\caption{Solid curves represent the best fit for the parametric model function.
The points with error bars indicate the observed abundances} 
\label{parametric_CEMP_rs}
\end{figure}

\noindent $\chi^{2}$ = $\sum\limits_{Z}$ $\frac{([X\rm_{Z}/Fe]\rm_{obs} - [X\rm_{Z}/Fe]\rm_{mod})^{2}}{\sigma^{2}\rm_{Z,obs}}$ \\

\noindent where [X$\rm_{Z}$/Fe]$\rm_{obs}$, [X$\rm_{Z}$/Fe]$\rm_{mod}$ the observed and model abundances of the element with atomic 
number Z and $\sigma^{2}\rm_{Z,obs}$, the observational uncertainty on [X$\rm_{Z}$/Fe]$\rm_{obs}$. 
The best fit model is selected in such a way that it shows least deviation from the observation, that is by minimizing 
the $\chi^{2}$.  The best fits obtained for the CEMP-r/s stars are shown in Figure \ref{parametric_CEMP_rs}. 
The neutron density responsible for the observed abundances in the stars BD$-$19 290, HE~1157$-$0518, and HE~1304$-$2111
are found to be n$\sim$ 10$^{11}$, 10$^{12}$, and 10$^{13}$ respectively.

\begin{figure}
\centering
\includegraphics[width=\columnwidth]{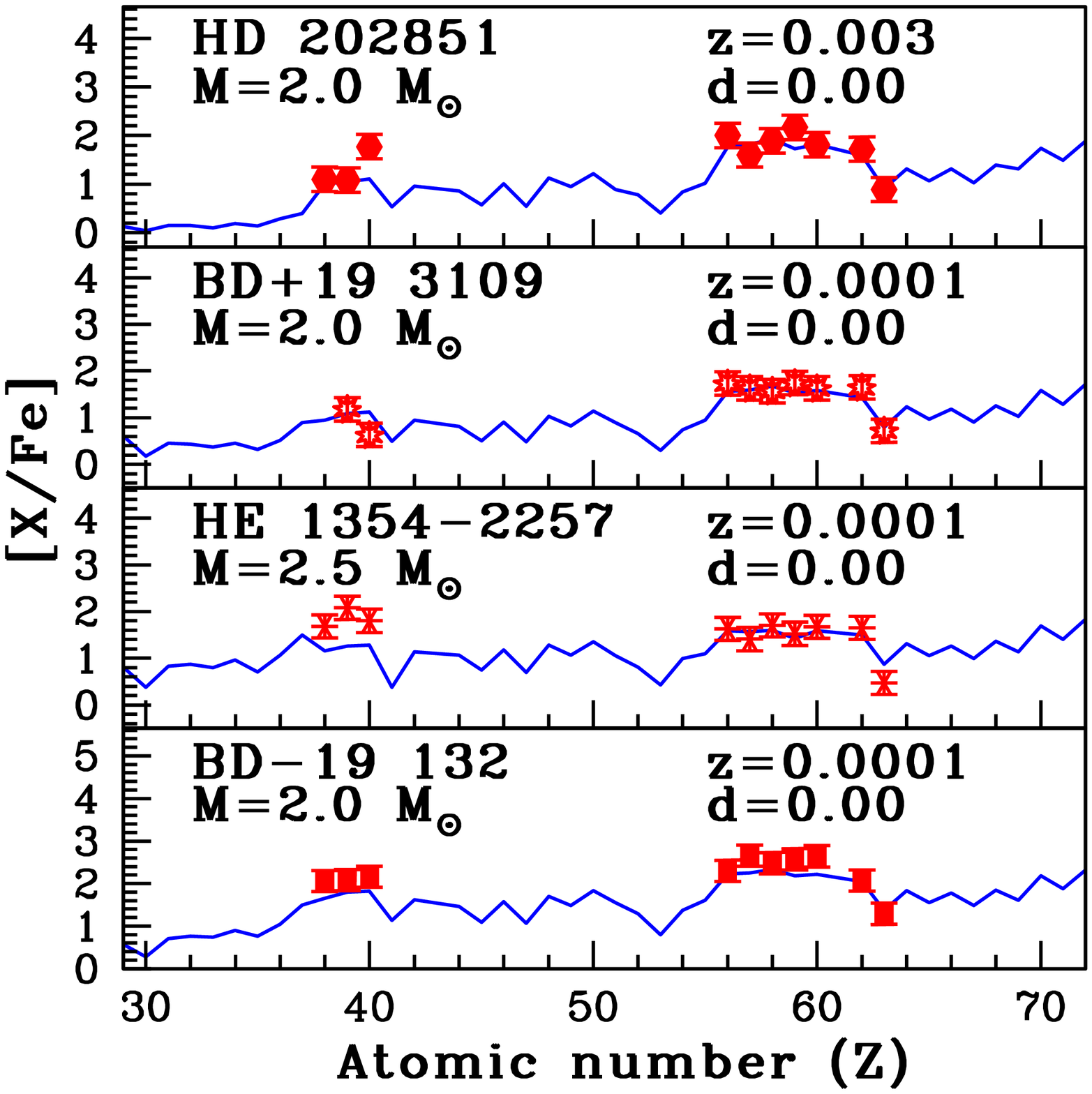}
\caption{Solid curves represent the best fit for the parametric model function.
The points with error bars indicate the observed abundances} 
\label{parametric_CH_CEMP_s}
\end{figure}

We have also conducted  a parametric model based analysis for the CEMP-s and CH stars in our sample to
find the mass of their AGB companion stars. The observed abundances in these stars are compared with the predicted
abundances for AGB stars  from FRUITY (FRANEC Repository of Updated Isotopic Tables \& Yields) models \citep{Cristallo_2009, Cristallo_2011, Cristallo_2015b}.
The best fitting mass of the companion AGB is derived by fitting the observed abundance with 
the parametric model function of \cite{Husti_2009} by minimizing the $\chi^{2}$ value. The detailed procedure 
is given in \cite{Shejeelammal_2020}. The best fits obtained for these stars are given in Figure \ref{parametric_CH_CEMP_s}.
The object HE~1354$-$2257 is found to have a companion of mass M $\sim$ 2.5 M$_{\odot}$
while the other three objects, BD$-$19 132, BD+19 3109, and HD~202851 have companions with mass M $\sim$ 2.0 M$_{\odot}$.

\subsection{Discussion on individual stars}
In this section we provide a discussion on the individual stars summarizing the results of the above analyses.\\

\noindent\textbf{BD$-$19 132, BD$-$19 290, BD+19 3109:} These object are listed in the CH star catalogue of
\cite{Bartkevicius_1996}. We present,  the first time  abundance analysis results for these objects. 
Our analysis have shown that BD$-$19 132 and BD+19 3109 are CEMP-s stars and
BD$-$19 290 is a CEMP-r/s star.
BD+19 3109 is a confirmed binary with a period of 
2129$\pm$13 days \citep{Sperauskas_2016}. We found that BD$-$19 290 and BD+19 3109 are high radial velocity
objects ($\mid$V$\rm_{r}$$\mid$$>$100 km s$^{-1}$) and BD$-$19 132 is a low radial 
velocity ($\mid$V$\rm_{r}$$\mid$$<$17 km s$^{-1}$) object. 
\cite{Yamashita_1975} classified BD+19 3109 as CH like star.
Kinematic analysis have shown that BD$-$19 132 belongs to the Galactic thin disk and BD$-$19 290 and BD+19 3109 
belong to the Galactic halo populations. From the neutron density dependent [Rb/Zr] ratio, 
these objects are found to have low-mass AGB companions. \\ 

\noindent\textbf{HE~1157$-$0518:} This object is a VMP star with [Fe/H]$\sim$$-$2.42.
This object is listed among the faint high latitude carbon  stars identified from 
the HES survey \citep{Christlieb_2001a}. \cite{Aoki_2007} could not determine 
the Eu abundance for this object, however, they have classified it as  a
CEMP-s star. Our analysis shows this star to be a CEMP-r/s star. 
Our estimates of atmospheric parameters 
are in agreement with that of \cite{Aoki_2007} except for log g which is 0.5 dex 
lower than our estimate. The abundances of C, N, Mg, Ca, Ti, and Cr also match  well  
 within the error limits. However, our estimate  [Ba/Fe]$\sim$1.75  is $\sim$0.40 dex
lower than their estimate. This object is found to be a member of Galactic halo with 
a probability of 100\%. It is a high spatial velocity and high radial velocity object. \\

\noindent\textbf{HE~1304$-$2111, HE~1354$-$2257:} Both these objects are found to be 
very metal poor stars with [Fe/H]$\sim$$-$2.34 and $-$2.11 respectively. Ours is the
first time abundance analysis for these objects. HE~1304$-$2111 is found to be a
CEMP-r/s star whereas the latter is a CEMP-s star.
While HE~1304$-$2111 is a disk object, HE~1354$-$2257 is found to belong to the Galactic 
halo population. \\

\noindent\textbf{HD~202851:} This object is listed in the CH star catalogue of
\cite{Bartkevicius_1996}. Our analysis shows  this star to be  a CH giant with 
[Fe/H]$\sim$$-$0.85. \cite{Sperauskas_2016} confirmed this object to be a 
binary with a period of 1295$\pm$6 days from radial velocity studies.
HD~202851 is found to be a low velocity object that belongs to the Galactic thin disk 
with 72\% probability. The estimated atmospheric parameters of this star by \cite{Sperauskas_2016}
(T$\rm_{eff}$, log g, $\zeta$, [Fe/H]) (4800 K, 2.1, 1.5, $-$0.70) are in close agreement with our estimates,
(4900 K, 2.2, 1.54, $-$0.85). \cite{Arentsen_2019} determined the atmospheric parameters of 
this star from medium-resolution (($\lambda/\delta\lambda$$\sim$ 10,000) spectra obtained 
with X-shooter spectrograph \citep{Vernet_2011} of Very Large Telescope (VLT). 
Their estimates for T$\rm_{eff}$, log g, and [Fe/H] are 4733 K, 1.6, and $-$0.88. 
The negative value obtained for the [Rb/Zr] ratio in this star confirms 
low-mass AGB companion for HD~202851.

\section{conclusions}
Detailed spectroscopic analyses of seven potential CH/CEMP stars are carried out 
based on the high-quality, high-resolution spectra obtained with
HCT/HESP and SUBARU/HDS. 
We have estimated the abundances of twenty eight elements including
both light and neutron-capture process elements and carbon isotopic ratio. 
We present the first time abundance analysis for the objects BD$-$19 132,
BD$-$19 290, HE~1304$-$2111, HE~1354$-$2257, and BD+19 3109. 

The objects BD+19 3109 and HD~202851 are confirmed binaries.
The difference in the estimated radial velocities of the stars 
BD$-$19 132, BD$-$19 290, and HE~1304$-$2111 from the Gaia values
may be a robust indication that they are likely binaries. 
In the A(C) - [Fe/H] diagram, all the program stars fall in  the 
region occupied by the binary stars. The kinematic analysis shows that 
the three low radial velocity objects
in our sample BD$-$19 132, HE~1304$-$2111, and HD~202851, are members of 
Galactic disk population and the rest four high radial velocity objects 
belong to the Galactic halo population. From the position on the HR diagram,
the objects BD$-$19 290, HE~1157$-$0518, and HD~202851 are found to be 
low mass objects ($\sim$1M$_{\odot}$). 

All the objects in our sample are bona-fide CEMP stars, except HD~202851
which is found be a CH star. Three objects BD$-$19 132, HE~1354$-$2257, 
and BD+19 3109 are CEMP-s stars, whereas BD$-$19 290, HE~1157$-$0518, 
and HE~1304$-$2111 are found to be CEMP-r/s stars. 
Our analysis based on different abundance ratios and abundance profiles
confirmed  low-mass  AGB companions of the program stars. The i-process models 
could successfully reproduce the observed abundances of the CEMP-r/s stars. 
The parametric model based analysis performed for the CEMP-s and CH stars confirms 
former low-mass AGB companions with mass M $\leq$ 2.5 M$_{\odot}$.

 \section{ACKNOWLEDGMENT}
We thank the staff at IAO and at the remote control station at 
CREST, Hosakotte for assisting during the observations.
Funding from the DST SERB project No. EMR/2016/005283 is gratefully 
acknowledged. We are thankful to  Partha Pratim
Goswami for generating the model fits used in Figure \ref{parametric_CEMP_rs}.
This work made use of the SIMBAD astronomical database, operated
at CDS, Strasbourg, France, and the NASA ADS, USA.
This work has made use of data from the European Space Agency (ESA) 
mission Gaia (\url{https://www.cosmos.esa.int/gaia}), processed by the Gaia 
Data Processing and Analysis Consortium 
(DPAC, \url{https://www.cosmos.esa.int/web/gaia/dpac/consortium}).
This research is based [in part] on data collected at Subaru Telescope,
which is operated by the National Astronomical Observatory of Japan.
We are honored and grateful for the opportunity of observing the Universe from Maunakea, 
which has the cultural, historical and natural significance in Hawaii. \\

\noindent
{\bf Data Availability}\\
The data underlying this article will be shared on reasonable request to the  authors.

\bibliography{Bibliography}{}
\bibliographystyle{aasjournal}

\appendix
\restartappendixnumbering
\section{Linelist}
{\footnotesize
\begin{table*}
\caption{Equivalent widths (in m\r{A}) of Fe lines used for deriving 
atmospheric parameters.} \label{linelist1}
\resizebox{\textwidth}{!}{\begin{tabular}{lccccccccccc}
\hline                       
Wavelength(\r{A}) & El & $E_{low}$(eV) & log gf & BD$-$19 132 & BD$-$19 290 & HE~1157$-$0518 & HE~1304$-$2111 & HE~1354$-$2257 & BD+19 3109 & HD~202851    \\ 
\hline 
4445.471	&	Fe I	&	0.087	&	$-$5.38	&	105.3(5.69)	&	34.9(4.75)	&	-	&	-	&	-	&	-	&	77.1(6.64)	&	\\
4466.551	&		&	2.832	&	$-$0.59	&	-	&	-	&	-	&	-	&	-	&	-	&	136.4(6.59)	&	\\
4484.22	&		&	3.603	&	$-$0.72	&	-	&	-	&	-	&	41.8(5.26)	&	-	&	-	&	-	&	\\
4489.739	&		&	0.121	&	$-$3.966	&	-	&	-	&	-	&	-	&	108.3(5.31)	&	-	&	128.3(6.53)	&	\\
4595.358	&		&	3.301	&	$-$1.72	&	56.0(5.73)	&	-	&	-	&	-	&	-	&	-	&	-	&	\\
4619.288	&		&	3.603	&	$-$1.12	&	63.5(5.65)	&	-	&	-	&	-	&	-	&	-	&	87.3(6.79)	&	\\
4625.045	&		&	3.241	&	$-$1.34	&	-	&	20.3(4.64)	&	-	&	36.6(5.25)	&	-	&	-	&	-	&	\\
4630.12	&		&	2.278	&	$-$2.6	&	-	&	-	&	-	&	28.4(5.03)	&	50.1(5.66)	&	-	&	-	&	\\
4635.846	&		&	2.845	&	$-$2.42	&	48.0(5.68)	&	-	&	-	&	-	&	-	&	24.0(5.17)	&	57.2(6.57)	&	\\
4637.503	&		&	3.283	&	$-$1.39	&	69.5(5.59)	&	-	&	-	&	-	&	-	&	-	&	-	&	\\
4643.464	&		&	3.654	&	$-$1.29	&	53.4(5.73)	&	-	&	-	&	-	&	-	&	38.2(5.40)	&	-	&	\\
4787.827	&		&	2.998	&	$-$2.77	&	-	&	-	&	-	&	-	&	-	&	-	&	39.4(6.77)	&	\\
4788.751	&		&	3.236	&	$-$1.81	&	52.5(5.61)	&	-	&	-	&	-	&	-	&	-	&	-	&	\\
4789.651	&		&	3.546	&	$-$0.91	&	-	&	-	&	-	&	-	&	33.0(5.20)	&	-	&	-	&	\\
4924.77	&		&	2.278	&	$-$2.22	&	-	&	-	&	22.1(5.20)	&	52.7(5.32)	&	48.8(5.26)	&	-	&	-	&	\\
4930.315	&		&	3.959	&	$-$1.35	&	-	&	-	&	-	&	-	&	-	&	-	&	44.7(6.53)	&	\\
4939.687	&		&	0.859	&	$-$3.34	&	-	&	128.5(4.67)	&	-	&	-	&	99.7(5.33)	&	-	&	-	&	\\
4967.89	&		&	4.191	&	$-$0.622	&	-	&	-	&	-	&	-	&	-	&	35.5(5.38)	&	-	&	\\
4985.253	&		&	3.93	&	$-$0.56	&	76.8(5.73)	&	-	&	-	&	-	&	-	&	-	&	96.6(6.75)	&	\\
5001.863	&		&	3.882	&	0.01	&	-	&	-	&	32.8(4.84)	&	-	&	78.9(5.33)	&	-	&	-	&	\\
5028.126	&		&	3.573	&	$-$1.474	&	41.8(5.59)	&	-	&	-	&	-	&	-	&	-	&	-	&	\\
5044.211	&		&	2.851	&	$-$2.15	&	63.3(5.62)	&	-	&	-	&	-	&	22.3(5.36)	&	-	&	-	&	\\
5049.819	&		&	2.278	&	$-$1.42	&	-	&	-	&	-	&	76.5(5.17)	&	-	&	-	&	-	&	\\
5079.223	&		&	2.2	&	$-$2.067	&	-	&	-	&	27.9(5.04)	&	-	&	-	&	-	&	112.6(6.54)	&	\\
5083.338	&		&	0.958	&	$-$2.958	&	-	&	-	&	-	&	81.4(5.11)	&	-	&	-	&	-	&	\\
5127.359	&		&	0.915	&	$-$3.307	&	-	&	-	&	32.8(4.89)	&	-	&	-	&	-	&	-	&	\\
5141.739	&		&	2.424	&	$-$2.15	&	-	&	-	&	-	&	-	&	-	&	78.1(5.14)	&	-	&	\\
5151.911	&		&	1.011	&	$-$3.32	&	-	&	107.8(4.63)	&	-	&	-	&	-	&	-	&	-	&	\\
5198.711	&		&	2.222	&	$-$2.135	&	-	&	-	&	-	&	-	&	58.3(5.18)	&	112.4(5.37)	&	-	&	\\
5215.179	&		&	3.266	&	$-$0.933	&	102.6(5.61)	&	-	&	-	&	-	&	-	&	91.7(5.31)	&	113.3(6.67)	&	\\
5242.491	&		&	3.634	&	$-$0.84	&	-	&	20.1(4.71)	&	-	&	-	&	46.0(5.54)	&	58.2(5.33)	&	94.3(6.71)	&	\\
5247.05	&		&	0.087	&	$-$4.946	&	139.3(5.64)	&	66.8(4.60)	&	-	&	55.6(5.01)	&	-	&	124.1(5.23)	&	115.2(6.79)	&	\\
5250.209	&		&	0.121	&	$-$4.938	&	-	&	58.7(4.52)	&	-	&	-	&	-	&	131.2(5.37)	&	-	&	\\
5263.305	&		&	3.265	&	$-$0.97	&	-	&	-	&	-	&	44.7(5.05)	&	75.5(5.55)	&	-	&	-	&	\\
5281.79	&		&	3.038	&	$-$1.02	&	-	&	-	&	41.2(4.96)	&	-	&	-	&	-	&	-	&	\\
5322.04	&		&	2.28	&	$-$2.84	&	72.0(5.57)	&	-	&	-	&	-	&	-	&	47.9(5.14)	&	68.9(6.42)	&	\\
5339.93	&		&	3.27	&	$-$0.68	&	-	&	-	&	55.0(5.26)	&	-	&	-	&	100.9(5.19)	&	-	&	\\
5365.399	&		&	3.573	&	$-$1.44	&	43.2(5.57)	&	-	&	-	&	-	&	-	&	-	&	-	&	\\
5367.479	&		&	4.415	&	0.35	&	-	&	34.7(4.66)	&	-	&	44.5(5.21)	&	-	&	69.9(5.25)	&	111.4(6.66)	&	\\
5369.961	&		&	4.37	&	0.35	&	87.1(5.58)	&	30.8(4.54)	&	-	&	-	&	-	&	80.2(5.36)	&	100.9(6.38)	&	\\
5383.369	&		&	4.312	&	0.5	&	-	&	-	&	-	&	-	&	-	&	80.7(5.13)	&	131.2(6.78)	&	\\
5445.042	&		&	4.39	&	$-$0.02	&	73.5(5.75)	&	-	&	-	&	27.6(5.07)	&	-	&	43.6(5.16)	&	98.2(6.70)	&	\\
5569.62	&		&	3.42	&	$-$0.49	&	-	&	-	&	-	&	-	&	99.8(5.67)	&	-	&	125.6(6.64)	&	\\
5576.09	&		&	3.43	&	$-$0.851	&	-	&	-	&	-	&	-	&	-	&	68.4(5.21)	&	-	&	\\
5586.756	&		&	3.368	&	$-$0.21	&	137.8(5.60)	&	-	&	-	&	68.9(5.10)	&	-	&	-	&	152.1(6.75)	&	\\
5615.644	&		&	3.332	&	$-$0.14	&	-	&	-	&	-	&	-	&	-	&	129.0(5.31)	&	-	&	\\
5701.544	&		&	2.559	&	$-$2.216	&	-	&	-	&	-	&	-	&	52.1(5.54)	&	80.1(5.37)	&	-	&	\\
5753.12	&		&	4.26	&	$-$0.76	&	-	&	-	&	-	&	-	&	-	&	23.1(5.33)	&	67.6(6.65)	&	\\
5883.813	&		&	3.959	&	$-$1.36	&	-	&	-	&	-	&	-	&	-	&	-	&	57.6(6.71)	&	\\
5956.692	&		&	0.859	&	$-$4.605	&	-	&	-	&	-	&	38.9(5.22)	&	-	&	-	&	86.5(6.69)	&	\\
6024.049	&		&	4.548	&	$-$0.12	&	-	&	-	&	-	&	-	&	-	&	35.7(5.30)	&	86.0(6.65)	&	\\
6082.71	&		&	2.222	&	$-$3.573	&	-	&	-	&	-	&	-	&	-	&	21.7(5.29)	&	57.5(6.83)	&	\\
6137.694	&		&	2.588	&	$-$1.403	&	-	&	105.3(4.73)	&	-	&	-	&	98.6(5.39)	&	135.7(5.38)	&	-	&	\\
6151.62	&		&	2.18	&	$-$3.29	&	-	&	-	&	-	&	20.4(5.24)	&	21.3(5.57)	&	39.1(5.30)	&	-	&	\\
6173.34	&		&	2.22	&	$-$2.88	&	-	&	-	&	-	&	-	&	30.6(5.41)	&	-	&	-	&	\\
6200.314	&		&	2.608	&	$-$2.437	&	-	&	-	&	-	&	-	&	36.6(5.54)	&	-	&	-	&	\\
6219.279	&		&	2.198	&	$-$2.433	&	-	&	-	&	25.2(5.27)	&	-	&	-	&	-	&	-	&	\\
6240.646	&		&	2.222	&	$-$3.38	&	-	&	-	&	-	&	-	&	27.9(5.64)	&	-	&	-	&	\\
6246.318	&		&	3.603	&	$-$0.96	&	75.7(5.60)	&	-	&	-	&	-	&	-	&	-	&	-	&	\\
6252.554	&		&	2.404	&	$-$1.687	&	-	&	87.3(4.60)	&	46.9(5.12)	&	-	&	82.5(5.19)	&	-	&	-	&	\\
6254.258	&		&	2.279	&	$-$2.48	&	-	&	-	&	-	&	-	&	73.6(5.62)	&	-	&	-	&	\\
6280.617	&		&	0.859	&	$-$4.39	&	-	&	-	&	-	&	52.1(5.26)	&	-	&	-	&	-	&	\\
6301.5	&		&	3.654	&	$-$0.672	&	-	&	-	&	-	&	37.6(5.06)	&	-	&	69.6(5.24)	&	101.8(6.51)	&	\\
6318.018	&		&	2.453	&	$-$2.33	&	-	&	-	&	-	&	-	&	41.2(5.31)	&	83.7(5.32)	&	-	&	\\
6322.69	&		&	2.588	&	$-$2.426	&	-	&	-	&	-	&	-	&	33.1(5.44)	&	-	&	-	&	\\
6335.328	&		&	2.198	&	$-$2.23	&	-	&	-	&	-	&	-	&	58.7(5.14)	&	105.0(5.15)	&	127.5(6.69)	&	\\
6336.823	&		&	3.686	&	$-$1.05	&	-	&	-	&	-	&	-	&	22.9(5.19)	&	-	&	-	&	\\
6393.602	&		&	2.432	&	$-$1.62	&	-	&	-	&	-	&	-	&	90.4(5.25)	&	-	&	-	&	\\
6416.933	&		&	4.796	&	$-$0.885	&	-	&	-	&	-	&	-	&	-	&	-	&	29.9(6.65)	&	\\
\hline
\end{tabular}}

The numbers in the parenthesis in columns 5 - 11 give the derived abundances from the respective line. 
\end{table*}
}

{\footnotesize
\begin{table*}
\resizebox{\textwidth}{!}{\begin{tabular}{lccccccccccc}
\hline                       
Wavelength(\r{A}) & El & $E_{low}$(eV) & log gf & BD$-$19 132 & BD$-$19 290 & HE~1157$-$0518 & HE~1304$-$2111 & HE~1354$-$2257 & BD+19 3109 & HD~202851    \\ 
\hline 
6421.349	&		&	2.278	&	$-$2.027	&	-	&	-	&	-	&	65.9(5.21)	&	80.7(5.32)	&	-	&	-	&	\\
6430.85	&		&	2.18	&	$-$2.01	&	-	&	-	&	-	&	-	&	107.9(5.55)	&	-	&	-	&	\\
6575.019	&		&	2.588	&	$-$2.82	&	52.0(5.67)	&	-	&	-	&	-	&	-	&	-	&	62.4(6.58)	&	\\
6592.91	&		&	2.72	&	$-$1.47	&	-	&	57.0(4.62)	&	-	&	-	&	56.6(5.12)	&	95.8(5.13)	&	124.5(6.61)	&	\\
6593.871	&		&	2.432	&	$-$2.422	&	-	&	-	&	-	&	-	&	-	&	-	&	105.1(6.71)	&	\\
6677.989	&		&	2.692	&	$-$1.47	&	-	&	-	&	45.0(5.17)	&	-	&	-	&	-	&	-	&	\\
4491.405	&	Fe II 	&	2.855	&	$-$2.7	&	-	&	-	&	-	&	-	&	33.7(5.21)	&	22.8(5.29)	&	-	&	\\
4508.288	&		&	2.855	&	$-$2.21	&	45.0(5.64)	&	29.5(4.55)	&	25.2(4.97)	&	-	&	-	&	40.4(5.27)	&	-	&	\\
4515.339	&		&	2.84	&	$-$2.48	&	-	&	-	&	-	&	-	&	-	&	30.3(5.26)	&	-	&	\\
4520.224	&		&	2.81	&	$-$2.6	&	32.7(5.65)	&	-	&	-	&	-	&	-	&	-	&	-	&	\\
4629.339	&		&	2.807	&	$-$2.28	&	-	&	-	&	-	&	31.5(5.16)	&	-	&	-	&	-	&	\\
4731.453	&		&	2.891	&	$-$3.36	&	-	&	-	&	-	&	-	&	20.5(5.58)	&	-	&	-	&	\\
5197.56	&		&	3.23	&	$-$2.25	&	-	&	-	&	-	&	24.0(5.23)	&	-	&	-	&	85.4(6.62)	&	\\
5234.62	&		&	3.22	&	$-$2.24	&	-	&	25.4(4.73)	&	25.4(5.20)	&	22.0(5.09)	&	52.3(5.35)	&	-	&	-	&	\\
5534.83	&		&	3.25	&	$-$2.77	&	-	&	-	&	-	&	-	&	-	&	-	&	48.3(6.64)	&	\\
6247.55	&		&	3.89	&	$-$2.34	&	-	&	-	&	-	&	-	&	-	&	-	&	46.4(6.72)	&	\\
6456.383	&		&	3.903	&	$-$2.075	&	-	&	-	&	-	&	-	&	19.9(5.40)	&	-	&	-	&	\\
\hline
\end{tabular}}

The numbers in the parenthesis in columns 5 - 11 give the derived abundances from the respective line. 
The line information are taken from Kurucz database of atomic line lists (\url{https://lweb.cfa.harvard.edu/amp/ampdata/kurucz23/sekur.html}).
\end{table*}
}

{\footnotesize
\begin{table*}
\caption{Equivalent widths (in m\r{A}) of lines used for deriving elemental abundances.} \label{linelist2}
\resizebox{\textwidth}{!}{\begin{tabular}{lccccccccccc}
\hline                       
Wavelength(\r{A}) & El & $E_{low}$(eV) & log gf & BD$-$19 132 & BD$-$19 290 & HE~1157$-$0518 & HE~1304$-$2111 & HE~1354$-$2257 & BD+19 3109 & HD~202851   \\ 
\hline 
5682.633	&	Na I	&	2.102	&	$-$0.7	&	120.0(5.52)	&	31.5(4.65)	&	-	&	129.8(6.65)	&	27.3(4.76)	&	21.5(4.14)	&	93.0(5.90)	&	\\
5688.205	&		&	2.1	&	$-$0.45	&	130.5(5.42)	&	35.4(4.47)	&	-	&	170.5(6.81)	&	35.5(4.67)	&	31.8(4.11)	&	104.3(5.84)	&	\\
6160.747	&		&	2.104	&	$-$1.26	&	-	&	-	&	-	&	-	&	-	&	-	&	66.4(6.00)	&	\\
4702.991	&	Mg I	&	4.346	&	$-$0.666	&	-	&	-	&	-	&	-	&	-	&	115.9(6.05)	&	151.9(6.95)	&	\\
4730.029	&		&	4.346	&	$-$2.523	&	37.1(6.30)	&	-	&	-	&	-	&	-	&	28.4(6.29)	&	-	&	\\
5528.405	&		&	4.346	&	$-$0.62	&	-	&	89.4(5.35)	&	67.0(5.45)	&	68.5(5.46)	&	-	&	141.5(6.06)	&	171.8(7.09)	&	\\
5711.088	&		&	4.346	&	$-$1.833	&	92.4(6.33)	&	-	&	-	&	-	&	12.4(5.47)	&	-	&	-	&	\\
5948.541	&	Si I	&	5.083	&	$-$1.23	&	-	&	110.0(6.99)	&	-	&	24.8(5.91)	&	25.7(6.01)	&	-	&	90.8(6.63)	&	\\
6237.319	&		&	5.613	&	$-$0.53	&	-	&	98.7(6.80)	&	-	&	27.0(5.92)	&	20.2(5.78)	&	-	&	-	&	\\
4425.437	&	Ca I	&	1.879	&	$-$0.385	&	-	&	-	&	66.0(4.60)	&	-	&	-	&	-	&	-	&	\\
4435.679	&		&	1.89	&	$-$0.52	&	-	&	57.8(4.00)	&	-	&	-	&	-	&	-	&	-	&	\\
4455.887	&		&	1.899	&	$-$0.51	&	-	&	-	&	38.8(4.03)	&	-	&	-	&	-	&	-	&	\\
5265.56	&		&	2.52	&	$-$0.11	&	-	&	40.4(3.90)	&	-	&	50.3(4.38)	&	-	&	-	&	-	&	\\
5581.965	&		&	2.523	&	$-$1.833	&	116.9(5.13)	&	-	&	-	&	-	&	-	&	56.7(4.36)	&	-	&	\\
5590.114	&		&	2.521	&	$-$0.71	&	-	&	-	&	12.4(4.22)	&	-	&	23.1(4.38)	&	47.4(4.21)	&	74.6(5.43)	&	\\
5594.462	&		&	2.523	&	$-$0.05	&	-	&	58.7(3.89)	&	31.8(3.91)	&	-	&	-	&	-	&	-	&	\\
5857.451	&		&	2.932	&	0.23	&	-	&	-	&	-	&	-	&	44.9(4.35)	&	59.7(4.01)	&	94.5(5.34)	&	\\
6102.723	&		&	1.879	&	$-$0.89	&	-	&	-	&	-	&	-	&	53.4(4.35)	&	98.2(4.12)	&	115.9(5.63)	&	\\
6162.173	&		&	1.899	&	0.1	&	-	&	137.6(3.92)	&	-	&	-	&	105.7(4.34)	&	-	&	-	&	\\
6166.439	&		&	2.521	&	$-$0.9	&	-	&	-	&	-	&	-	&	-	&	37.5(4.16)	&	72.1(5.54)	&	\\
6169.042	&		&	2.523	&	$-$0.55	&	-	&	-	&	-	&	-	&	36.2(4.47)	&	-	&	-	&	\\
6169.563	&		&	2.523	&	$-$0.27	&	-	&	-	&	-	&	-	&	-	&	-	&	96.0(5.37)	&	\\
6439.075	&		&	2.525	&	0.47	&	-	&	-	&	-	&	-	&	103.7(4.66)	&	137.5(4.24)	&	155.9(5.61)	&	\\
6449.808	&		&	2.523	&	$-$0.55	&	138.0(5.12)	&	-	&	-	&	-	&	-	&	-	&	92.6(5.55)	&	\\
6471.662	&		&	2.525	&	$-$0.59	&	152.7(5.36)	&	-	&	-	&	-	&	42.9(4.62)	&	68.0(4.28)	&	-	&	\\
6493.781	&		&	2.521	&	0.14	&	-	&	-	&	69.1(4.12)	&	-	&	-	&	-	&	112.0(5.23)	&	\\
4453.312	&	Ti I	&	1.43	&	$-$0.051	&	-	&	-	&	-	&	-	&	-	&	-	&	59.2(4.34)	&	\\
4533.239	&		&	0.848	&	0.476	&	-	&	-	&	-	&	-	&	62.2(2.60)	&	-	&	-	&	\\
4617.269	&		&	1.749	&	0.389	&	-	&	-	&	-	&	-	&	-	&	-	&	81.0(4.25)	&	\\
4778.255	&		&	2.24	&	$-$0.22	&	-	&	-	&	-	&	-	&	-	&	-	&	38.1(4.54)	&	\\
4820.41	&		&	1.503	&	$-$0.441	&	-	&	-	&	-	&	25.4(3.17)	&	-	&	-	&	-	&	\\
4840.874	&		&	0.899	&	$-$0.509	&	163.9(3.80)	&	-	&	-	&	-	&	-	&	-	&	89.1(4.27)	&	\\
4926.147	&		&	0.818	&	$-$2.17	&	78.8(3.83)	&	-	&	-	&	-	&	-	&	-	&	-	&	\\
4999.5	&		&	0.83	&	0.31	&	-	&	-	&	28.4(2.36)	&	82.8(3.13)	&	-	&	-	&	133.6(4.31)	&	\\
5024.842	&		&	0.818	&	$-$0.602	&	-	&	-	&	-	&	-	&	-	&	-	&	94.3(4.33)	&	\\
5039.96	&		&	0.02	&	$-$1.13	&	-	&	68.4(2.32)	&	-	&	-	&	-	&	-	&	104.5(4.12)	&	\\
5210.386	&		&	0.047	&	$-$0.884	&	-	&	84.8(2.24)	&	-	&	77.6(2.90)	&	-	&	-	&	-	&	\\
5471.197	&		&	1.443	&	$-$1.44	&	-	&	-	&	-	&	-	&	-	&	32.6(3.45)	&	-	&	\\
5739.982	&		&	2.236	&	$-$0.67	&	-	&	-	&	-	&	-	&	-	&	-	&	23.6(4.62)	&	\\
5918.535	&		&	1.07	&	$-$1.46	&	103.4(3.64)	&	-	&	-	&	-	&	-	&	-	&	36.6(4.30)	&	\\
5922.11	&		&	1.046	&	$-$1.466	&	95.3(3.50)	&	-	&	-	&	-	&	-	&	-	&	45.0(4.44)	&	\\
5937.811	&		&	1.067	&	$-$1.89	&	-	&	-	&	-	&	-	&	-	&	38.6(3.36)	&	-	&	\\
5941.751	&		&	1.05	&	$-$1.51	&	95.8(3.56)	&	-	&	-	&	-	&	-	&	-	&	31.7(4.24)	&	\\
\hline
\end{tabular}}

The numbers in the parenthesis in columns 5 - 11 give the derived abundances from the respective line. \\
\end{table*}
}

{\footnotesize
\begin{table*}
\resizebox{\textwidth}{!}{\begin{tabular}{lccccccccccc}
\hline                       
Wavelength(\r{A}) & El & $E_{low}$(eV) & log gf & BD$-$19 132 & BD$-$19 290 & HE~1157$-$0518 & HE~1304$-$2111 & HE~1354$-$2257 & BD+19 3109 & HD~202851    \\ 
\hline 
5965.828	&		&	1.879	&	$-$0.409	&	-	&	-	&	-	&	-	&	-	&	57.0(3.38)	&	-	&	\\
5978.543	&		&	1.873	&	$-$0.496	&	-	&	-	&	-	&	-	&	-	&	52.6(3.39)	&	-	&	\\
6064.629	&		&	1.046	&	$-$1.944	&	-	&	-	&	-	&	-	&	-	&	-	&	26.8(4.56)	&	\\
6303.757	&		&	1.443	&	$-$1.566	&	-	&	-	&	-	&	-	&	-	&	43.5(3.64)	&	-	&	\\
6556.062	&		&	1.46	&	$-$1.074	&	89.1(3.59)	&	-	&	-	&	-	&	-	&	-	&	42.5(4.45)	&	\\
6599.133	&		&	0.9	&	$-$2.085	&	-	&	-	&	-	&	-	&	-	&	60.2(3.49)	&	-	&	\\
4161.535	&	Ti II	&	1.084	&	$-$2.36	&	-	&	-	&	-	&	-	&	-	&	-	&	92.8(4.35)	&	\\
4443.794	&		&	1.08	&	$-$0.7	&	-	&	-	&	-	&	-	&	-	&	-	&	157.1(4.15)	&	\\
4493.51	&		&	1.08	&	$-$2.73	&	60.0(3.34)	&	-	&	-	&	-	&	-	&	-	&	-	&	\\
4563.761	&		&	1.22	&	$-$0.96	&	139.8(3.33)	&	114.3(2.20)	&	-	&	-	&	-	&	124.4(3.00)	&	-	&	\\
4568.314	&		&	1.22	&	$-$2.65	&	61.1(3.47)	&	-	&	-	&	-	&	-	&	42.1(3.03)	&	-	&	\\
4571.96	&		&	1.571	&	$-$0.53	&	123.1(4.73)	&	-	&	-	&	-	&	-	&	-	&	-	&	\\
4636.32	&		&	1.164	&	$-$2.855	&	-	&	-	&	-	&	-	&	-	&	-	&	67.0(4.42)	&	\\
4708.665	&		&	1.236	&	$-$2.21	&	-	&	-	&	-	&	-	&	-	&	-	&	86.9(4.29)	&	\\
4764.526	&		&	1.236	&	$-$2.77	&	62.0(3.57)	&	-	&	-	&	-	&	22.2(2.80)	&	-	&	76.8(4.61)	&	\\
4779.985	&		&	2.048	&	$-$1.37	&	-	&	20.9(2.16)	&	18.3(2.75)	&	-	&	-	&	57.0(3.14)	&	97.5(4.64)	&	\\
4798.521	&		&	1.08	&	$-$2.43	&	-	&	296(2.14)	&	-	&	-	&	42.2(2.68)	&	-	&	82.4(4.21)	&	\\
4805.085	&		&	2.061	&	$-$1.1	&	87.9(3.56)	&	-	&	-	&	33.2(2.54)	&	-	&	-	&	-	&	\\
4865.612	&		&	1.116	&	$-$2.61	&	73.7(3.43)	&	-	&	-	&	-	&	31.8(2.70)	&	-	&	80.6(4.38)	&	\\
5185.9	&		&	1.89	&	$-$1.35	&	-	&	-	&	21.8(2.61)	&	99.50(5.36)	&	48.4(2.64)	&	87.2(3.32)	&	-	&	\\
5226.543	&		&	1.566	&	$-$1.3	&	-	&	-	&	32.1(2.42)	&	-	&	-	&	-	&	-	&	\\
5336.771	&		&	1.58	&	$-$1.7	&	81.4(3.23)	&	-	&	-	&	-	&	-	&	-	&	107.4(4.56)	&	\\
5381.015	&		&	1.566	&	$-$2.08	&	-	&	-	&	-	&	-	&	-	&	69.9(3.20)	&	-	&	\\
5418.751	&		&	1.581	&	$-$1.999	&	-	&	50.8(4.42)	&	-	&	-	&	-	&	-	&	-	&	\\
4351.05	&	Cr I	&	0.97	&	$-$1.45	&	-	&	-	&	17.3(3.26)	&	-	&	-	&	-	&	-	&	\\
4600.748	&		&	1.004	&	$-$1.26	&	-	&	-	&	-	&	87.0(4.52)	&	-	&	-	&	-	&	\\
4626.173	&		&	0.968	&	$-$1.32	&	-	&	-	&	-	&	-	&	-	&	73.7(2.71)	&	-	&	\\
4737.347	&		&	3.087	&	$-$0.099	&	-	&	-	&	-	&	29.9(4.29)	&	-	&	-	&	73.5(5.31)	&	\\
4829.372	&		&	2.544	&	$-$0.81	&	-	&	-	&	-	&	-	&	-	&	-	&	58.8(5.08)	&	\\
4870.801	&		&	3.079	&	0.05	&	-	&	-	&	-	&	42.0(4.51)	&	-	&	-	&	-	&	\\
4942.49	&		&	0.941	&	$-$2.294	&	-	&	-	&	-	&	-	&	-	&	33.9(2.91)	&	99.5(5.55)	&	\\
5247.565	&		&	0.961	&	$-$1.64	&	-	&	-	&	-	&	-	&	-	&	77.3(2.86)	&	-	&	\\
5296.691	&		&	0.982	&	$-$1.4	&	-	&	-	&	-	&	-	&	-	&	-	&	114.3(5.18)	&	\\
5298.277	&		&	0.983	&	$-$1.15	&	-	&	-	&	-	&	100.0(4.39)	&	-	&	-	&	125.6(5.19)	&	\\
5300.744	&		&	0.982	&	$-$2.12	&	-	&	-	&	-	&	-	&	-	&	46.1(2.92)	&	-	&	\\
5312.871	&		&	3.45	&	$-$0.562	&	-	&	-	&	-	&	-	&	-	&	-	&	28.6(5.23)	&	\\
5345.801	&		&	1.003	&	$-$0.98	&	-	&	-	&	-	&	-	&	-	&	122.9(2.91)	&	145.7(5.16)	&	\\
5348.312	&		&	1.003	&	$-$1.29	&	-	&	-	&	18.8(3.10)	&	-	&	-	&	95.6(2.81)	&	-	&	\\
5409.772	&		&	1.03	&	$-$0.72	&	-	&	-	&	-	&	-	&	-	&	125.4(2.70)	&	168.4(5.25)	&	\\
5787.965	&		&	3.323	&	$-$0.083	&	-	&	-	&	-	&	-	&	-	&	-	&	65.9(5.32)	&	\\
4634.07	&	Cr II	&	4.072	&	$-$1.24	&	-	&	25.9(3.87)	&	-	&	-	&	44.6(4.24)	&	-	&	-	&	\\
4812.337	&		&	3.864	&	$-$1.8	&	-	&	39.4(3.73)	&	-	&	-	&	-	&	-	&	-	&	\\
4848.25	&		&	3.864	&	$-$1.14	&	-	&	-	&	-	&	-	&	35.9(4.36)	&	-	&	-	&	\\
5305.853	&		&	3.827	&	$-$2.357	&	-	&	-	&	-	&	-	&	-	&	-	&	36.1(5.40)	&	\\
5308.404	&		&	4.072	&	$-$1.81	&	-	&	-	&	-	&	-	&	-	&	-	&	37.8(5.16)	&	\\
5334.869	&		&	4.073	&	$-$1.562	&	-	&	-	&	-	&	-	&	-	&	-	&	43.5(5.04)	&	\\
4470.472	&	Ni I	&	3.399	&	$-$0.31	&	30.8(4.19)	&	-	&	-	&	-	&	-	&	-	&	-	&	\\
4686.207	&		&	3.597	&	$-$0.64	&	-	&	-	&	-	&	-	&	-	&	44.7(4.59)	&	75.0(5.77)	&	\\
4703.803	&		&	3.658	&	$-$0.735	&	-	&	-	&	-	&	-	&	-	&	-	&	-	&	\\
4714.408	&		&	3.38	&	0.23	&	-	&	-	&	-	&	-	&	30.5(3.59)	&	-	&	-	&	\\
4731.793	&		&	3.833	&	$-$0.85	&	-	&	-	&	-	&	-	&	-	&	-	&	61.0(5.92)	&	\\
4752.415	&		&	3.658	&	$-$0.7	&	-	&	-	&	-	&	-	&	-	&	-	&	76.4(5.91)	&	\\
4756.51	&		&	3.48	&	$-$0.34	&	-	&	-	&	-	&	-	&	-	&	-	&	90.8(5.69)	&	\\
4980.166	&		&	3.606	&	$-$0.11	&	-	&	-	&	-	&	28.10(4.00)	&	-	&	-	&	-	&	\\
5035.357	&		&	3.635	&	0.29	&	78.9(4.21)	&	-	&	-	&	-	&	-	&	-	&	-	&	\\
5081.107	&		&	3.847	&	0.3	&	52.8(4.02)	&	-	&	-	&	38.7(4.23)	&	-	&	-	&	-	&	\\
5082.35	&		&	3.657	&	$-$0.54	&	-	&	-	&	-	&	32.2(4.06)	&	-	&	-	&	-	&	\\
5084.089	&		&	3.678	&	0.03	&	75.0(4.45)	&	-	&	15.6(3.92)	&	-	&	20.7(3.87)	&	-	&	-	&	\\
5099.927	&		&	3.678	&	$-$0.1	&	-	&	-	&	-	&	-	&	-	&	-	&	97.4(5.78)	&	\\
5102.96	&		&	1.676	&	$-$2.62	&	-	&	44.9(4.07)	&	-	&	-	&	-	&	89.0(4.45)	&	-	&	\\
5115.389	&		&	3.834	&	$-$0.11	&	-	&	-	&	11.7(4.09)	&	-	&	-	&	-	&	83.3(5.64)	&	\\
5146.48	&		&	3.706	&	0.12	&	-	&	55.4(4.15)	&	-	&	-	&	-	&	-	&	-	&	\\
6175.36	&		&	4.089	&	$-$0.53	&	-	&	-	&	-	&	-	&	-	&	-	&	52.0(5.64)	&	\\
6176.807	&		&	4.088	&	$-$0.26	&	-	&	-	&	-	&	-	&	-	&	-	&	70.5(6.01)	&	\\
6204.6	&		&	4.088	&	$-$1.13	&	-	&	-	&	-	&	-	&	-	&	-	&	31.3(5.82)	&	\\
6327.593	&		&	1.676	&	$-$3.15	&	-	&	-	&	-	&	-	&	-	&	66.8(4.34)	&	75.7(5.84)	&	\\
6378.247	&		&	4.154	&	$-$0.89	&	-	&	-	&	-	&	-	&	-	&	-	&	37.8(5.79)	&	\\
6643.629	&		&	1.676	&	$-$2.3	&	-	&	102.2(4.14)	&	14.6(3.89)	&	61.3(4.31)	&	34.8(3.98)	&	-	&	-	&	\\
4722.15	&	Zn I	&	4.029	&	$-$0.37	&	-	&	27.4(2.33)	&	-	&	-	&	29.9(2.50)	&	-	&	-	&	\\
4810.53	&		&	4.08	&	$-$0.17	&	48.3(3.00)	&	-	&	-	&	-	&	-	&	29.6(2.48)	&	64.1(3.48)	&	\\
4883.684	&	Y II	&	1.084	&	0.07	&	-	&	-	&	73.5(0.67)	&	-	&	-	&	-	&	169.0(2.97)	&	\\
5087.416	&		&	1.084	&	$-$0.17	&	-	&	-	&	-	&	-	&	-	&	-	&	164.9(3.12)	&	\\
5119.112	&		&	0.992	&	$-$1.36	&	-	&	-	&	-	&	-	&	-	&	89.2(1.55)	&	104.3(3.11)	&	\\
5200.406	&		&	0.992	&	$-$1.36	&	-	&	-	&	35.5(0.55)	&	-	&	-	&	-	&	-	&	\\
\hline
\end{tabular}}

The numbers in the parenthesis in columns 5 - 11 give the derived abundances from the respective line. \\
\end{table*}
}

{\footnotesize
\begin{table*}
\resizebox{\textwidth}{!}{\begin{tabular}{lccccccccccc}
\hline                       
Wavelength(\r{A}) & El & $E_{low}$(eV) & log gf & BD$-$19 132 & BD$-$19 290 & HE~1157$-$0518 & HE~1304$-$2111 & HE~1354$-$2257 & BD+19 3109 & HD~202851    \\ 
\hline 
5205.724	&		&	1.033	&	$-$0.34	&	-	&	-	&	41.8(0.48)	&	-	&	-	&	-	&	-	&	\\
5289.815	&		&	1.033	&	$-$1.85	&	-	&	-	&	-	&	-	&	-	&	44.4(1.23)	&	78.2(2.99)	&	\\
5402.774	&		&	1.839	&	$-$0.51	&	-	&	25.9(0.46)	&	-	&	-	&	-	&	-	&	-	&	\\
5544.611	&		&	1.738	&	$-$1.09	&	101.0(2.64)	&	-	&	-	&	14.1(0.83)	&	-	&	-	&	-	&	\\
5546.009	&		&	1.748	&	$-$1.11	&	-	&	-	&	-	&	-	&	-	&	-	&	-	&	\\
5662.925	&		&	1.944	&	0.16	&	156.8(2.64)	&	52.7(0.31)	&	22.7(0.61)	&	-	&	-	&	-	&	-	&	\\
6613.733	&		&	1.748	&	$-$1.1	&	123.5(2.85)	&	-	&	-	&	12.1(0.67)	&	-	&	54.5(1.47)	&	-	&	\\
4739.48	&	Zr I	&	0.651	&	0.23	&	-	&	8.3(0.95)	&	-	&	30.9(1.80)	&	-	&	-	&	-	&	\\
4772.323	&		&	0.623	&	0.04	&	-	&	8.3(1.10)	&	-	&	-	&	-	&	-	&	-	&	\\
6134.585	&		&	0	&	$-$1.28	&	-	&	-	&	-	&	19.2(1.75)	&	-	&	-	&	-	&	\\
4317.321	&	Zr II	&	0.713	&	$-$1.38	&	98.3(2.24)	&	-	&	-	&	-	&	-	&	53.7(1.28)	&	-	&	\\
5112.297	&		&	1.665	&	$-$0.59	&	88.0(2.34)	&	-	&	-	&	-	&	-	&	-	&	-	&	\\
4336.244	&	Ce II	&	0.704	&	$-$0.564	&	-	&	47.3(0.64)	&	-	&	-	&	-	&	-	&	-	&	\\
4364.653	&		&	0.495	&	$-$0.201	&	-	&	-	&	-	&	-	&	-	&	94.5(1.08)	&	-	&	\\
4407.273	&		&	0.701	&	$-$0.741	&	-	&	-	&	-	&	21.6(0.67)	&	-	&	-	&	-	&	\\
4483.893	&		&	0.864	&	0.01	&	-	&	-	&	38.5(0.99)	&	-	&	-	&	-	&	-	&	\\
4497.846	&		&	0.958	&	$-$0.349	&	-	&	-	&	-	&	-	&	48.2(1.11)	&	-	&	80.5(2.90)	&	\\
4508.079	&		&	0.621	&	$-$1.238	&	-	&	-	&	-	&	-	&	-	&	36.1(1.04)	&	57.4(2.58)	&	\\
4560.28	&		&	0.91	&	0	&	-	&	-	&	34.4(0.96)	&	-	&	-	&	-	&	-	&	\\
4562.359	&		&	0.477	&	0.081	&	-	&	-	&	62.9(0.87)	&	-	&	97.8(1.20)	&	-	&	103.5(2.30)	&	\\
4628.161	&		&	0.516	&	0.008	&	174.3(2.31)	&	-	&	-	&	-	&	-	&	105.1(0.95)	&	115.7(2.69)	&	\\
4725.069	&		&	0.521	&	$-$1.204	&	-	&	32.1(0.65)	&	-	&	-	&	-	&	40.6(0.87)	&	-	&	\\
4747.167	&		&	0.32	&	$-$1.246	&	-	&	-	&	-	&	-	&	-	&	53.7(0.86)	&	-	&	\\
4773.941	&		&	0.924	&	$-$0.498	&	-	&	-	&	-	&	-	&	-	&	42.3(0.77)	&	-	&	\\
4873.999	&		&	1.107	&	$-$0.892	&	-	&	-	&	-	&	12.5(0.84)	&	-	&	-	&	53.6(2.67)	&	\\
5187.458	&		&	1.211	&	$-$0.104	&	-	&	60.9(0.78)	&	-	&	-	&	58.5(1.25)	&	-	&	82.9(2.68)	&	\\
5274.229	&		&	1.044	&	$-$0.323	&	-	&	89.2(0.46)	&	62.2(1.19)	&	-	&	-	&	89.2(0.84)	&	-	&	\\
5330.556	&		&	0.869	&	$-$0.76	&	103.2(2.06)	&	-	&	-	&	-	&	39.8(1.12)	&	-	&	70.9(2.60)	&	\\
6034.205	&		&	1.458	&	$-$1.019	&	53.6(2.23)	&	-	&	-	&	-	&	-	&	-	&	35.2(2.74)	&	\\
5188.217	&	Pr II	&	0.922	&	$-$1.145	&	53.4(1.52)	&	-	&	-	&	-	&	-	&	-	&	36.0(2.17)	&	\\
5219.045	&		&	0.795	&	$-$0.24	&	112.1(1.47)	&	-	&	11.9(0.31)	&	21.3(-0.06)	&	24.0(0.06)	&	60.3(0.37)	&	-	&	\\
5259.728	&		&	0.633	&	$-$0.682	&	93.4(1.30)	&	-	&	-	&	-	&	-	&	-	&	72.7(2.15)	&	\\
5292.619	&		&	0.648	&	$-$0.3	&	-	&	53.1(-0.06)	&	10.8(0.15)	&	-	&	33.3(0.13)	&	54.5(0.10)	&	77.3(1.90)	&	\\
5322.772	&		&	0.482	&	$-$0.315	&	-	&	55.9(-0.23)	&	12.4(0.04)	&	-	&	49.5(0.23)	&	79.3(0.28)	&	86.9(1.94)	&	\\
6165.891	&		&	0.923	&	$-$0.205	&	-	&	-	&	8.70(0.21)	&	-	&	-	&	50.3(0.18)	&	-	&	\\
4446.384	&	Nd II	&	0.204	&	$-$0.59	&	-	&	-	&	49.4(0.84)	&	-	&	-	&	93.6(0.75)	&	-	&	\\
4451.563	&		&	0.38	&	$-$0.04	&	-	&	-	&	-	&	-	&	-	&	112.6(0.84)	&	119.8(2.52)	&	\\
4556.133	&		&	0.064	&	$-$1.61	&	-	&	-	&	-	&	-	&	-	&	58.1(0.76)	&	-	&	\\
4556.133	&		&	0.064	&	$-$1.61	&	123.1(2.14)	&	-	&	17.0(0.99)	&	-	&	-	&	-	&	-	&	\\
4703.572	&		&	0.38	&	$-$1.07	&	-	&	58.8(0.50)	&	-	&	-	&	-	&	-	&	-	&	\\
4706.543	&		&	0	&	$-$0.88	&	-	&	-	&	-	&	-	&	-	&	-	&	-	&	\\
4797.153	&		&	0.559	&	$-$0.95	&	-	&	-	&	-	&	29.1(0.67)	&	41.7(0.82)	&	-	&	-	&	\\
4811.342	&		&	0.064	&	$-$1.14	&	148.9(2.05)	&	90.7(0.50)	&	-	&	-	&	79.9(1.15)	&	85.3(0.68)	&	98.4(2.59)	&	\\
4825.478	&		&	0.182	&	$-$0.86	&	-	&	-	&	43.0(0.90)	&	-	&	-	&	-	&	-	&	\\
4859.039	&		&	0.32	&	$-$0.83	&	-	&	-	&	41.1(0.99)	&	-	&	63.6(0.80)	&	-	&	87.5(2.29)	&	\\
4947.02	&		&	0.559	&	$-$1.25	&	109.0(2.08)	&	28.9(0.40)	&	-	&	-	&	29.5(0.85)	&	56.0(0.96)	&	62.3(2.35)	&	\\
4961.387	&		&	0.631	&	$-$0.71	&	-	&	58.7(0.40)	&	39.0(1.18)	&	-	&	63.0(1.04)	&	-	&	-	&	\\
5130.59	&		&	1.3	&	0.1	&	-	&	-	&	37.0(1.09)	&	-	&	-	&	-	&	-	&	\\
5212.361	&		&	0.204	&	$-$0.87	&	-	&	123.9(0.70)	&	60.1(1.17)	&	-	&	87.5(1.13)	&	103.3(0.80)	&	99.6(2.41)	&	\\
5255.506	&		&	0.204	&	$-$0.82	&	-	&	-	&	41.9(0.82)	&	-	&	83.5(0.98)	&	105.9(0.78)	&	-	&	\\
5276.869	&		&	0.859	&	$-$0.44	&	-	&	-	&	22.4(0.82	&	-	&	-	&	65.8(0.67)	&	-	&	\\
5287.133	&		&	0.744	&	$-$1.3	&	101.5(2.17)	&	-	&	-	&	-	&	-	&	42.4(0.96)	&	49.5(2.31)	&	\\
5293.163	&		&	0.822	&	$-$0.06	&	-	&	-	&	46.9(0.86)	&	-	&	-	&	-	&	-	&	\\
5311.453	&		&	0.986	&	$-$0.42	&	-	&	-	&	-	&	28.2(0.71)	&	49.8(1.04)	&	58.1(0.84)	&	66.8(2.23)	&	\\
5319.815	&		&	0.55	&	$-$0.21	&	-	&	-	&	63.0(0.94)	&	-	&	-	&	-	&	-	&	\\
5356.967	&		&	1.264	&	$-$0.25	&	126.0(2.34)	&	-	&	-	&	-	&	30.90(2.51)	&	-	&	72.2(2.38)	&	\\
5361.51	&		&	0.68	&	$-$0.4	&	-	&	106.9(0.65)	&	51.3(1.10)	&	-	&	-	&	94.2(0.85)	&	-	&	\\
5371.927	&		&	1.412	&	0.003	&	-	&	-	&	-	&	-	&	-	&	-	&	-	&	\\
5442.264	&		&	0.68	&	$-$0.91	&	143.8(2.39)	&	-	&	28.1(1.19)	&	-	&	-	&	71.0(0.92)	&	-	&	\\
5718.118	&		&	1.41	&	$-$0.34	&	-	&	-	&	-	&	-	&	-	&	37.4(0.78)	&	-	&	\\
5825.857	&		&	1.08	&	$-$0.76	&	-	&	-	&	-	&	-	&	-	&	49.4(0.93)	&	59.3(2.33)	&	\\
4424.337	&	Sm II	&	0.485	&	$-$0.26	&	-	&	-	&	-	&	34.7(0.08)	&	-	&	-	&	-	&	\\
4434.318	&		&	0.378	&	$-$0.576	&	108.3(1.14)	&	52.8(-0.21)	&	-	&	-	&	-	&	80.1(0.50)	&	-	&	\\
4458.509	&		&	0.104	&	$-$1.11	&	-	&	44.3(-0.19)	&	-	&	28.0(0.05)	&	56.1(0.51)	&	67.7(0.34)	&	71.8(1.74)	&	\\
4499.475	&		&	0.248	&	$-$1.413	&	81.6(1.18)	&	-	&	16.6(0.73)	&	-	&	28.3(0.42)	&	51.2(0.52)	&	59.4(1.89)	&	\\
4519.63	&		&	0.543	&	$-$0.751	&	-	&	-	&	-	&	-	&	-	&	59.6(0.46)	&	-	&	\\
4615.444	&		&	0.544	&	$-$1.262	&	-	&	-	&	-	&	-	&	-	&	25.0(0.21)	&	53.9(1.94)	&	\\
4674.593	&		&	0.184	&	$-$1.055	&	-	&	-	&	-	&	-	&	-	&	66.3(0.28)	&	-	&	\\
4676.902	&		&	0.04	&	$-$1.407	&	94.2(1.04)	&	-	&	16.3(0.46)	&	-	&	-	&	-	&	-	&	\\
4704.4	&		&	0	&	$-$1.562	&	99.7(1.23)	&	-	&	-	&	-	&	-	&	-	&	-	&	\\
4726.026	&		&	0.333	&	$-$1.849	&	-	&	-	&	-	&	-	&	-	&	22.6(0.40)	&	-	&	\\
4791.58	&		&	0.104	&	$-$1.846	&	84.7(1.35)	&	-	&	-	&	-	&	25.6(0.57)	&	-	&	42.5(1.73)	&	\\
4815.805	&		&	0.185	&	$-$1.501	&	-	&	23.2(-0.17)	&	-	&	-	&	-	&	45.9(0.30)	&	-	&	\\
4844.209	&		&	0.277	&	$-$1.558	&	-	&	-	&	-	&	-	&	-	&	-	&	53.2(1.87)	&	\\
4854.368	&		&	0.379	&	$-$1.873	&	44.4(1.07)	&	-	&	-	&	-	&	-	&	-	&	29.1(1.79)	&	\\
\hline
\end{tabular}}

The numbers in the parenthesis in columns 5 - 11 give the derived abundances from the respective line. 
The line information are taken from Kurucz database of atomic line lists (\url{https://lweb.cfa.harvard.edu/amp/ampdata/kurucz23/sekur.html}).
\end{table*}
}

\end{document}